# $G_2$ Matrix Manifold: A Software Construct

## Version 1.0

Dara O Shayda
dara@lossofgenerality.com

July 2012

## Abstract

An ensemble of symbolic, numeric and graphic computations developed to construct the Octonionic and compact $G_2$ structures in *Mathematica* 8.0. Cayley-Dickenson Construction symbolically applied from Reals to Octonions. Baker–Campbell–Hausdorff formula (BCH) in bracket form verified for Octonions. Algorithms for both exponentiation and logarithm of Octonions developed. Exclusive validity of vector Product verified for $0, 1, 3$ and $7$ dimensions. Symbolic exponential computations carried out for two distinct $\mathbf{g}_2$ basis(s) and arbitrary precision BCH for $G_2$ was coded.

Example and counter-example Maximal Torus for $G_2$ was uncovered. Densely coiled shapes of actions of $G_2$ rendered. Kolmogorov Complexity for BCH investigated and upper bounds computed: *Complexity of non-commutative non-associative algebraic expression is at most the Complexity of corresponding commutative associative algebra plus K(BCH).*

# Contents

# Introduction

The data in laboratories ranging from the orbital telescopes to deciphering of the genome to the planetary sciences trumpet the fact that the universe is more and more software-like and less and less mathematical.

During the author's research for understanding the Octonions and their $G_2$ automorphism group and holonomy, it has become clear that these mathematical structures are software-like and supposed to be used in software-like fashion.

Octonionic mathematical descriptions are beyond the reach of most minds, no matter how intelligent and or well trained, and author proposes the teaching of Octonions to commence from high schools!

$G_2$ manifolds should be stable computational space of the undergraduates, and they should use software libraries and discuss algorithms for how such spaces and their Octonionic algebras describe the processes of the nature.

In order to start such a journey a small library of functions has been produced to empower computations as well as speedy grasp of the underlying theories. Learning by code examples is encouraged and actually works well. Many proofs are turned into actual symbolic computations to avoid maintaining difficult rigor without loss of discipline.

Reals, Complex numbers and Quaternions are treated as special cases of the Octonions. This makes it easier to learn the theory. However the construction from Reals to Octonions is symbolically carried in the code. However, arithmetic functions specific to different number types are provided.

As the code was being developed several new computational aspects were uncovered that might shed light on new areas of theory:

1. Kolmogorov Complexity was deployed for understanding how non-commutative and non-associative algebras are mapped to their corresponding commutative and associative algebras

2. Better understanding of vector product and their exclusive validity in dimensions 0, 1, 3, and 7. [8]

3. BCH bracket form for Octonions experimentally verified.

4. A Forgetful Representation for powers of Octonions has been uncovered to map them into Complex numbers for easier visualization and computations.

5. Shapes of actions of $G_2$ indicate how complex $G_2$ manifolds are and mathematical approach alone will not be able to shed light on their complexities.

Finally, any serious practical development of a quantum computers requires the intricate interplay of $G_2$ manifolds and Octonions, not just as theoretical thoughts, but as computational sturctures.

# 1. Software Setup

## How To Read This Document?

There are three entities in this document:

1. **Code**, which is up-to-date and runs well with no known bugs (huh!), if you have the *Mathematica* license you should just run the code and experiment with it, plug in numbers or new equations and variables and see how output changes. Run the code line by line make sure you understand what it does.
2. **Symbolic Calculations**: output from the code; browse them carefully to understand the theory, you can grab the symbolics and plug in numbers and plot new geometries and see how the symbolics tie-in with the geometry and computation
3. **Graphics**: Understand what the renderings convey

English and Mathematical expressions are slim and are so intentionally not to drown the reader into the abyss of theory and oblivion of imprecise language.

The output follows the code without any separation markings, but the relationship and separation is obvious.

The variable names and typeset of the equations are faithful to the original papers.

## Programmer's Manuals

For any and all *Mathematica* 8.0 help please use the following official link:

http://reference.wolfram.com/mathematica/guide/Mathematica.html

## License

All code and documentation unless specified otherwise are under the Apache 2.0 license:

http://www.apache.org/licenses/LICENSE-2.0.html

## SVN, Trac and Files

***SVN***
https://svn17.devzing.com/lossofgenerality/MatrixManifold   anonymous user * with no password

***Trac***
http://trac.devzing.com/lossofgenerality/MatrixManifold

***Files***
http://lossofgenerality.devzing.com/files/

## Package

MatrixManifolds.m is the *Mathematica* 8.0 Package provided for the $G_2$ computations. Load package as follows:

```
<< "Desktop/Groupoid/MatrixManifolds.m"
```

## Initialization

Currently two basis for Lie Algebra of **g**$_2$is provided, one from [1] :

```
(* 14 matrices forming a basis for the g2 are initialized, C1-C14 *)
gbasis = G2[];
gbasis[[14]] // MatrixForm
C14 // MatrixForm
```

$$\begin{pmatrix} 0 & 0 & 0 & 0 & 0 & 0 & -\frac{2}{\sqrt{3}} \\ 0 & 0 & 0 & \frac{1}{\sqrt{3}} & 0 & 0 & 0 \\ 0 & 0 & 0 & 0 & -\frac{1}{\sqrt{3}} & 0 & 0 \\ 0 & -\frac{1}{\sqrt{3}} & 0 & 0 & 0 & 0 & 0 \\ 0 & 0 & \frac{1}{\sqrt{3}} & 0 & 0 & 0 & 0 \\ 0 & 0 & 0 & 0 & 0 & 0 & 0 \\ \frac{2}{\sqrt{3}} & 0 & 0 & 0 & 0 & 0 & 0 \end{pmatrix}$$

$$\begin{pmatrix} 0 & 0 & 0 & 0 & 0 & 0 & -\frac{2}{\sqrt{3}} \\ 0 & 0 & 0 & \frac{1}{\sqrt{3}} & 0 & 0 & 0 \\ 0 & 0 & 0 & 0 & -\frac{1}{\sqrt{3}} & 0 & 0 \\ 0 & -\frac{1}{\sqrt{3}} & 0 & 0 & 0 & 0 & 0 \\ 0 & 0 & \frac{1}{\sqrt{3}} & 0 & 0 & 0 & 0 \\ 0 & 0 & 0 & 0 & 0 & 0 & 0 \\ \frac{2}{\sqrt{3}} & 0 & 0 & 0 & 0 & 0 & 0 \end{pmatrix}$$

Second basis from the PhD Thesis of Rarenas:

```
gbasis2 = G2Rarenas[];

(* two sets of 7 vectors *)
gbasis2[[1]][[4]] // MatrixForm
X4 // MatrixForm

gbasis2[[2]][[7]] // MatrixForm
Y7 // MatrixForm

(* or an index of all 14 vectors suitable for loops *)
gbasis2[[3]][[14]] // MatrixForm
```

$$\begin{pmatrix} 0 & 0 & 0 & 0 & -1 & 0 & 0 \\ 0 & 0 & 0 & 0 & 0 & 1 & 0 \\ 0 & 0 & 0 & 0 & 0 & 0 & 0 \\ 0 & 0 & 0 & 0 & 0 & 0 & 0 \\ 1 & 0 & 0 & 0 & 0 & 0 & 0 \\ 0 & -1 & 0 & 0 & 0 & 0 & 0 \\ 0 & 0 & 0 & 0 & 0 & 0 & 0 \end{pmatrix}$$

$$\begin{pmatrix} 0 & 0 & 0 & 0 & -1 & 0 & 0 \\ 0 & 0 & 0 & 0 & 0 & 1 & 0 \\ 0 & 0 & 0 & 0 & 0 & 0 & 0 \\ 0 & 0 & 0 & 0 & 0 & 0 & 0 \\ 1 & 0 & 0 & 0 & 0 & 0 & 0 \\ 0 & -1 & 0 & 0 & 0 & 0 & 0 \\ 0 & 0 & 0 & 0 & 0 & 0 & 0 \end{pmatrix}$$

$$\begin{pmatrix} 0 & 0 & 0 & 0 & 0 & 0 & 0 \\ 0 & 0 & 0 & 0 & -1 & 0 & 0 \\ 0 & 0 & 0 & 1 & 0 & 0 & 0 \\ 0 & 0 & -1 & 0 & 0 & 0 & 0 \\ 0 & 1 & 0 & 0 & 0 & 0 & 0 \\ 0 & 0 & 0 & 0 & 0 & 0 & 0 \\ 0 & 0 & 0 & 0 & 0 & 0 & 0 \end{pmatrix}$$

$$\begin{pmatrix} 0 & 0 & 0 & 0 & 0 & 0 & 0 \\ 0 & 0 & 0 & 0 & -1 & 0 & 0 \\ 0 & 0 & 0 & 1 & 0 & 0 & 0 \\ 0 & 0 & -1 & 0 & 0 & 0 & 0 \\ 0 & 1 & 0 & 0 & 0 & 0 & 0 \\ 0 & 0 & 0 & 0 & 0 & 0 & 0 \\ 0 & 0 & 0 & 0 & 0 & 0 & 0 \end{pmatrix}$$

$$\begin{pmatrix} 0 & 0 & 0 & 0 & 0 & 0 & 0 \\ 0 & 0 & 0 & 0 & -1 & 0 & 0 \\ 0 & 0 & 0 & 1 & 0 & 0 & 0 \\ 0 & 0 & -1 & 0 & 0 & 0 & 0 \\ 0 & 1 & 0 & 0 & 0 & 0 & 0 \\ 0 & 0 & 0 & 0 & 0 & 0 & 0 \\ 0 & 0 & 0 & 0 & 0 & 0 & 0 \end{pmatrix}$$

# 2. Octonions

Simplest and most correct way of thinking about Octonions are arrays of 8 Reals, their computable product is another array of 8 Reals with properties as follows.

**Remark**: *How the Octonions were found is of lesser importance, the idea is to learn how Octonions work as we learned how simple arithmetic worked way back in early schooling years. In other words, by lots of examples, understanding their properties through practice, and at the end one may endeavor to understand how they were theoretically derived.*

Calculations by hand are nearly impossible and time consuming for the experts, so the code below assists in the calculations to make room for practice and trial and error.

## Multiplication

```
A = {a1, a2, a3, a4, a5, a6, a7, a8};
B = {b1, b2, b3, b4, b5, b6, b7, b8};

OctMult[A, B]
{a1 b1 – a2 b2 – a3 b3 – a4 b4 – a5 b5 – a6 b6 – a7 b7 – a8 b8,
 a2 b1 + a1 b2 – a4 b3 + a3 b4 – a6 b5 + a5 b6 + a8 b7 – a7 b8,
 a3 b1 + a4 b2 + a1 b3 – a2 b4 – a7 b5 – a8 b6 + a5 b7 + a6 b8,
 a4 b1 – a3 b2 + a2 b3 + a1 b4 – a8 b5 + a7 b6 – a6 b7 + a5 b8,
 a5 b1 + a6 b2 + a7 b3 + a8 b4 + a1 b5 – a2 b6 – a3 b7 – a4 b8,
 a6 b1 – a5 b2 + a8 b3 – a7 b4 + a2 b5 + a1 b6 + a4 b7 – a3 b8,
 a7 b1 – a8 b2 – a5 b3 + a6 b4 + a3 b5 – a4 b6 + a1 b7 + a2 b8,
 a8 b1 + a7 b2 – a6 b3 – a5 b4 + a4 b5 + a3 b6 – a2 b7 + a1 b8}
```

Let's build a multiplication table:

```
(* e0 is +1 and e1 is the complex i *)
e0 = {1, 0, 0, 0, 0, 0, 0, 0};
e1 = {0, 1, 0, 0, 0, 0, 0, 0};
e2 = {0, 0, 1, 0, 0, 0, 0, 0};
e3 = {0, 0, 0, 1, 0, 0, 0, 0};
e4 = {0, 0, 0, 0, 1, 0, 0, 0};
e5 = {0, 0, 0, 0, 0, 1, 0, 0};
e6 = {0, 0, 0, 0, 0, 0, 1, 0};
e7 = {0, 0, 0, 0, 0, 0, 0, 1};
ei = {e0, e1, e2, e3, e4, e5, e6, e7};

(* generate the table by pairwise mutiplication *)
mults = Table[OctMult[ei[[i]], ei[[j]]], {i, 1, 8}, {j, 1, 8}];

(*map the symbols *)
mults = mults /. {e0 → "1", e1 → "e1", e2 → "e2", e3 → "e3", e4 → "e4",
    e5 → "e5", e6 → "e6", e7 → "e7", -e0 → "-1", -e1 → "-e1", -e2 → "-e2",
    -e3 → "-e3", -e4 → "-e4", -e5 → "-e5", -e6 → "-e6", -e7 → "-e7"};

(* plot a table *)
TableForm[mults, TableHeadings → {{"e0", "e1", "e2", "e3", "e4", "e5", "e6", "e7"},
    {"e0", "e1", "e2", "e3", "e4", "e5", "e6", "e7"}}]
```

|     | e0  | e1   | e2   | e3   | e4   | e5   | e6   | e7   |
| --- | --- | ---- | ---- | ---- | ---- | ---- | ---- | ---- |
| e0  | 1   | e1   | e2   | e3   | e4   | e5   | e6   | e7   |
| e1  | e1  | −1   | e3   | −e2  | e5   | −e4  | −e7  | e6   |
| e2  | e2  | −e3  | −1   | e1   | e6   | e7   | −e4  | −e5  |
| e3  | e3  | e2   | −e1  | −1   | e7   | −e6  | e5   | −e4  |
| e4  | e4  | −e5  | −e6  | −e7  | −1   | e1   | e2   | e3   |
| e5  | e5  | e4   | −e7  | e6   | −e1  | −1   | −e3  | e2   |
| e6  | e6  | e7   | e4   | −e5  | −e2  | e3   | −1   | −e1  |
| e7  | e7  | −e6  | e5   | e4   | −e3  | −e2  | e1   | −1   |

By looking at the table we can deduce: ei ej = −ej ei

```
OctMult[e3, e6]
OctMult[e6, e3]
```

{0, 0, 0, 0, 0, 1, 0, 0}

{0, 0, 0, 0, 0, −1, 0, 0}

By looking at the table we can also deduce: (ei ej) ek = −ei (ej ek)

```
OctMult[OctMult[e3, e5], e7]
OctMult[e3, OctMult[e5, e7]]
```

{0, 1, 0, 0, 0, 0, 0, 0}

{0, −1, 0, 0, 0, 0, 0, 0}

Real Multiplication

```
A = {a1, 0, 0, 0, 0, 0, 0, 0};
B = {b1, 0, 0, 0, 0, 0, 0, 0};

OctMult[A, B]
```
{a1 b1, 0, 0, 0, 0, 0, 0, 0}

```
A = {a1, 0, 0, 0, 0, 0, 0, 0};
B = {b1, b2, b3, b4, b5, b6, b7, b8};

OctMult[A, B]
```

{a1 b1, a1 b2, a1 b3, a1 b4, a1 b5, a1 b6, a1 b7, a1 b8}

Complex Multiplication
```
A = {a1, a2, 0, 0, 0, 0, 0, 0};
B = {b1, b2, 0, 0, 0, 0, 0, 0};

OctMult[A, B]
```
{a1 b1 – a2 b2, a2 b1 + a1 b2, 0, 0, 0, 0, 0, 0}

## 2.1 Non-Commutative

```
A = {a1, a2, a3, a4, a5, a6, a7, a8};
B = {b1, b2, b3, b4, b5, b6, b7, b8};
AB = OctMult[A, B]
BA = OctMult[B, A]

(* === is a conditional test to see if both sides are equal or not *)
AB === BA
```

```
{a1 b1 – a2 b2 – a3 b3 – a4 b4 – a5 b5 – a6 b6 – a7 b7 – a8 b8,
 a2 b1 + a1 b2 – a4 b3 + a3 b4 – a6 b5 + a5 b6 + a8 b7 – a7 b8,
 a3 b1 + a4 b2 + a1 b3 – a2 b4 – a7 b5 – a8 b6 + a5 b7 + a6 b8,
 a4 b1 – a3 b2 + a2 b3 + a1 b4 – a8 b5 + a7 b6 – a6 b7 + a5 b8,
 a5 b1 + a6 b2 + a7 b3 + a8 b4 + a1 b5 – a2 b6 – a3 b7 – a4 b8,
 a6 b1 – a5 b2 + a8 b3 – a7 b4 + a2 b5 + a1 b6 + a4 b7 – a3 b8,
 a7 b1 – a8 b2 – a5 b3 + a6 b4 + a3 b5 – a4 b6 + a1 b7 + a2 b8,
 a8 b1 + a7 b2 – a6 b3 – a5 b4 + a4 b5 + a3 b6 – a2 b7 + a1 b8}

{a1 b1 – a2 b2 – a3 b3 – a4 b4 – a5 b5 – a6 b6 – a7 b7 – a8 b8,
 a2 b1 + a1 b2 + a4 b3 – a3 b4 + a6 b5 – a5 b6 – a8 b7 + a7 b8,
 a3 b1 – a4 b2 + a1 b3 + a2 b4 + a7 b5 + a8 b6 – a5 b7 – a6 b8,
 a4 b1 + a3 b2 – a2 b3 + a1 b4 + a8 b5 – a7 b6 + a6 b7 – a5 b8,
 a5 b1 – a6 b2 – a7 b3 – a8 b4 + a1 b5 + a2 b6 + a3 b7 + a4 b8,
 a6 b1 + a5 b2 – a8 b3 + a7 b4 – a2 b5 + a1 b6 – a4 b7 + a3 b8,
 a7 b1 + a8 b2 + a5 b3 – a6 b4 – a3 b5 + a4 b6 + a1 b7 – a2 b8,
 a8 b1 – a7 b2 + a6 b3 + a5 b4 – a4 b5 – a3 b6 + a2 b7 + a1 b8}
```
False

## 2.2 Non-Associative

```
A = {a1, a2, a3, a4, a5, a6, a7, a8};
B = {b1, b2, b3, b4, b5, b6, b7, b8};
c = {c1, c2, c3, c4, c5, c6, c7, c8};

(*    (AB)C      *)
ABparanC = FullSimplify[OctMult[OctMult[A, B], c]];
(*    A(BC)      *)
AparanBC = FullSimplify[OctMult[A, OctMult[B, c]]];

(* === is a conditional test to see if both sides are equal or not *)
ABparanC === AparanBC
```

```
False
```

## 2.3 Commutator

Commutator measures how far from commutativity two Octonions are:

```
A = {a1, a2, a3, a4, a5, a6, a7, a8};
B = {b1, b2, b3, b4, b5, b6, b7, b8};
OctCommutator[A, B]
```

```
{0, - 2 a4 b3 + 2 a3 b4 - 2 a6 b5 + 2 a5 b6 + 2 a8 b7 - 2 a7 b8,
 2 a4 b2 - 2 a2 b4 - 2 a7 b5 - 2 a8 b6 + 2 a5 b7 + 2 a6 b8,
 - 2 a3 b2 + 2 a2 b3 - 2 a8 b5 + 2 a7 b6 - 2 a6 b7 + 2 a5 b8,
 2 a6 b2 + 2 a7 b3 + 2 a8 b4 - 2 a2 b6 - 2 a3 b7 - 2 a4 b8,
 - 2 a5 b2 + 2 a8 b3 - 2 a7 b4 + 2 a2 b5 + 2 a4 b7 - 2 a3 b8,
 - 2 a8 b2 - 2 a5 b3 + 2 a6 b4 + 2 a3 b5 - 2 a4 b6 + 2 a2 b8,
 2 a7 b2 - 2 a6 b3 - 2 a5 b4 + 2 a4 b5 + 2 a3 b6 - 2 a2 b7}
```

Real numbers commute

```
A = {a1, 0, 0, 0, 0, 0, 0, 0};
B = {b1, b2, 0, 0, 0, 0, 0, 0};
OctCommutator[A, B]
```

```
{0, 0, 0, 0, 0, 0, 0, 0}
```

Complex Numbers Commutator

```
A = {a1, a2, 0, 0, 0, 0, 0, 0};
B = {b1, b2, 0, 0, 0, 0, 0, 0};
OctCommutator[A, B]
```

```
{0, 0, 0, 0, 0, 0, 0, 0}
```

Quaternions as special case of Octonions

```
(* use Mathematica's package to check our code *)
<< Quaternions`

(* Quaternions are the special case of
  the Octonions with the last 4 numbers set to 0 *)
A = {a1, a2, a3, a4, 0, 0, 0, 0};
B = {b1, b2, b3, b4, 0, 0, 0, 0};

OctMult[A, B]
Quaternion[a1, a2, a3, a4] ** Quaternion[b1, b2, b3, b4]
```

{a1 b1 − a2 b2 − a3 b3 − a4 b4, a2 b1 + a1 b2 − a4 b3 + a3 b4,
 a3 b1 + a4 b2 + a1 b3 − a2 b4, a4 b1 − a3 b2 + a2 b3 + a1 b4, 0, 0, 0, 0}

Quaternion[a1 b1 − a2 b2 − a3 b3 − a4 b4, a2 b1 + a1 b2 − a4 b3 + a3 b4,
 a3 b1 + a4 b2 + a1 b3 − a2 b4, a4 b1 − a3 b2 + a2 b3 + a1 b4]

Quaternions do not commute

```
A = {a1, a2, a3, a4, 0, 0, 0, 0};
B = {b1, b2, b3, b4, 0, 0, 0, 0};

OctCommutator[A, B]
```

{0, −2 a4 b3 + 2 a3 b4, 2 a4 b2 − 2 a2 b4, −2 a3 b2 + 2 a2 b3, 0, 0, 0, 0}

## 2.4 Associator

Associator measures how far from associativity three Octonions are:

```
A = {a1, a2, a3, a4, a5, a6, a7, a8};
B = {b1, b2, b3, b4, b5, b6, b7, b8};
c = {c1, c2, c3, c4, c5, c6, c7, c8};

(* (AB)A − A(BA) *)
FullSimplify[OctAssociator[A, B, c]] === 0
(* === is a conditional test to see if both sides are equal or not *)
```

False

Some well-known cases for 0-associator

```
A = {a1, a2, a3, a4, a5, a6, a7, a8};
B = {b1, b2, b3, b4, b5, b6, b7, b8};
c = {c1, c2, c3, c4, c5, c6, c7, c8};

(* (AB)A − A(BA) *)
FullSimplify[OctAssociator[A, B, A]]
FullSimplify[OctAssociator[B, B, A]]
FullSimplify[OctAssociator[A, B, B]]
```

{0, 0, 0, 0, 0, 0, 0, 0}

{0, 0, 0, 0, 0, 0, 0, 0}

{0, 0, 0, 0, 0, 0, 0, 0}

Complex Numbers and Reals Associate

```
A = {a1, a2, 0, 0, 0, 0, 0, 0};
B = {b1, b2, 0, 0, 0, 0, 0, 0};
c = {c1, c2, 0, 0, 0, 0, 0, 0};

(* A(BC) = (AB)C  for complex numbers *)
FullSimplify[OctAssociator[A, B, c]]
```

```
{0, 0, 0, 0, 0, 0, 0, 0}
```

Quaternions Associate

```
A = {a1, a2, a3, a4, 0, 0, 0, 0};
B = {b1, b2, b3, b4, 0, 0, 0, 0};
c = {c1, c2, c3, c4, 0, 0, 0, 0};

FullSimplify[OctAssociator[A, B, c]]
```
```
{0, 0, 0, 0, 0, 0, 0, 0}
```

## 2.5 Cayley-Dickenson Construction

Instead of going though the skull numbing algebra of the mathematical construction proofs for the Octonions and Quaternions, rather check the important equations with the current definition of the Octonions in our code. In other words, start from Reals and construct Complex numbers, then try to DUPLICATE or REPEATE the same process to get Quaternions and Octonions:

**See Also**: *http://math.ucr.edu/home/baez/octonions/node5.html*

**Complex Numbers based upon Reals**

```
realA = {a1, 0, 0, 0, 0, 0, 0, 0};
realB = {b1, 0, 0, 0, 0, 0, 0, 0};
realC = {c1, 0, 0, 0, 0, 0, 0, 0};
realD = {d1, 0, 0, 0, 0, 0, 0, 0};
AB = {a1, b1, 0, 0, 0, 0, 0, 0};
CD = {c1, d1, 0, 0, 0, 0, 0, 0};
```

```
expr1 = realMult[realA, realC] - realMult[realConjugate[realD], realB]
expr2 = realMult[realD, realA] + realMult[realB, realConjugate[realC]]
```

{a1 c1 – b1 d1, 0, 0, 0, 0, 0, 0, 0}

{b1 c1 + a1 d1, 0, 0, 0, 0, 0, 0, 0}

```
(* Construct (AC-D*B, DA+BC*) by concatenating 1 Real from AC-
 D*B and 1 Real from DA+BC* *)
CayleyDickensonConstruct = {expr1[[1]], expr2[[1]], 0, 0, 0, 0, 0, 0}
```

```
realsMult = complexMult[AB, CD]
```

```
CayleyDickensonConstruct === realsMult
```

{a1 c1 – b1 d1, b1 c1 + a1 d1, 0, 0, 0, 0, 0, 0}

{a1 c1 – b1 d1, b1 c1 + a1 d1, 0, 0, 0, 0, 0, 0}

True

**Quaternions based upon Complex numbers**

1. Take two complex numbers A and B, form a tuple q1= (A, B), to represent a Quaternion
2. Take another two complex numbers C and D, form a tuple q2 = (A, B), to represent another Quaternion
3. Define multiplication of q1*q2 = (AC – D* B, DA + BC*) another Quaternion
4. Check and see if Octonion multiplication and addition will get the same results i.e. treat A, B, C and D as Quaternions with padded 0s

```
complexA = {a1, a2, 0, 0, 0, 0, 0, 0};
complexB = {b1, b2, 0, 0, 0, 0, 0, 0};
complexC = {c1, c2, 0, 0, 0, 0, 0, 0};
complexD = {d1, d2, 0, 0, 0, 0, 0, 0};
AB = {a1, a2, b1, b2, 0, 0, 0, 0};
CD = {c1, c2, d1, d2, 0, 0, 0, 0};
```

```
expr1 = complexMult[complexA, complexC] -
   complexMult[complexConjugate[complexD], complexB]
expr2 = complexMult[complexD, complexA +
   complexMult[complexB, complexConjugate[complexC]]
```

{a1 c1 − a2 c2 − b1 d1 − b2 d2, a2 c1 + a1 c2 − b2 d1 + b1 d2, 0, 0, 0, 0, 0, 0}

{b1 c1 + b2 c2 + a1 d1 − a2 d2, b2 c1 − b1 c2 + a2 d1 + a1 d2, 0, 0, 0, 0, 0, 0}

Compare the results and it is obvious the extension of the Complex numbers into Quaternions work:

```
(* Construct (AC-D*B, DA+BC*) by concatenating 2 Reals from AC-
 D*B and 2 Reals from DA+BC* *)
CayleyDickensonConstruct = {expr1[[1]], expr1[[2]], expr2[[1]], expr2[[2]], 0, 0, 0, 0}

QuaternionMult = quaternionMult[AB, CD]

CayleyDickensonConstruct === QuaternionMult
```

{a1 c1 − a2 c2 − b1 d1 − b2 d2, a2 c1 + a1 c2 − b2 d1 + b1 d2,
 b1 c1 + b2 c2 + a1 d1 − a2 d2, b2 c1 − b1 c2 + a2 d1 + a1 d2, 0, 0, 0, 0}

{a1 c1 − a2 c2 − b1 d1 − b2 d2, a2 c1 + a1 c2 − b2 d1 + b1 d2,
 b1 c1 + b2 c2 + a1 d1 − a2 d2, b2 c1 − b1 c2 + a2 d1 + a1 d2, 0, 0, 0, 0}

```
True
```

Let's define the Quaternion Conjugation as follows:

$(A, \ B)^* = (A^*, \ -B)$

Then multiply as Complex numbers, along the lines of the scheme above:

```
complexC = complexConjugate[complexA];
complexD = - complexB;
(* (A, B) *)
AB = {a1, a2, b1, b2, 0, 0, 0, 0};
(* (A*, -B) *)
ABconjugate = {complexC[[1]], complexC[[2]], complexD [[1]], complexD [[2]], 0, 0, 0, 0};

expr1 =
   complexMult[complexA, complexC] - complexMult[complexConjugate[complexD], complexB];
expr2 = complexMult[complexD, complexA] +
    complexMult[complexB, complexConjugate[complexC]];

(* Construct (AC-D*B, DA+BC*) by concatenating 2 Reals from AC-
 D*B and 2 Reals from DA+BC* *)
CayleyDickensonConstruct = {expr1[[1]], expr1[[2]], expr2[[1]], expr2[[2]], 0, 0, 0, 0}

QuaternionMult = quaternionMult[AB, ABconjugate]

CayleyDickensonConstruct === QuaternionMult
```

$\{a1^2 + a2^2 + b1^2 + b2^2, 0, 0, 0, 0, 0, 0, 0\}$

$\{a1^2 + a2^2 + b1^2 + b2^2, 0, 0, 0, 0, 0, 0, 0\}$

True

## Octonions based upon Quaternions

Let's REPEAT the Cayley-Dickenson PROCESS for Quaternions:

```
complexA = {a1, a2, a3, a4, 0, 0, 0, 0};
complexB = {b1, b2, b3, b4, 0, 0, 0, 0};
complexC = {c1, c2, c3, c4, 0, 0, 0, 0};
complexD = {d1, d2, d3, d4, 0, 0, 0, 0};
AB = {a1, a2, a3, a4, b1, b2, b3, b4};
CD = {c1, c2, c3, c4, d1, d2, d3, d4};

expr1 = OctMult[complexA, complexC] - OctMult[OctConjugate[complexD], complexB]
expr2 = OctMult[complexD, complexA] + OctMult[complexB, OctConjugate[complexC]]
```

$\{a1\,c1 - a2\,c2 - a3\,c3 - a4\,c4 - b1\,d1 - b2\,d2 - b3\,d3 - b4\,d4,$
$\ a2\,c1 + a1\,c2 - a4\,c3 + a3\,c4 - b2\,d1 + b1\,d2 + b4\,d3 - b3\,d4,$
$\ a3\,c1 + a4\,c2 + a1\,c3 - a2\,c4 - b3\,d1 - b4\,d2 + b1\,d3 + b2\,d4,$
$\ a4\,c1 - a3\,c2 + a2\,c3 + a1\,c4 - b4\,d1 + b3\,d2 - b2\,d3 + b1\,d4, 0, 0, 0, 0\}$

$\{b1\,c1 + b2\,c2 + b3\,c3 + b4\,c4 + a1\,d1 - a2\,d2 - a3\,d3 - a4\,d4,$
$\ b2\,c1 - b1\,c2 + b4\,c3 - b3\,c4 + a2\,d1 + a1\,d2 + a4\,d3 - a3\,d4,$
$\ b3\,c1 - b4\,c2 - b1\,c3 + b2\,c4 + a3\,d1 - a4\,d2 + a1\,d3 + a2\,d4,$
$\ b4\,c1 + b3\,c2 - b2\,c3 - b1\,c4 + a4\,d1 + a3\,d2 - a2\,d3 + a1\,d4, 0, 0, 0, 0\}$

```
(* Construct (AC-D*B, DA+BC*) by concatenating 4 Reals from AC-
 D*B and 4 Reals from DA+BC* *)
CayleyDickensonConstruct = {expr1[[1]], expr1[[2]], expr1[[3]],
   expr1[[4]], expr2[[1]], expr2[[2]], expr2[[3]], expr2[[4]]}

OctonionMult = OctMult[AB, CD]

CayleyDickenson Construct === OctonionMult
```

```
{a1 c1 − a2 c2 − a3 c3 − a4 c4 − b1 d1 − b2 d2 − b3 d3 − b4 d4,
 a2 c1 + a1 c2 − a4 c3 + a3 c4 − b2 d1 + b1 d2 + b4 d3 − b3 d4,
 a3 c1 + a4 c2 + a1 c3 − a2 c4 − b3 d1 − b4 d2 + b1 d3 + b2 d4,
 a4 c1 − a3 c2 + a2 c3 + a1 c4 − b4 d1 + b3 d2 − b2 d3 + b1 d4,
 b1 c1 + b2 c2 + b3 c3 + b4 c4 + a1 d1 − a2 d2 − a3 d3 − a4 d4,
 b2 c1 − b1 c2 + b4 c3 − b3 c4 + a2 d1 + a1 d2 + a4 d3 − a3 d4,
 b3 c1 − b4 c2 − b1 c3 + b2 c4 + a3 d1 − a4 d2 + a1 d3 + a2 d4,
 b4 c1 + b3 c2 − b2 c3 − b1 c4 + a4 d1 + a3 d2 − a2 d3 + a1 d4}

{a1 c1 − a2 c2 − a3 c3 − a4 c4 − b1 d1 − b2 d2 − b3 d3 − b4 d4,
 a2 c1 + a1 c2 − a4 c3 + a3 c4 − b2 d1 + b1 d2 + b4 d3 − b3 d4,
 a3 c1 + a4 c2 + a1 c3 − a2 c4 − b3 d1 − b4 d2 + b1 d3 + b2 d4,
 a4 c1 − a3 c2 + a2 c3 + a1 c4 − b4 d1 + b3 d2 − b2 d3 + b1 d4,
 b1 c1 + b2 c2 + b3 c3 + b4 c4 + a1 d1 − a2 d2 − a3 d3 − a4 d4,
 b2 c1 − b1 c2 + b4 c3 − b3 c4 + a2 d1 + a1 d2 + a4 d3 − a3 d4,
 b3 c1 − b4 c2 − b1 c3 + b2 c4 + a3 d1 − a4 d2 + a1 d3 + a2 d4,
 b4 c1 + b3 c2 − b2 c3 − b1 c4 + a4 d1 + a3 d2 − a2 d3 + a1 d4}

True
```

Let's define the Octonion Conjugation as follows:

$$(A, \ B)^* = (A^*, \ -B)$$

Then multiply as Quaternions, along the lines of the scheme above:

```
complexC = OctConjugate[complexA];
complexD = - complexB;

expr1 = OctMult[complexA, complexC] - OctMult[OctConjugate[complexD], complexB];
expr2 = OctMult[complexD, complexA] + OctMult[complexB, OctConjugate[complexC]];

(* Construct (AC-D*B, DA+BC*) by concatenating 4 Reals from AC-
 D*B and 4 Reals from DA+BC* *)
CayleyDickensonConstruct = {expr1[[1]], expr1[[2]], expr1[[3]],
   expr1[[4]], expr2[[1]], expr2[[2]], expr2[[3]], expr2[[4]]}

OctonionMult = OctMult[AB, OctConjugate[AB]]

CayleyDickensonConstruct === OctonionMult
```

$$\{a1^2 + a2^2 + a3^2 + a4^2 + b1^2 + b2^2 + b3^2 + b4^2, \ 0, \ 0, \ 0, \ 0, \ 0, \ 0, \ 0\}$$

$$\{a1^2 + a2^2 + a3^2 + a4^2 + b1^2 + b2^2 + b3^2 + b4^2, \ 0, \ 0, \ 0, \ 0, \ 0, \ 0, \ 0\}$$

```
True
```

## 2.6 Matrix Representation

```
z = {{a, -b}, {b, a}};
z // MatrixForm
```

$$\begin{pmatrix} a & -b \\ b & a \end{pmatrix}$$

z is in matrix form and the complex number in tuple/array form (c, d), we get the same outcome as the algebraic multiplication and addition and expansion:

```
z.{c, d}
ComplexExpand[(a + i * b) * (c + i * d)]
```

$\{a\,c - b\,d,\, b\,c + a\,d\}$

$a\,c - b\,d + i\,(b\,c + a\,d)$

Multiply two matrix representations:

```
z1 = {{a, -b}, {b, a}};
z2 = {{c, -d}, {d, c}};
```

```
z1.z2 // MatrixForm
z2.z1 // MatrixForm
```

$$\begin{pmatrix} a\,c - b\,d & -b\,c - a\,d \\ b\,c + a\,d & a\,c - b\,d \end{pmatrix}$$

$$\begin{pmatrix} a\,c - b\,d & -b\,c - a\,d \\ b\,c + a\,d & a\,c - b\,d \end{pmatrix}$$

The resultant matrix times 1 gives the same algebraic product of two complex numbers:

```
z = z1.z2;
```

```
z.{1, 0}
```

$\{a\,c - b\,d,\, b\,c + a\,d\}$

The following matrices give the representation for Left hand side multiplication for Quaternions:

```
quaternionMatrices[];
Lq1 // MatrixForm
Lq2 // MatrixForm
Lq3 // MatrixForm
Lq4 // MatrixForm
```

$$\begin{pmatrix} 1 & 0 & 0 & 0 \\ 0 & 1 & 0 & 0 \\ 0 & 0 & 1 & 0 \\ 0 & 0 & 0 & 1 \end{pmatrix}$$

$$\begin{pmatrix} 0 & -1 & 0 & 0 \\ 1 & 0 & 0 & 0 \\ 0 & 0 & 0 & -1 \\ 0 & 0 & 1 & 0 \end{pmatrix}$$

$$\begin{pmatrix} 0 & 0 & -1 & 0 \\ 0 & 0 & 0 & 1 \\ 1 & 0 & 0 & 0 \\ 0 & -1 & 0 & 0 \end{pmatrix}$$

$$\begin{pmatrix} 0 & 0 & 0 & -1 \\ 0 & 0 & -1 & 0 \\ 0 & 1 & 0 & 0 \\ 1 & 0 & 0 & 0 \end{pmatrix}$$

```
Clear[a1, a2, a3, a4];
z1 = a1 * Lq1 + a2 * Lq2 + a3 * Lq3 + a4 * Lq4;
z1 // MatrixForm
```

$$\begin{pmatrix} a1 & -a2 & -a3 & -a4 \\ a2 & a1 & -a4 & a3 \\ a3 & a4 & a1 & -a2 \\ a4 & -a3 & a2 & a1 \end{pmatrix}$$

```
z2 = b1 * Lq1 + b2 * Lq2 + b3 * Lq3 + b4 * Lq4;
z2 // MatrixForm
```

$$\begin{pmatrix} b1 & -b2 & -b3 & -b4 \\ b2 & b1 & -b4 & b3 \\ b3 & b4 & b1 & -b2 \\ b4 & -b3 & b2 & b1 \end{pmatrix}$$

Notice z1 is multiplied from left:

```
p1 = (z1.z2).{1, 0, 0, 0}
p2 = Drop[quaternionMult[{a1, a2, a3, a4}, {b1, b2, b3, b4}], -4]

p1 === p2
```

{a1 b1 – a2 b2 – a3 b3 – a4 b4, a2 b1 + a1 b2 – a4 b3 + a3 b4,
 a3 b1 + a4 b2 + a1 b3 – a2 b4, a4 b1 – a3 b2 + a2 b3 + a1 b4}

{a1 b1 – a2 b2 – a3 b3 – a4 b4, a2 b1 + a1 b2 – a4 b3 + a3 b4,
 a3 b1 + a4 b2 + a1 b3 – a2 b4, a4 b1 – a3 b2 + a2 b3 + a1 b4}

True

```
quaternionMatrices[];
Rq1 // MatrixForm
Rq2 // MatrixForm
Rq3 // MatrixForm
Rq4 // MatrixForm
```

$$\begin{pmatrix} 1 & 0 & 0 & 0 \\ 0 & 1 & 0 & 0 \\ 0 & 0 & 1 & 0 \\ 0 & 0 & 0 & 1 \end{pmatrix}$$

$$\begin{pmatrix} 0 & -1 & 0 & 0 \\ 1 & 0 & 0 & 0 \\ 0 & 0 & 0 & 1 \\ 0 & 0 & -1 & 0 \end{pmatrix}$$

$$\begin{pmatrix} 0 & 0 & -1 & 0 \\ 0 & 0 & 0 & -1 \\ 1 & 0 & 0 & 0 \\ 0 & 1 & 0 & 0 \end{pmatrix}$$

$$\begin{pmatrix} 0 & 0 & 0 & -1 \\ 0 & 0 & 1 & 0 \\ 0 & -1 & 0 & 0 \\ 1 & 0 & 0 & 0 \end{pmatrix}$$

```
Clear[a1, a2, a3, a4];
z1 = a1 * Rq1 + a2 * Rq2 + a3 * Rq3 + a4 * Rq4;
z1 // MatrixForm
```

$$\begin{pmatrix} a1 & -a2 & -a3 & -a4 \\ a2 & a1 & a4 & -a3 \\ a3 & -a4 & a1 & a2 \\ a4 & a3 & -a2 & a1 \end{pmatrix}$$

```
z2 = b1 * Rq1 + b2 * Rq2 + b3 * Rq3 + b4 * Rq4;
z2 // MatrixForm
```

$$\begin{pmatrix} b1 & -b2 & -b3 & -b4 \\ b2 & b1 & b4 & -b3 \\ b3 & -b4 & b1 & b2 \\ b4 & b3 & -b2 & b1 \end{pmatrix}$$

Notice that z1 is multiplied from right:

```
p1 = (z2.z1).{1, 0, 0, 0}
p2 = Drop[quaternionMult[{a1, a2, a3, a4}, {b1, b2, b3, b4}], -4]

p1 === p2
{a1 b1 - a2 b2 - a3 b3 - a4 b4, a2 b1 + a1 b2 - a4 b3 + a3 b4,
 a3 b1 + a4 b2 + a1 b3 - a2 b4, a4 b1 - a3 b2 + a2 b3 + a1 b4}

{a1 b1 - a2 b2 - a3 b3 - a4 b4, a2 b1 + a1 b2 - a4 b3 + a3 b4,
 a3 b1 + a4 b2 + a1 b3 - a2 b4, a4 b1 - a3 b2 + a2 b3 + a1 b4}

True
```

Also note that the right and hand left representation did not change the outcome of the multiplication.

Left matrix representation for Octonions [4] :

```
z = {a1, a2, a3, a4, a5, a6, a7, a8};
LOctMultMatrix[z] // MatrixForm
```

$$
\begin{pmatrix}
a1 & -a2 & -a3 & -a4 & -a5 & -a6 & -a7 & -a8 \\
a2 & a1 & -a4 & a3 & -a6 & a5 & a8 & -a7 \\
a3 & a4 & a1 & -a2 & -a7 & -a8 & a5 & a6 \\
a4 & -a3 & a2 & a1 & -a8 & a7 & -a6 & a5 \\
a5 & a6 & a7 & a8 & a1 & -a2 & -a3 & -a4 \\
a6 & -a5 & a8 & -a7 & a2 & a1 & a4 & -a3 \\
a7 & -a8 & -a5 & a6 & a3 & -a4 & a1 & a2 \\
a8 & a7 & -a6 & -a5 & a4 & a3 & -a2 & a1
\end{pmatrix}
$$

Right matrix representation for Octonions:

```
z = {a1, a2, a3, a4, a5, a6, a7, a8};
ROctMultMatrix[z] // MatrixForm
```

$$
\begin{pmatrix}
a1 & -a2 & -a3 & -a4 & -a5 & -a6 & -a7 & -a8 \\
a2 & a1 & a4 & -a3 & a6 & -a5 & -a8 & a7 \\
a3 & -a4 & a1 & a2 & a7 & a8 & -a5 & -a6 \\
a4 & a3 & -a2 & a1 & a8 & -a7 & a6 & -a5 \\
a5 & -a6 & -a7 & -a8 & a1 & a2 & a3 & a4 \\
a6 & a5 & -a8 & a7 & -a2 & a1 & -a4 & a3 \\
a7 & a8 & a5 & -a6 & -a3 & a4 & a1 & -a2 \\
a8 & -a7 & a6 & a5 & -a4 & -a3 & a2 & a1
\end{pmatrix}
$$

Let's verify by symbolic multiplication:

```
z1 = {a1, a2, a3, a4, a5, a6, a7, a8};
z2 = {b1, b2, b3, b4, b5, b6, b7, b8};
p1 = LOctMultMatrix[z1] .z2;
p2 = ROctMultMatrix[z2] .z1;

p1 === p2
```

```
True
```

## 2.7 Powers

```
(* imaginary part to power 2-8,
substitution (a2²+a3²+a4²+a5²+a6²+a7²+a8²)→u for abbreviation *)
A0 = {0, a2, a3, a4, a5, a6, a7, a8};
Simplify[OctPower[A0, 2]] /.
  {(a2² + a3² + a4² + a5² + a6² + a7² + a8²) → u, -(a2² + a3² + a4² + a5² + a6² + a7² + a8²) → -u}
Simplify[OctPower[A0, 3]] /. {(a2² + a3² + a4² + a5² + a6² + a7² + a8²) → u}
Simplify[OctPower[A0, 4]] /. {(a2² + a3² + a4² + a5² + a6² + a7² + a8²) → u}
Simplify[OctPower[A0, 5]] /. {(a2² + a3² + a4² + a5² + a6² + a7² + a8²) → u}
Simplify[OctPower[A0, 6]] /. {(a2² + a3² + a4² + a5² + a6² + a7² + a8²) → u}
Simplify[OctPower[A0, 7]] /. {(a2² + a3² + a4² + a5² + a6² + a7² + a8²) → u}
Simplify[OctPower[A0, 8]] /. {(a2² + a3² + a4² + a5² + a6² + a7² + a8²) → u}
```

$\{-u, 0, 0, 0, 0, 0, 0, 0\}$

$\{0, -a2\,u, -a3\,u, -a4\,u, -a5\,u, -a6\,u, -a7\,u, -a8\,u\}$

$\{u^2, 0, 0, 0, 0, 0, 0, 0\}$

$\{0, a2\,u^2, a3\,u^2, a4\,u^2, a5\,u^2, a6\,u^2, a7\,u^2, a8\,u^2\}$

$\{-u^3, 0, 0, 0, 0, 0, 0, 0\}$

$\{0, -a2\,u^3, -a3\,u^3, -a4\,u^3, -a5\,u^3, -a6\,u^3, -a7\,u^3, -a8\,u^3\}$

$\{u^4, 0, 0, 0, 0, 0, 0, 0\}$

```
(* A general Octonion to power 2-7,
substitution (a2²+a3²+a4²+a5²+a6²+a7²+a8²)→u for abbreviation  *)
A = {a1, a2, a3, a4, a5, a6, a7, a8};
Simplify[OctPower[A, 2]] /.
   {(a2² + a3² + a4² + a5² + a6² + a7² + a8²) → u, -(a2² + a3² + a4² + a5² + a6² + a7² + a8²) → -u}
Simplify[OctPower[A, 3]] /. {(a2² + a3² + a4² + a5² + a6² + a7² + a8²) → u}
Simplify[OctPower[A, 4]] /. {(a2² + a3² + a4² + a5² + a6² + a7² + a8²) → u}
Simplify[OctPower[A, 5]] /. {(a2² + a3² + a4² + a5² + a6² + a7² + a8²) → u}
Simplify[OctPower[A, 6]] /. {(a2² + a3² + a4² + a5² + a6² + a7² + a8²) → u}
Simplify[OctPower[A, 7]] /. {(a2² + a3² + a4² + a5² + a6² + a7² + a8²) → u}
```

$\{a1^2 - u,\ 2\ a1\ a2,\ 2\ a1\ a3,\ 2\ a1\ a4,\ 2\ a1\ a5,\ 2\ a1\ a6,\ 2\ a1\ a7,\ 2\ a1\ a8\}$

$\{a1\ (a1^2 - 3\ u),\ -a2\ (-3\ a1^2 + u),\ -a3\ (-3\ a1^2 + u),\ -a4\ (-3\ a1^2 + u),$
$\ -a5\ (-3\ a1^2 + u),\ -a6\ (-3\ a1^2 + u),\ -a7\ (-3\ a1^2 + u),\ -a8\ (-3\ a1^2 + u)\}$

$\{a1^4 - 6\ a1^2\ u + u^2,\ -4\ a1\ a2\ (-a1^2 + u),\ -4\ a1\ a3\ (-a1^2 + u),\ -4\ a1\ a4\ (-a1^2 + u),$
$\ -4\ a1\ a5\ (-a1^2 + u),\ -4\ a1\ a6\ (-a1^2 + u),\ -4\ a1\ a7\ (-a1^2 + u),\ -4\ a1\ a8\ (-a1^2 + u)\}$

$\{a1\ (a1^4 - 10\ a1^2\ u + 5\ u^2),\ a2\ (5\ a1^4 - 10\ a1^2\ u + u^2),$
$\ a3\ (5\ a1^4 - 10\ a1^2\ u + u^2),\ a4\ (5\ a1^4 - 10\ a1^2\ u + u^2),\ a5\ (5\ a1^4 - 10\ a1^2\ u + u^2),$
$\ a6\ (5\ a1^4 - 10\ a1^2\ u + u^2),\ a7\ (5\ a1^4 - 10\ a1^2\ u + u^2),\ a8\ (5\ a1^4 - 10\ a1^2\ u + u^2)\}$

$\{a1^6 - 15\ a1^4\ u + 15\ a1^2\ u^2 - u^3,\ 2\ a1\ a2\ (3\ a1^4 - 10\ a1^2\ u + 3\ u^2),$
$\ 2\ a1\ a3\ (3\ a1^4 - 10\ a1^2\ u + 3\ u^2),\ 2\ a1\ a4\ (3\ a1^4 - 10\ a1^2\ u + 3\ u^2),\ 2\ a1\ a5\ (3\ a1^4 - 10\ a1^2\ u + 3\ u^2),$
$\ 2\ a1\ a6\ (3\ a1^4 - 10\ a1^2\ u + 3\ u^2),\ 2\ a1\ a7\ (3\ a1^4 - 10\ a1^2\ u + 3\ u^2),\ 2\ a1\ a8\ (3\ a1^4 - 10\ a1^2\ u + 3\ u^2)\}$

$\{a1\ (a1^6 - 21\ a1^4\ u + 35\ a1^2\ u^2 - 7\ u^3),\ -a2\ (-7\ a1^6 + 35\ a1^4\ u - 21\ a1^2\ u^2 + u^3),$
$\ -a3\ (-7\ a1^6 + 35\ a1^4\ u - 21\ a1^2\ u^2 + u^3),\ -a4\ (-7\ a1^6 + 35\ a1^4\ u - 21\ a1^2\ u^2 + u^3),$
$\ -a5\ (-7\ a1^6 + 35\ a1^4\ u - 21\ a1^2\ u^2 + u^3),\ -a6\ (-7\ a1^6 + 35\ a1^4\ u - 21\ a1^2\ u^2 + u^3),$
$\ -a7\ (-7\ a1^6 + 35\ a1^4\ u - 21\ a1^2\ u^2 + u^3),\ -a8\ (-7\ a1^6 + 35\ a1^4\ u - 21\ a1^2\ u^2 + u^3)\}$

## 2.8 Exponentiation

Definition is the usual Taylor series expansion:

$$e^x = \sum_{n=0}^{\infty} \frac{x^n}{n!} \qquad x \in \mathbb{O} \qquad\qquad 2.8.1$$

This series has a closed form expression [5] :

$$e^x = e^{\mathrm{Re}\,x} \left( \mathrm{Cos}(\|\,\mathrm{Im}\,x\,\|) + \left(\frac{\mathrm{Im}\,x}{\|\mathrm{Im}\,x\|}\right) \mathrm{Sin}(\|\mathrm{Im}\,x\,\|) \right) \quad x \in \mathbb{O} \qquad\qquad 2.8.2$$

Set u = ‖Im x‖ and this is how exponentiation looks like in *Mathematica*:

```
(* √(Abs[o2]^2+Abs[o3]^2+Abs[o4]^2+Abs[o5]^2+Abs[o6]^2+Abs[o7]^2+Abs[o8]^2)→u *)
Clear[o, o1, o2, o3, o4, o5, o6, o7, o8];
o = {o1, o2, o3, o4, o5, o6, o7, o8};
Element[o, Reals];
OctExp[o] /.
  {√(Abs[o2]^2 + Abs[o3]^2 + Abs[o4]^2 + Abs[o5]^2 + Abs[o6]^2 + Abs[o7]^2 + Abs[o8]^2) → u,
   1 / √(Abs[o2]^2 + Abs[o3]^2 + Abs[o4]^2 + Abs[o5]^2 + Abs[o6]^2 + Abs[o7]^2 + Abs[o8]^2) → 1 / u}
```

$$\left\{ e^{o1} \text{Cos}[u], \ \frac{e^{o1} \, o2 \, \text{Sin}[u]}{u}, \ \frac{e^{o1} \, o3 \, \text{Sin}[u]}{u}, \ \frac{e^{o1} \, o4 \, \text{Sin}[u]}{u}, \right.$$
$$\left. \frac{e^{o1} \, o5 \, \text{Sin}[u]}{u}, \ \frac{e^{o1} \, o6 \, \text{Sin}[u]}{u}, \ \frac{e^{o1} \, o7 \, \text{Sin}[u]}{u}, \ \frac{e^{o1} \, o8 \, \text{Sin}[u]}{u} \right\}$$

Let's see if the Taylor's series expansion converges to the above formula:

```
x = {1, 2, 9, 4, 3, 6, 5, 4};

(* calculate the Taylor expansion up to n term *)
(* n is a bit large since x is not unit vector due to e^(Re x) factor *)

n = 34;

taylorseries = N[OctExpSeries[x, n]];

N[Norm[OctExp[x] - taylorseries]]
0.583791
```

Let's increase n:

```
n = 40;

taylorseries = N[OctExpSeries[x, n]];

N[Norm[OctExp[x] - taylorseries]]
0.00121418
```

And more, to see that Taylor's series approaches towards the Octonion exponential formula mentioned earlier:

```
n = 50;

taylorseries = N[OctExpSeries[x, n]];

N[Norm[OctExp[x] - taylorseries]]
6.22669 × 10^-9
```

## 2.9 Logarithm

Let's assume an Octonion y is represented by exponentiation

$$y = e^x \qquad y \neq \mathbf{0}, \ x \in \mathbb{O}$$

$$x = \ln(\|y\|) + \text{Im}\left(\frac{y}{\|y\|}\right) \left( \text{ArcCos}\left( \text{Re}\left(\frac{y}{\|y\|}\right) \right) + 2 \pi k \right) \Big/ \text{Sin}\left( \text{ArcCos}\left( \text{Re}\left( \left(\frac{y}{\|y\|}\right) \right) \right) \right) \qquad k \in \mathbb{Z}$$

$$y = e^x \qquad y \neq 0, \; x \in \mathbb{O} \qquad\qquad 2.9.1$$

$$x = \ln(\|y\|) + \text{Im}\left(\frac{y}{\|y\|}\right)\left(\text{ArcCos}\left(\text{Re}\left(\frac{y}{\|y\|}\right)\right) + 2\,\pi\,k\right)\Big/\text{Sin}\left(\text{ArcCos}\left(\text{Re}\left(\left(\frac{y}{\|y\|}\right)\right)\right)\right) \qquad k \in \mathbb{Z} \qquad 2.9.2$$

It is easy to verify for non-zero Octonions this is always possible and it is not unique, actually there are countably infinite such x.

```
Clear[y];

y = {-3, -12, 2, -100, 9, -6, 3, 13};

(* Normalize y *)
yu = y / Norm[y];

k = 30; (* degeneracy for ArcCos, has to be integer *)
ux = ArcCos[OctReal[yu]] + 2 * Pi * k;

(* Try to reverse the exponentiation formular *)
x = N[Prepend[OctIm7[yu] * ux / Sin[ux], Log[Norm[y]]]]

(* let's verify by exponentiating x we get y back *)
N[OctExp[x]]
```

```
{4.62727, -22.3224, 3.7204, -186.02, 16.7418, -11.1612, 5.5806, 24.1826}

{-3., -12., 2., -100., 9., -6., 3., 13.}
```

Let's make the code above into a function and pass the degenerate second argument as an integer:

```
x = OctLog[y, 10]
OctExp[x]
```
```
{4.62727, -7.56607, 1.26101, -63.0506, 5.67455, -3.78303, 1.89152, 8.19657}

{-3., -12., 2., -100., 9., -6., 3., 13.}
```

**FIXME 2.9** : *The formula above does not work well around 1 :*

```
y = {1, 0, 0, 0, 0, 0, 0, 0} + 0.00000000000001;
```

```
x = OctLog[y, 3001]
OctExp[x]
```

```
{9.99201 × 10⁻¹⁵, -526.045, -526.045, -526.045, -526.045, -526.045, -526.045, -526.045}

{-0.998277, 0.0221811, 0.0221811, 0.0221811, 0.0221811, 0.0221811, 0.0221811, 0.0221811}
```

## 2.10 Forgetful Representation

Octonions as in array of 8 Reals are fairly difficult and useless to compute with, instead if they are represented as the exponentiation of another Octonion then better uses and computations are possible.

And no matter which way to represent the Octonions the problem is their visualization and manipulations would require vectors in $\mathbb{R}^7$ or $\mathbb{R}^8$.

However it seems a Complex variables **Forgetful Map** might help in alternative representations and computations:

$$e^x \longrightarrow e^{\mathbf{Re}\,x}\, e^{i\,\|Im\,x\|} \qquad x \in \mathbb{O} \qquad 2.10.1$$

Basically the $\left(\frac{Im\,x}{\|Im\,x\|}\right)$ which is variable is mapped into the constant $i$ :

$$e^x \;=\; e^{\mathrm{Re}\,x}\left(\mathrm{Cos}(\|\,Im\,x\,\|) + \left(\frac{Im\,x}{\|Im\,x\|}\right)\mathrm{Sin}(\|Im\,x\|)\right) \longrightarrow e^{\mathrm{Re}\,x}\left(\mathrm{Cos}(\|\,Im\,x\,\|) + i\,\mathrm{Sin}(\|Im\,x\|)\right) \qquad 2.10.2$$

**Forgetful Representation Preserved Under $G_2$ Action**

Since the actions of $G_2$ preserve the norm and the Real part of the Octonions from the above Forgetful Map:

$$y = g\,x \Longrightarrow e^{\mathrm{Re}\,y}\left(\mathrm{Cos}(\|\,Im\,y\,\|) + i\,\mathrm{Sin}(\|Im\,y\|)\right) \;=\; e^{\mathrm{Re}\,x}\left(\mathrm{Cos}(\|\,Im\,x\,\|) + i\,\mathrm{Sin}(\|Im\,x\|)\right) \qquad 2.10.3$$

Let's verify:

```
x = {-3, -12, 2, -100, 9, -6, 3, 13};
forgetfulrepx = N[Exp[x[[1]]] * (Cos[Norm[OctIm7[x]]] + i * Sin[Norm[OctIm7[x]]]])]

g = MatrixExp[2 * C4 + 4 * C14];

y = N[actionG2[g, x]];

forgetfulrepgx = N[Exp[y[[1]]] * (Cos[Norm[OctIm7[y]]] + i * Sin[Norm[OctIm7[y]]]])]
```

$-0.00443692 + 0.049589\,i$

$-0.00443692 + 0.049589\,i$

Let's turn the latter into a function call:

```
N[OctForget[x]]
OctForget[N[actionG2[g, x]]]
```

$-0.00443692 + 0.049589\,i$

$-0.00443692 + 0.049589\,i$

### 2.11 BCH Formula for Exp: $\mathbb{O} \longrightarrow \mathbb{O}$

In general

$e^x \, e^y \neq e^{x+y} \quad x, y \in \mathbb{O}$      2.11.1

 However an additional series of terms can set the equality, these series of terms are called **Baker–Campbell–Hausdorff** (BCH) formula [6].

```
Clear[o1, o2, o3, o4, o5, o6, o7, o8, O1,
  O2, O3, O4, O5, O6, O7, O8, oo1, oo2, exp1, exp2];
oo1 = {o1, o2, o3, o4, o5, o6, o7, o8};
oo2 = {O1, O2, O3, O4, O5, O6, O7, O8};
Element[oo1, Reals];
Element[oo2, Reals];

exp1 = OctExp[oo1 + oo2];
exp2 = OctMult[OctExp[oo1], OctExp[oo2]];

(* === is a conditional test to see if both sides are equal or not *)
exp1 - exp2 === 0
False
```

Let's look at one element of the Octonion exponentiation and multiplications to see they are not equal:

```
exp1[[4]] /.
  {√(Abs[o2 + O2]² + Abs[o3 + O3]² + Abs[o4 + O4]² + Abs[o5 + O5]² + Abs[o6 + O6]² + Abs[o7 + O7]² +
      Abs[o8 + O8]²) → uoo12, 1 / √(Abs[o2 + O2]² + Abs[o3 + O3]² + Abs[o4 + O4]² +
        Abs[o5 + O5]² + Abs[o6 + O6]² + Abs[o7 + O7]² + Abs[o8 + O8]²) → 1 / uoo12}
exp2[[4]] /. {√(Abs[O2]² + Abs[O3]² + Abs[O4]² + Abs[O5]² + Abs[O6]² + Abs[O7]² + Abs[O8]²) →
    uoo2, 1 / √(Abs[O2]² + Abs[O3]² + Abs[O4]² + Abs[O5]² + Abs[O6]² + Abs[O7]² + Abs[O8]²) → 1 /
      uoo2, √(Abs[o2]² + Abs[o3]² + Abs[o4]² + Abs[o5]² + Abs[o6]² + Abs[o7]² + Abs[o8]²) → uoo1,
  1 / √(Abs[o2]² + Abs[o3]² + Abs[o4]² + Abs[o5]² + Abs[o6]² + Abs[o7]² + Abs[o8]²) → 1 / uoo1}
```

$$\frac{e^{o1+O1} \, (o4 + O4) \, \text{Sin}[uoo12]}{uoo12}$$

$$\frac{e^{o1+O1} \, o4 \, \text{Cos}[uoo2] \, \text{Sin}[uoo1]}{uoo1} + \frac{e^{o1+O1} \, O4 \, \text{Cos}[uoo1] \, \text{Sin}[uoo2]}{uoo2} -$$

$$\frac{e^{o1+O1} \, O2 \, o3 \, \text{Sin}[uoo1] \, \text{Sin}[uoo2]}{uoo1 \, uoo2} + \frac{e^{o1+O1} \, o2 \, O3 \, \text{Sin}[uoo1] \, \text{Sin}[uoo2]}{uoo1 \, uoo2} +$$

$$\frac{e^{o1+O1} \, O6 \, o7 \, \text{Sin}[uoo1] \, \text{Sin}[uoo2]}{uoo1 \, uoo2} - \frac{e^{o1+O1} \, o6 \, O7 \, \text{Sin}[uoo1] \, \text{Sin}[uoo2]}{uoo1 \, uoo2} -$$

$$\frac{e^{o1+O1} \, O5 \, o8 \, \text{Sin}[uoo1] \, \text{Sin}[uoo2]}{uoo1 \, uoo2} + \frac{e^{o1+O1} \, o5 \, O8 \, \text{Sin}[uoo1] \, \text{Sin}[uoo2]}{uoo1 \, uoo2}$$

Let's so sample numerical calculations, as the numbers show

$e^{xO} \, e^{yO} \neq e^{xO + yO}$

$e^{x0} e^{y0} \neq e^{x0 + y0}$      2.11.2

```
xO = {10, 2, 3, 1, 3, 9, 19, 6};
yO = {90, 3, 1, 10, 30, 12, 13, 10};

(*Normalize or else the accuracy for BCH
 drops due to the exponential of the real part *)
xO = xO / Norm[xO];
yO = yO / Norm[yO];

N[OctExp[(xO + yO)]]
N[OctMult[OctExp[xO], OctExp[yO]]]
```

{1.39813, 0.331078, 0.390902, 0.422278, 1.26683, 1.44455, 2.67708, 1.02344}

{1.44118, −0.106594, −0.313587, 0.29343, 1.10879, 1.19137, 2.87643, 1.05158}

However the addition of BCH term makes the exponentiation equal i.e.

$e^{x0} e^{y0} = e^{x0 + y0 + BCH[x0, y0]}$      2.11.3

**TODO**: *The BCH formula used is the usual bracket one but we need a formal proof to make sure this BCH correctly holds the equality. Currently the author made a guess that it does and tested it with numerous cases and it seems to hold fine.*

```
N[OctExp[(xO + yO) + OctBCH[xO, yO]]]
N[OctMult[OctExp[xO], OctExp[yO]]]
```

{1.43759, −0.10386, −0.309476, 0.294924, 1.11196, 1.1937, 2.87621, 1.05216}

{1.44118, −0.106594, −0.313587, 0.29343, 1.10879, 1.19137, 2.87643, 1.05158}

## Special Case of BCH [x, y] = 0 : Complex Numbers

Exponentiation of the Octonions agrees with the special case exponentiation of the Complex numbers, in other words the Octonion exponentiation is the generalization of the de Moivre's formula:

```
Clear[oo1, oo2, o1, o2, O1, O2];
oo1 = {o1, o2, 0, 0, 0, 0, 0, 0};
oo2 = {O1, O2, 0, 0, 0, 0, 0, 0};
Element[oo1, Reals];
Element[oo2, Reals];
OctBCH[oo1, oo2]
exp1 = OctExp[oo1 + oo2]
exp2 = OctMult[OctExp[oo1], OctExp[oo2]]
```

$\{0., 0., 0., 0., 0., 0., 0., 0.\}$

$$\left\{ e^{o1+O1} \, \text{Cos}[\text{Abs}[o2 + O2]], \; \frac{e^{o1+O1} \, (o2 + O2) \, \text{Sin}[\text{Abs}[o2 + O2]]}{\text{Abs}[o2 + O2]}, \; 0, 0, 0, 0, 0, 0 \right\}$$

$$\left\{ e^{o1+O1} \, \text{Cos}[\text{Abs}[o2]] \, \text{Cos}[\text{Abs}[O2]] - \frac{e^{o1+O1} \, o2 \, O2 \, \text{Sin}[\text{Abs}[o2]] \, \text{Sin}[\text{Abs}[O2]]}{\text{Abs}[o2] \, \text{Abs}[O2]}, \right.$$
$$\left. \frac{e^{o1+O1} \, o2 \, \text{Cos}[\text{Abs}[O2]] \, \text{Sin}[\text{Abs}[o2]]}{\text{Abs}[o2]} + \frac{e^{o1+O1} \, O2 \, \text{Cos}[\text{Abs}[o2]] \, \text{Sin}[\text{Abs}[O2]]}{\text{Abs}[O2]}, \; 0, 0, 0, 0, 0, 0 \right\}$$

Let's verify if the outputs are the same:

```
Clear[o2, O2, g]
Element[o2, Reals];
Element[O2, Reals];
```

$$g = \text{TrigFactor}\left[ \text{Sin}[o2 + O2] - \frac{(o2 + O2) \, \text{Sin}[\text{Abs}[o2 + O2]]}{\text{Abs}[o2 + O2]} \right];$$

```
Reduce[g == 0, {o2, O2}, Reals]
```

$o2 + O2 \neq 0$

```
Clear[o2, O2, g]
Element[o2, Reals];
Element[O2, Reals];
```

$$g = \text{TrigFactor}\left[ \text{Cos}[o2 + O2] - \left( \text{Cos}[\text{Abs}[o2]] \, \text{Cos}[\text{Abs}[O2]] - \frac{o2 \, O2 \, \text{Sin}[\text{Abs}[o2]] \, \text{Sin}[\text{Abs}[O2]]}{\text{Abs}[o2] \, \text{Abs}[O2]} \right) \right];$$

```
Reduce[g == 0, {o2, O2}, Reals]
```

$o2 \neq 0 \, \&\& \, O2 \neq 0$

```
Clear[o2, O2, g]
Element[o2, Reals];
Element[O2, Reals];
```

$$g = \text{TrigFactor}\left[ \text{Sin}[o2 + O2] - \left( \frac{o2 \, \text{Cos}[\text{Abs}[O2]] \, \text{Sin}[\text{Abs}[o2]]}{\text{Abs}[o2]} + \frac{O2 \, \text{Cos}[\text{Abs}[o2]] \, \text{Sin}[\text{Abs}[O2]]}{\text{Abs}[O2]} \right) \right];$$

```
Reduce[g == 0, {o2, O2}, Reals]
```

$o2 \neq 0 \, \&\& \, O2 \neq 0$

# 3. $G_2$ Structure

**Remark 3.1**: *By Structure we mean that which makes the spaces similar*.

A theorem of Berger in 1955 [9,10]:

**Theorem**: *Let M be a simply-connected, n-dimensional manifold, and g an irreducible, non-symmetric Riemannian metric on M. Then either*

(i)   *Hol(g) = **SO**(n)*,
(ii)  *n = 2m and Hol(g) = **SU**(m) or **U**(m)*,
(iii) *n = 4m and Hol(g) = **Sp**(m) or **Sp**(m)**Sp**(1)*,
(iv)  *n = 7 and Hol(g) = **$G_2$**, or*
(v)   *n = 8 and Hol(g) = **Spin**(7)*.

Moreover the Categorial proof from [8] :

**Proposition 3**: *Any d-dimensional finite-dimensional vector space with a an inner product admitting vector product algebra has the following relationship:*

$$d\,(d-1)\,(d-3)\,(d-7) \;=\; 0. \qquad\qquad 3.1$$

In other words its dimensions are bound to 0, 1, 3 and 7.

Summarizing the two results:

**The most general Riemannian manifold admitting vector product (on its tangent spaces) has to be at most 7-dimensional and for that matter its holonomy group ought to be $G_2$.**

In particular the significance of this combination is that in order to have an electromagnetic theory e.g. describing light, the underlying space has to be a 7-dimensional manifold with $G_2$ holonomy group [7]. This is so due to the Maxwell-Heaveside-Lorenz equation:

$$F \;=\; q(E \,+\, v{\times}B) \qquad\qquad 3.2$$

'×' is the vector product. This version of the equation or the others all use the vector product, therefore the vector product algebra has to be in the core of the theory of electricity and magnetism and thus the above said $G_2$ structure a necessity.

FIG 3.1

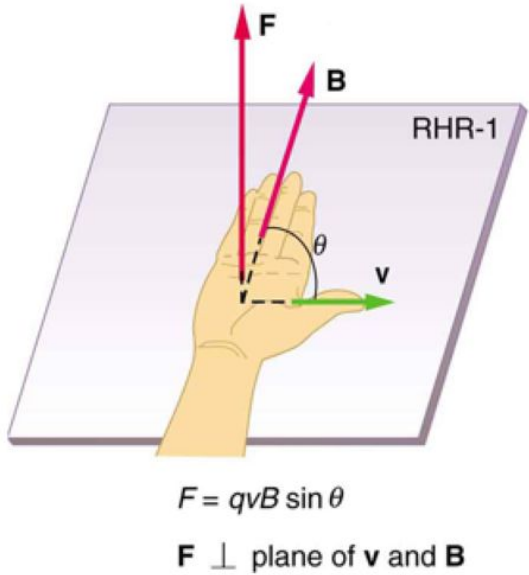

$$F = qvB \sin\theta$$

**F** ⊥ plane of **v** and **B**

**Remark**: *The vector product algebra as such also provides another non-electro-magnetic concept: Handedness! In other words the vector product algebra properties define a computational structure for the concepts of Right-hand vs. Left-hand. Again the Berger theorem indicates that Handedness i.e. can only happen in spaces of dimensions 0, 1, 3, 7.*

$G_2$ acts on Octonions i.e. it is used as a left-multiplicand:

```
gexp =.;
(* grab any Octonion, simply make an array of 8 Reals *)
xO = {10, 220, 340, 1, 0, 98, 190, 60}
```

{10, 220, 340, 1, 0, 98, 190, 60}

$G_2$ members are represented by 7x7 matrices of Reals, usually obtained by exponentiation of vectors from the Lie Algebra $\mathfrak{g}_2$:

```
(* grab any element of G2,
simply make a linear combination of basis vectors C1-C14 and calculate
  the EXP for it which is the 7x7 matrix exponential *)
gexp = N[MatrixExp[(0.1 * C1 + 2.3 * C11 + Pi * C14)]];
gexp // MatrixForm
```

$$\begin{pmatrix} -0.21375 & 0.0375759 & 0. & 0.586461 & 0. & 0. & 0.780361 \\ -0.0375759 & -0.625896 & 0. & 0.632147 & 0. & 0. & -0.455229 \\ 0. & 0. & -0.625079 & 0. & -0.584403 & 0.517445 & 0. \\ -0.566736 & -0.624927 & 0. & -0.481804 & 0. & 0. & 0.236943 \\ 0. & 0. & 0.669813 & 0. & -0.0612176 & 0.740002 & 0. \\ 0. & 0. & -0.400783 & 0. & 0.809151 & 0.429707 & 0. \\ -0.794802 & 0.465091 & 0. & 0.155946 & 0. & 0. & -0.357298 \end{pmatrix}$$

The $G_2$ action simply leaves the Real part of the Octonion unaltered and uses only the Imaginary part, for example the Real part being 10 is preserved:

*G2*

```
actionG2[gexp, xO]
```

```
{10, 12.5724, -248.385, 40.4178, -322.94, 135.271, 160.54, -38.1633}
```

This is the action code from the MatrixManifold Package, see the simplicity of the matrix multiplication from left as the core computation for such action.

**Remark 3.2**: *Do not execute this code!*

```
actionG2[g0_, x0_] := Module[{x = x0, g = g0, b},
```

```
(* Simply mulitpl the G2 element as if a 7x7 matrice, and do so from left *)
b = g.OctIm7[x];
```

```
(* take the 7 numbers from that multiplication and concatenate the original Real part
of the Octonion *)
PrependTo[b, OctReal[x]]]
```

# 3.1 Octonion Multiplication Preserved

$$g(x\,y) = g(x)\,g(y) \qquad x, y \in \mathbb{O},\ \boldsymbol{g} \in G_2 \qquad\qquad 3.1.1$$

***actionG2[g, x]*** is g(x).

```
gexp = .;
(* grab any two Octonions, simply make an array of 8 Reals *)
xO = {10, 220, 340, 1, 0, 98, 190, 60};
yO = {20, 0, 100, 10, 30, 120, 130, 1};
```

```
(* grab any element of G2,
simply make a linear combination of basis vectors C1-C14 and calculate
  the EXP for it which is the 7x7 matrix exponential *)
gexp = N[MatrixExp[(0.1 * C1 + 2.3 * C11 + Pi * C14)]];
```

```
actionG2[gexp, OctMult[xO, yO]]
OctMult[actionG2[gexp, xO], actionG2[gexp, yO]]
```

```
{-70 330, -29 808.3, -29 664.6, -18 491.3, 22 780.6, 31 634.5, 5907.71, -22 388.5}
```

```
{-70 330., -29 808.3, -29 664.6, -18 491.3, 22 780.6, 31 634.5, 5907.71, -22 388.5}
```

# 3.2 Norm Preserved

Action of $G_2$ on Octonions preserves length

$$\|g(x\,y)\| = \|g(x)\|\,\|g(y)\| \qquad x, y \in \mathbb{O},\ \boldsymbol{g} \in G_2 \qquad\qquad 3.2.1$$

$$g\,(x)$$

$G_2$

$$\|g(x\,y)\| = \|g(x)\|\,\|g(y)\| \qquad x, y \in \mathbb{O}, \ \ g \in G_2$$

**Remark 1.1**: *'Action' means for example* $g(x)$ *and in this case it is nothing more than the left multiplication of a 7×7 matrix times the Im part of the Octonion and* **Re** *part left unchanged.*

```
gexp =.;
(* grab any two Octonions, simply make an array of 8 Reals *)
xO = {10, 220, 340, 1, 0, 98, 190, 60};
yO = {20, 0, 100, 10, 30, 120, 130, 1};

(* grab any element of G2,
simply make a linear combination of basis vectors C1-C14 and calculate
  the EXP for it which is the 7x7 matrix exponential *)
gexp = N[MatrixExp[(0.1 * C1 + 2.3 * C11 + Pi * C14)]];

Norm[actionG2[gexp, OctMult[xO, yO]]]
Norm[OctMult[actionG2[gexp, xO], actionG2[gexp, yO]]]
```

95 460.

95 460.

Lets do the same computation using the second $\mathbf{g}_2$ basis [3]:

```
gexp =.;
(* grab any two Octonions, simply make an array of 8 Reals *)
xO = {10, 220, 340, 1, 0, 98, 190, 60};
yO = {20, 0, 100, 10, 30, 120, 130, 1};

(* grab any element of G2,
simply make a linear combination of basis vectors X1-X7 and Y1-Y7 and calculate
  the EXP for it which is the 7x7 matrix exponential *)
gexp = N[MatrixExp[(0.1 * Y2 + 4 * X2 + 8 * X7 + 0.001 * Y6)]];

Norm[actionG2[gexp, OctMult[xO, yO]]]
Norm[OctMult[actionG2[gexp, xO], actionG2[gexp, yO]]]
```

95 460.

95 460.

# 3.3 Vector Product Preserved

In general SO(7) does not preserve the vector product:

```
so7basis = SO[7];

(* this vector does not preserve the vector product *)
v = so7basis[[1]] + so7basis[[16]];

xO = {50, 220, -30, 1.02, 0.1, -8, -90, 160};
yO = {20, -0.9, 600, 10, -20, 120, 130, -0.01};

gexp = MatrixExp[v];

actionG2[gexp, OctCommutator[xO, yO] / 2]

(* Apply the action two each Octonion and then compute the corss product *)
OctCommutator[actionG2[gexp, xO], actionG2[gexp, yO]] / 2
```

{0., 38 369.9, 11 923.4, 125 413., -74 892.8, 92 632.4, 479.4, 7770.83}

{0., 10 983.8, -30 674.5, 4675.5, -27 118.3, 109 253., -121 217., 56 076.}

However some members of its subspaces do preserve the vector product:

```
so7basis = SO[7];

(* this vector preserves the corss product *)
v = so7basis[[1]] + so7basis[[15]];

xO = {50, 220, -30, 1.02, 0.1, -8, -90, 160};
yO = {20, -0.9, 600, 10, -20, 120, 130, -0.01};

gexp = MatrixExp[v];

actionG2[gexp, OctCommutator[xO, yO] / 2]

(* Apply the action two each Octonion and then compute the corss product *)
OctCommutator[actionG2[gexp, xO], actionG2[gexp, yO]] / 2
```

{0., 30 177., 4081.44, 125 413., -74 892.8, 49 646.1, 78 206.5, -27 340.4}

{0., 30 177., 4081.44, 125 413., -74 892.8, 49 646.1, 78 206.5, -27 340.4}

In specific, action of $G_2$ on Octonions preserves vector product

$x \times y = \frac{1}{2}[x, \ y] = \frac{1}{2}(x \ y - y \ x)$     $x, y \in \mathbb{O}$     3.3.1

$g(x \times y) = g(x) \times g(y)$        $g \in G_2$         3.3.2

**Remark**: *7-dimensional* vector *products are not preserved under the rotations of SO(7)! Therefore rotations in SO(7) that preserve the* vector *product are in* $G_2$.

```
gexp =.;
(* grab any two Octonions, simply make an array of 8 Reals *)
xO = {50, 220, -30, 1.02, 0.1, -8, -90, 160};
yO = {20, -0.9, 600, 10, -20, 120, 130, -0.01};

(* grab any element of G2,
simply make a linear combination of basis vectors C1-C14 and calculate
  the EXP for it which is the 7x7 matrix exponential *)
gexp = N[MatrixExp[(0.1 * C4 + 2.3 * C10 + Pi * C14)]];

(* calculate the action on the vector product of two Octonions *)
actionG2[gexp, OctCommutator[xO, yO] / 2]

(* Apply the action two each Octonion and then compute the corss product *)
OctCommutator[actionG2[gexp, xO], actionG2[gexp, yO]] / 2
```

{0., 45 472.6, -43 372.1, -137 804., 61 109.3, 29 486.3, -12 569.8, 62 391.}

{0., 45 472.6, -43 372.1, -137 804., 61 109.3, 29 486.3, -12 569.8, 62 391.}

Let's check the same preservation via the Rarenas matrices [3]:

```
gexp =.;
(* grab any two Octonions, simply make an array of 8 Reals *)
xO = {50, 220, -30, 1.02, 0.1, -8, -90, 160};
yO = {20, -0.9, 600, 10, -20, 120, 130, -0.01};

(* grab any element of G2,
simply make a linear combination of basis vectors C1-C14 and calculate
  the EXP for it which is the 7x7 matrix exponential *)
gexp = N[MatrixExp[0.1 * Y7 + Pi * X2 + 1000 * Y4 - 180 * X6]];

(* calculate the action on the vector product of two Octonions *)
actionG2[gexp, OctCommutator[xO, yO] / 2]

(* Apply the action two each Octonion and then compute the corss product *)
OctCommutator[actionG2[gexp, xO], actionG2[gexp, yO]] / 2
```

{0., -1359.59, -8027.32, -1800.04, -68 012.2, 92 659., 38 397.1, 129 761.}

{0., -1359.59, -8027.32, -1800.04, -68 012.2, 92 659., 38 397.1, 129 761.}

Let's do some symbolic verifications for the 14 basis vectors of $\mathbf{g}_2$:

```
(* Loop through all the basis matrices Ci,
calculate Exp[zCi] and see if the Exp[zCi] preserves the vector product SYMBOLICALLY,
if so output TRUE *)
(* Note that x and y are aribitrary Octonions *)

Clear [x, y, z]
x = {x1, x2, x3, x4, x5, x6, x7, x8};
y = {y1, y2, y3, y4, y5, y6, y7, y8};

gbasis = G2[];

Table[
 gexp = MatrixExp[z * gbasis[[i]]];

 (* calculate the action on the vector product of two Octonions *)
 cross1 = FullSimplify[actionG2[gexp, OctCommutator[x, y] / 2]];

 (* Apply the action two each Octonion and then compute the vector product *)
 cross2 = FullSimplify[OctCommutator[actionG2[gexp, x], actionG2[gexp, y]] / 2 ];

 cross1 === cross2,

 {i, 1, 14}]

{True, True, True, True, True, True, True, True, True, True, True, True, True, True}
```

Some sample symbolic computation from the latter code:

```
(* Sample for i = 9 *)
i = 9;
gexp = MatrixExp[z * gbasis[[i]]];

gexp // MatrixForm
(* calculate the action on the vector product of two Octonions *)
cross1 = FullSimplify[actionG2[gexp, OctCommutator[x, y] / 2]];

cross1 // MatrixForm

(* Apply the action two each Octonion and then compute the vector product *)
cross2 = FullSimplify[OctCommutator[actionG2[gexp, x], actionG2[gexp, y]] / 2];

cross2 // MatrixForm

cross1 === cross2
```

$$
\begin{pmatrix}
\cos\left[\frac{2z}{\sqrt{3}}\right] & -\sin\left[\frac{2z}{\sqrt{3}}\right] & 0 & 0 & 0 & 0 & 0 \\
\sin\left[\frac{2z}{\sqrt{3}}\right] & \cos\left[\frac{2z}{\sqrt{3}}\right] & 0 & 0 & 0 & 0 & 0 \\
0 & 0 & 1 & 0 & 0 & 0 & 0 \\
0 & 0 & 0 & \cos\left[\frac{z}{\sqrt{3}}\right] & 0 & 0 & \sin\left[\frac{z}{\sqrt{3}}\right] \\
0 & 0 & 0 & 0 & \cos\left[\frac{z}{\sqrt{3}}\right] & -\sin\left[\frac{z}{\sqrt{3}}\right] & 0 \\
0 & 0 & 0 & 0 & \sin\left[\frac{z}{\sqrt{3}}\right] & \cos\left[\frac{z}{\sqrt{3}}\right] & 0 \\
0 & 0 & 0 & -\sin\left[\frac{z}{\sqrt{3}}\right] & 0 & 0 & \cos\left[\frac{z}{\sqrt{3}}\right]
\end{pmatrix}
$$

$$
\begin{pmatrix}
0 \\
(-x4\,y3 + x3\,y4 - x6\,y5 + x5\,y6 + x8\,y7 - x7\,y8)\cos\left[\frac{2z}{\sqrt{3}}\right] + (-x4\,y2 + x2\,y4 + x7\,y5 + x8\,y6 - x5\,y7 - x6\,y8)\,S \\
(x4\,y2 - x2\,y4 - x7\,y5 - x8\,y6 + x5\,y7 + x6\,y8)\cos\left[\frac{2z}{\sqrt{3}}\right] + (-x4\,y3 + x3\,y4 - x6\,y5 + x5\,y6 + x8\,y7 - x7\,y8)\,S \\
-x3\,y2 + x2\,y3 - x8\,y5 + x7\,y6 - x6\,y7 + x5\,y8 \\
(x6\,y2 + x7\,y3 + x8\,y4 - x2\,y6 - x3\,y7 - x4\,y8)\cos\left[\frac{z}{\sqrt{3}}\right] + (x7\,y2 - x6\,y3 - x5\,y4 + x4\,y5 + x3\,y6 - x2\,y7)\,Si \\
(-x5\,y2 + x8\,y3 - x7\,y4 + x2\,y5 + x4\,y7 - x3\,y8)\cos\left[\frac{z}{\sqrt{3}}\right] + (x8\,y2 + x5\,y3 - x6\,y4 - x3\,y5 + x4\,y6 - x2\,y8)\,S \\
(-x8\,y2 - x5\,y3 + x6\,y4 + x3\,y5 - x4\,y6 + x2\,y8)\cos\left[\frac{z}{\sqrt{3}}\right] + (-x5\,y2 + x8\,y3 - x7\,y4 + x2\,y5 + x4\,y7 - x3\,y8)\,S \\
(x7\,y2 - x6\,y3 - x5\,y4 + x4\,y5 + x3\,y6 - x2\,y7)\cos\left[\frac{z}{\sqrt{3}}\right] + (-x6\,y2 - x7\,y3 - x8\,y4 + x2\,y6 + x3\,y7 + x4\,y8)\,S
\end{pmatrix}
$$

$$
\begin{pmatrix}
0 \\
(-x4\,y3 + x3\,y4 - x6\,y5 + x5\,y6 + x8\,y7 - x7\,y8)\cos\left[\frac{2z}{\sqrt{3}}\right] + (-x4\,y2 + x2\,y4 + x7\,y5 + x8\,y6 - x5\,y7 - x6\,y8)\,S \\
(x4\,y2 - x2\,y4 - x7\,y5 - x8\,y6 + x5\,y7 + x6\,y8)\cos\left[\frac{2z}{\sqrt{3}}\right] + (-x4\,y3 + x3\,y4 - x6\,y5 + x5\,y6 + x8\,y7 - x7\,y8)\,S \\
-x3\,y2 + x2\,y3 - x8\,y5 + x7\,y6 - x6\,y7 + x5\,y8 \\
(x6\,y2 + x7\,y3 + x8\,y4 - x2\,y6 - x3\,y7 - x4\,y8)\cos\left[\frac{z}{\sqrt{3}}\right] + (x7\,y2 - x6\,y3 - x5\,y4 + x4\,y5 + x3\,y6 - x2\,y7)\,Si \\
(-x5\,y2 + x8\,y3 - x7\,y4 + x2\,y5 + x4\,y7 - x3\,y8)\cos\left[\frac{z}{\sqrt{3}}\right] + (x8\,y2 + x5\,y3 - x6\,y4 - x3\,y5 + x4\,y6 - x2\,y8)\,S \\
(-x8\,y2 - x5\,y3 + x6\,y4 + x3\,y5 - x4\,y6 + x2\,y8)\cos\left[\frac{z}{\sqrt{3}}\right] + (-x5\,y2 + x8\,y3 - x7\,y4 + x2\,y5 + x4\,y7 - x3\,y8)\,S \\
(x7\,y2 - x6\,y3 - x5\,y4 + x4\,y5 + x3\,y6 - x2\,y7)\cos\left[\frac{z}{\sqrt{3}}\right] + (-x6\,y2 - x7\,y3 - x8\,y4 + x2\,y6 + x3\,y7 + x4\,y8)\,S
\end{pmatrix}
$$

```
True
```

### 3.3.1 Vector Product Valid Only for 0, 1, 3, 7 Dimensions!!!

See the Categorial proof from [8] :

Proposition: Any d Dimensional finite-dimensional vector space with a an inner product admitting vector product algebra has the following relationship:

$$d\,(d-1)\,(d-3)\,(d-7) = 0 \qquad\qquad 3.3.1.1$$

In other words its dimensions are bound to 0, 1, 3 and 7.

```
A = {a1, a2, a3, a4, a5, a6, a7, a8};
B = {b1, b2, b3, b4, b5, b6, b7, b8};

(* definition of vector product,
as you can seee the first elment or the Re is always 0 *)
OctCommutator[A, B] / 2
```

$$\left\{0, \frac{1}{2}\,(-2\,a4\,b3 + 2\,a3\,b4 - 2\,a6\,b5 + 2\,a5\,b6 + 2\,a8\,b7 - 2\,a7\,b8),\right.$$

$$\frac{1}{2}\,(2\,a4\,b2 - 2\,a2\,b4 - 2\,a7\,b5 - 2\,a8\,b6 + 2\,a5\,b7 + 2\,a6\,b8),$$

$$\frac{1}{2}\,(-2\,a3\,b2 + 2\,a2\,b3 - 2\,a8\,b5 + 2\,a7\,b6 - 2\,a6\,b7 + 2\,a5\,b8),$$

$$\frac{1}{2}\,(2\,a6\,b2 + 2\,a7\,b3 + 2\,a8\,b4 - 2\,a2\,b6 - 2\,a3\,b7 - 2\,a4\,b8),$$

$$\frac{1}{2}\,(-2\,a5\,b2 + 2\,a8\,b3 - 2\,a7\,b4 + 2\,a2\,b5 + 2\,a4\,b7 - 2\,a3\,b8),$$

$$\frac{1}{2}\,(-2\,a8\,b2 - 2\,a5\,b3 + 2\,a6\,b4 + 2\,a3\,b5 - 2\,a4\,b6 + 2\,a2\,b8),$$

$$\left.\frac{1}{2}\,(2\,a7\,b2 - 2\,a6\,b3 - 2\,a5\,b4 + 2\,a4\,b5 + 2\,a3\,b6 - 2\,a2\,b7)\right\}$$

```
(* Let's instead only look at the Im part *)
OctIm7[OctCommutator[A, B] / 2]
```

$$\left\{\frac{1}{2}\,(-2\,a4\,b3 + 2\,a3\,b4 - 2\,a6\,b5 + 2\,a5\,b6 + 2\,a8\,b7 - 2\,a7\,b8),\right.$$

$$\frac{1}{2}\,(2\,a4\,b2 - 2\,a2\,b4 - 2\,a7\,b5 - 2\,a8\,b6 + 2\,a5\,b7 + 2\,a6\,b8),$$

$$\frac{1}{2}\,(-2\,a3\,b2 + 2\,a2\,b3 - 2\,a8\,b5 + 2\,a7\,b6 - 2\,a6\,b7 + 2\,a5\,b8),$$

$$\frac{1}{2}\,(2\,a6\,b2 + 2\,a7\,b3 + 2\,a8\,b4 - 2\,a2\,b6 - 2\,a3\,b7 - 2\,a4\,b8),$$

$$\frac{1}{2}\,(-2\,a5\,b2 + 2\,a8\,b3 - 2\,a7\,b4 + 2\,a2\,b5 + 2\,a4\,b7 - 2\,a3\,b8),$$

$$\frac{1}{2}\,(-2\,a8\,b2 - 2\,a5\,b3 + 2\,a6\,b4 + 2\,a3\,b5 - 2\,a4\,b6 + 2\,a2\,b8),$$

$$\left.\frac{1}{2}\,(2\,a7\,b2 - 2\,a6\,b3 - 2\,a5\,b4 + 2\,a4\,b5 + 2\,a3\,b6 - 2\,a2\,b7)\right\}$$

**Remark 3.3.1.1**: *Dim 0 is just one vector and there is only one* vector *product i.e.* $\vec{0} \times \vec{0} = \vec{0}$.

Let's what this theorem actually means, say we start with Dim 1 i.e. only the first element of the Im part of the Octonion:

```
A = {0, a1, 0, 0, 0, 0, 0, 0};
B = {0, b1, 0, 0, 0, 0, 0, 0};

(* definition of vector product,
as you can seee the first element or the Re is always 0 *)
OctIm7[OctCommutator[A, B] / 2]
{0, 0, 0, 0, 0, 0, 0}
```

This means all vector products of Reals are 0 and indeed they fit in the first element of the Im part of the Octonion. Therefore the Dim 1 has a trivial 0 vector product that fits in one dimension.

Now Let's play with Dim 2:

```
A = {0, a1, a2, 0, 0, 0, 0, 0};
B = {0, b1, b2, 0, 0, 0, 0, 0};

(* definition of vector product,
as you can seee the first element or the Re is always 0 *)
OctIm7[OctCommutator[A, B] / 2]
```

$$\left\{ 0, 0, \frac{1}{2} \ (-2 \ a2 \ b1 + 2 \ a1 \ b2), \ 0, 0, 0, 0 \right\}$$

Now there is a problem here, we started with Dim 2 and the resultant vector product has non-zero coordinates in Dim 3 ! In other words if we limit the dimension to 2 there is no vector product fitting in 2 dimensions, we need an additional dimension.

What about the Dim 3:

```
A = {0, a1, a2, a3, 0, 0, 0, 0};
B = {0, b1, b2, b3, 0, 0, 0, 0};

(* definition of vector product,
as you can seee the first element or the Re is always 0 *)
OctIm7[OctCommutator[A, B] / 2]
```

$$\left\{ \frac{1}{2} \ (-2 \ a3 \ b2 + 2 \ a2 \ b3), \ \frac{1}{2} \ (2 \ a3 \ b1 - 2 \ a1 \ b3), \ \frac{1}{2} \ (-2 \ a2 \ b1 + 2 \ a1 \ b2), \ 0, 0, 0, 0 \right\}$$

WOW! vector product of vectors with 3 dimensions fits in 3 dimensions i.e. for Dim 3 there is a vector product.

Dim 4 needs all the way to the 7th element of the Im part of the Octonion, therefore there is no vector product in Dim 4:

```
A = {0, a1, a2, a3, a4, 0, 0, 0};
B = {0, b1, b2, b3, b4, 0, 0, 0};

(* definition of vector product,
as you can seee the first elment or the Re is always 0 *)
OctIm7[OctCommutator[A, B] / 2]
```

$$\left\{ \frac{1}{2} \, (-2 \, a3 \, b2 + 2 \, a2 \, b3), \; \frac{1}{2} \, (2 \, a3 \, b1 - 2 \, a1 \, b3), \; \frac{1}{2} \, (-2 \, a2 \, b1 + 2 \, a1 \, b2), \right.$$

$$\left. 0, \; \frac{1}{2} \, (-2 \, a4 \, b1 + 2 \, a1 \, b4), \; \frac{1}{2} \, (-2 \, a4 \, b2 + 2 \, a2 \, b4), \; \frac{1}{2} \, (-2 \, a4 \, b3 + 2 \, a3 \, b4) \right\}$$

**Remark 3.3.1.2** : *Note that the* vector *product at Dim 4 i.e. the Quaternions has a 0 at coordinate 5 of the Octonion, and the whole vector is equivalent to two 3D vectors!*

Dim 5 also needs more than 5 coordinates and spreads all the way to the 7th element:

```
A = {0, a1, a2, a3, a4, a5, 0, 0};
B = {0, b1, b2, b3, b4, b5, 0, 0};

(* definition of vector product,
as you can seee the first elment or the Re is always 0 *)
OctIm7[OctCommutator[A, B] / 2]
```

$$\left\{ \frac{1}{2} \, (-2 \, a3 \, b2 + 2 \, a2 \, b3 - 2 \, a5 \, b4 + 2 \, a4 \, b5), \; \frac{1}{2} \, (2 \, a3 \, b1 - 2 \, a1 \, b3), \right.$$

$$\frac{1}{2} \, (-2 \, a2 \, b1 + 2 \, a1 \, b2), \; \frac{1}{2} \, (2 \, a5 \, b1 - 2 \, a1 \, b5), \; \frac{1}{2} \, (-2 \, a4 \, b1 + 2 \, a1 \, b4),$$

$$\left. \frac{1}{2} \, (-2 \, a4 \, b2 + 2 \, a5 \, b3 + 2 \, a2 \, b4 - 2 \, a3 \, b5), \; \frac{1}{2} \, (-2 \, a5 \, b2 - 2 \, a4 \, b3 + 2 \, a3 \, b4 + 2 \, a2 \, b5) \right\}$$

Same situation Dim 6 has no vector product:

```
A = {0, a1, a2, a3, a4, a5, a6, 0};
B = {0, b1, b2, b3, b4, b5, b6, 0};

(* definition of vector product,
as you can seee the first elment or the Re is always 0 *)
OctIm7[OctCommutator[A, B] / 2]
```

$$\left\{ \frac{1}{2} \, (-2 \, a3 \, b2 + 2 \, a2 \, b3 - 2 \, a5 \, b4 + 2 \, a4 \, b5), \; \frac{1}{2} \, (2 \, a3 \, b1 - 2 \, a1 \, b3 - 2 \, a6 \, b4 + 2 \, a4 \, b6), \right.$$

$$\frac{1}{2} \, (-2 \, a2 \, b1 + 2 \, a1 \, b2 + 2 \, a6 \, b5 - 2 \, a5 \, b6), \; \frac{1}{2} \, (2 \, a5 \, b1 + 2 \, a6 \, b2 - 2 \, a1 \, b5 - 2 \, a2 \, b6),$$

$$\frac{1}{2} \, (-2 \, a4 \, b1 - 2 \, a6 \, b3 + 2 \, a1 \, b4 + 2 \, a3 \, b6), \; \frac{1}{2} \, (-2 \, a4 \, b2 + 2 \, a5 \, b3 + 2 \, a2 \, b4 - 2 \, a3 \, b5),$$

$$\left. \frac{1}{2} \, (2 \, a6 \, b1 - 2 \, a5 \, b2 - 2 \, a4 \, b3 + 2 \, a3 \, b4 + 2 \, a2 \, b5 - 2 \, a1 \, b6) \right\}$$

Dim 7 uses all the 7 and no other coordinates are required, therefore the Octonions have vector product in 7th dimension:

```
A = {0, a1, a2, a3, a4, a5, a6, a7};
B = {0, b1, b2, b3, b4, b5, b6, b7};

(* definition of vector product,
as you can seee the first elment or the Re is always 0 *)
OctIm7[OctCommutator[A, B] / 2]
```

$$\left\{ \frac{1}{2} \left( -2\,a3\,b2 + 2\,a2\,b3 - 2\,a5\,b4 + 2\,a4\,b5 + 2\,a7\,b6 - 2\,a6\,b7 \right), \right.$$

$$\frac{1}{2} \left( 2\,a3\,b1 - 2\,a1\,b3 - 2\,a6\,b4 - 2\,a7\,b5 + 2\,a4\,b6 + 2\,a5\,b7 \right),$$

$$\frac{1}{2} \left( -2\,a2\,b1 + 2\,a1\,b2 - 2\,a7\,b4 + 2\,a6\,b5 - 2\,a5\,b6 + 2\,a4\,b7 \right),$$

$$\frac{1}{2} \left( 2\,a5\,b1 + 2\,a6\,b2 + 2\,a7\,b3 - 2\,a1\,b5 - 2\,a2\,b6 - 2\,a3\,b7 \right),$$

$$\frac{1}{2} \left( -2\,a4\,b1 + 2\,a7\,b2 - 2\,a6\,b3 + 2\,a1\,b4 + 2\,a3\,b6 - 2\,a2\,b7 \right),$$

$$\frac{1}{2} \left( -2\,a7\,b1 - 2\,a4\,b2 + 2\,a5\,b3 + 2\,a2\,b4 - 2\,a3\,b5 + 2\,a1\,b7 \right),$$

$$\left. \frac{1}{2} \left( 2\,a6\,b1 - 2\,a5\,b2 - 2\,a4\,b3 + 2\,a3\,b4 + 2\,a2\,b5 - 2\,a1\,b6 \right) \right\}$$

# 4. Symbolic Computation for Exp: $\mathfrak{g}_2 \longrightarrow G_2$

Just to get an idea how complicated is the Exp map, and also to see how extremely 'winded' or 'coiled' the outcome is in $G_2$ , the basis for vectors for $\mathfrak{g}_2$ were exponentiated and matrix-multiplied back to back. Note that the BCH series were not added, if so a lot more complexity is introduced.

```
a = {a1, a2, a3, a4, a5, a6, a7, a8, a9, a10, a11, a12, a13, a14};
gexp = IdentityMatrix[7];
Table[gexp = gexp.MatrixExp[(a[[i]] * gbasis[[i]])], {i, 1, 14}];

Style[gexp[[7]][[6]], FontSize → 7]
```

$-\text{Sin}\left[\frac{2\,a13}{\sqrt{3}}\right]$

$\left(\text{Sin}\left[\frac{2\,a12}{\sqrt{3}}\right]\left(\text{Cos}\left[\frac{a10}{\sqrt{3}}\right]\left(\text{Cos}\left[\frac{a9}{\sqrt{3}}\right]\left(\text{Cos}\left[\frac{a8}{\sqrt{3}}\right](\text{Cos}[a7]\;(\text{Cos}[a6]\;(\text{Cos}[a1]\;\text{Cos}[a3]\;\text{Sin}[a2] - \text{Cos}[a2]\;\text{Sin}[a1]\;\text{Sin}[a3]) + (-\text{Cos}[a5]\;(\text{Cos}[a1]\;\text{Cos}[a2]\;\text{Cos}[a3] - \right.\right.\right.\right.\right.$

$\text{Sin}[a1]\;\text{Sin}[a2]\;\text{Sin}[a3])\;\text{Sin}[a4] + \text{Cos}[a4]\;(\text{Cos}[a3]\;\text{Sin}[a1]\;\text{Sin}[a2] + \text{Cos}[a1]\;\text{Cos}[a2]\;\text{Sin}[a3])\;\text{Sin}[a5])\;\text{Sin}[a6]) -$

$\text{Cos}[a6]\;(-\text{Cos}[a5]\;(\text{Cos}[a3]\;\text{Sin}[a1]\;\text{Sin}[a2] + \text{Cos}[a1]\;\text{Cos}[a2]\;\text{Sin}[a3])\;\text{Sin}[a4] + \text{Cos}[a4]\;(\text{Cos}[a1]\;\text{Cos}[a2]\;\text{Cos}[a3] -$

$\text{Sin}[a1]\;\text{Sin}[a2]\;\text{Sin}[a3])\;\text{Sin}[a5]) + (\text{Cos}[a2]\;\text{Cos}[a3]\;\text{Sin}[a1] + \text{Cos}[a1]\;\text{Sin}[a2]\;\text{Sin}[a3])\;\text{Sin}[a6])\;\text{Sin}[a7]) +$

$(\text{Cos}[a7]\;(\text{Cos}[a6]\;(\text{Cos}[a2]\;\text{Cos}[a3]\;\text{Sin}[a1] + \text{Cos}[a1]\;\text{Sin}[a2]\;\text{Sin}[a3]) - (-\text{Cos}[a5]\;(\text{Cos}[a3]\;\text{Sin}[a1]\;\text{Sin}[a2] + \text{Cos}[a1]\;\text{Cos}[a2]\;\text{Sin}[a3])$

$\text{Sin}[a4] - \text{Cos}[a4]\;(\text{Cos}[a1]\;\text{Cos}[a2]\;\text{Cos}[a3] - \text{Sin}[a1]\;\text{Sin}[a2]\;\text{Sin}[a3])\;\text{Sin}[a5])\;\text{Sin}[a6]) -$

$(\text{Cos}[a6]\;(-\text{Cos}[a5]\;(\text{Cos}[a1]\;\text{Cos}[a2]\;\text{Cos}[a3] - \text{Sin}[a1]\;\text{Sin}[a2]\;\text{Sin}[a3])\;\text{Sin}[a4] + \text{Cos}[a4]\;(\text{Cos}[a3]\;\text{Sin}[a1]\;\text{Sin}[a2] +$

$\text{Cos}[a1]\;\text{Cos}[a2]\;\text{Sin}[a3])\;\text{Sin}[a5]) - (\text{Cos}[a1]\;\text{Cos}[a3]\;\text{Sin}[a2] - \text{Cos}[a2]\;\text{Sin}[a1]\;\text{Sin}[a3])\;\text{Sin}[a6])\;\text{Sin}[a7])\;\text{Sin}\left[\frac{a8}{\sqrt{3}}\right]\right) +$

$\left(\text{Cos}\left[\frac{a8}{\sqrt{3}}\right](\text{Cos}[a4]\;\text{Cos}[a5]\;(\text{Cos}[a3]\;\text{Sin}[a1]\;\text{Sin}[a2] + \text{Cos}[a1]\;\text{Cos}[a2]\;\text{Sin}[a3]) + (\text{Cos}[a1]\;\text{Cos}[a2]\;\text{Cos}[a3] - \text{Sin}[a1]\;\text{Sin}[a2]\;\text{Sin}[a3])\right.$

$\text{Sin}[a4]\;\text{Sin}[a5]) + (\text{Cos}[a4]\;\text{Cos}[a5]\;(\text{Cos}[a1]\;\text{Cos}[a2]\;\text{Cos}[a3] - \text{Sin}[a1]\;\text{Sin}[a2]\;\text{Sin}[a3]) -$

$(\text{Cos}[a3]\;\text{Sin}[a1]\;\text{Sin}[a2] + \text{Cos}[a1]\;\text{Cos}[a2]\;\text{Sin}[a3])\;\text{Sin}[a4]\;\text{Sin}[a5])\;\text{Sin}\left[\frac{a8}{\sqrt{3}}\right]\right)\;\text{Sin}\left[\frac{a9}{\sqrt{3}}\right]\right) +$

$\text{Sin}\left[\frac{a10}{\sqrt{3}}\right]\left(\text{Cos}\left[\frac{a9}{\sqrt{3}}\right]\left(\text{Cos}\left[\frac{a8}{\sqrt{3}}\right](\text{Cos}[a4]\;\text{Cos}[a5]\;(\text{Cos}[a1]\;\text{Cos}[a2]\;\text{Cos}[a3] - \text{Sin}[a1]\;\text{Sin}[a2]\;\text{Sin}[a3]) - \right.\right.$

$(\text{Cos}[a3]\;\text{Sin}[a1]\;\text{Sin}[a2] + \text{Cos}[a1]\;\text{Cos}[a2]\;\text{Sin}[a3])\;\text{Sin}[a4]\;\text{Sin}[a5]) - (\text{Cos}[a4]\;\text{Cos}[a5]$

$(\text{Cos}[a3]\;\text{Sin}[a1]\;\text{Sin}[a2] + \text{Cos}[a1]\;\text{Cos}[a2]\;\text{Sin}[a3]) + (\text{Cos}[a1]\;\text{Cos}[a2]\;\text{Cos}[a3] - \text{Sin}[a1]\;\text{Sin}[a2]\;\text{Sin}[a3])\;\text{Sin}[a4]\;\text{Sin}[a5])\;\text{Sin}\left[\frac{a8}{\sqrt{3}}\right]\right) +$

$\left(\text{Cos}\left[\frac{a8}{\sqrt{3}}\right](\text{Cos}[a7]\;(\text{Cos}[a6]\;(\text{Cos}[a2]\;\text{Cos}[a3]\;\text{Sin}[a1] + \text{Cos}[a1]\;\text{Sin}[a2]\;\text{Sin}[a3]) - (-\text{Cos}[a5]\;(\text{Cos}[a3]\;\text{Sin}[a1]\;\text{Sin}[a2] + \right.$

$\text{Cos}[a1]\;\text{Cos}[a2]\;\text{Sin}[a3])\;\text{Sin}[a4] - \text{Cos}[a4]\;(\text{Cos}[a1]\;\text{Cos}[a2]\;\text{Cos}[a3] - \text{Sin}[a1]\;\text{Sin}[a2]\;\text{Sin}[a3])\;\text{Sin}[a5])\;\text{Sin}[a6]) -$

$(\text{Cos}[a6]\;(-\text{Cos}[a5]\;(\text{Cos}[a1]\;\text{Cos}[a2]\;\text{Cos}[a3] - \text{Sin}[a1]\;\text{Sin}[a2]\;\text{Sin}[a3])\;\text{Sin}[a4] + \text{Cos}[a4]\;(\text{Cos}[a3]\;\text{Sin}[a1]\;\text{Sin}[a2] +$

$\text{Cos}[a1]\;\text{Cos}[a2]\;\text{Sin}[a3])\;\text{Sin}[a5]) - (\text{Cos}[a1]\;\text{Cos}[a3]\;\text{Sin}[a2] - \text{Cos}[a2]\;\text{Sin}[a1]\;\text{Sin}[a3])\;\text{Sin}[a6])\;\text{Sin}[a7]) +$

$(\text{Cos}[a7]\;(\text{Cos}[a6]\;(\text{Cos}[a1]\;\text{Cos}[a3]\;\text{Sin}[a2] - \text{Cos}[a2]\;\text{Sin}[a1]\;\text{Sin}[a3]) + (-\text{Cos}[a5]\;(\text{Cos}[a1]\;\text{Cos}[a2]\;\text{Cos}[a3] - \text{Sin}[a1]\;\text{Sin}[a2]\;\text{Sin}[a3])$

$\text{Sin}[a4] + \text{Cos}[a4]\;(\text{Cos}[a3]\;\text{Sin}[a1]\;\text{Sin}[a2] + \text{Cos}[a1]\;\text{Cos}[a2]\;\text{Sin}[a3])\;\text{Sin}[a5])\;\text{Sin}[a6]) -$

$(\text{Cos}[a6]\;(-\text{Cos}[a5]\;(\text{Cos}[a3]\;\text{Sin}[a1]\;\text{Sin}[a2] + \text{Cos}[a1]\;\text{Cos}[a2]\;\text{Sin}[a3])\;\text{Sin}[a4] + \text{Cos}[a4]\;(\text{Cos}[a1]\;\text{Cos}[a2]\;\text{Cos}[a3] -$

$\text{Sin}[a1]\;\text{Sin}[a2]\;\text{Sin}[a3])\;\text{Sin}[a5]) + (\text{Cos}[a2]\;\text{Cos}[a3]\;\text{Sin}[a1] + \text{Cos}[a1]\;\text{Sin}[a2]\;\text{Sin}[a3])\;\text{Sin}[a6])\;\text{Sin}[a7])\;\text{Sin}\left[\frac{a8}{\sqrt{3}}\right]\right)\;\text{Sin}\left[\frac{a9}{\sqrt{3}}\right]\right)\right) +$

$\text{Cos}\left[\frac{2\,a12}{\sqrt{3}}\right]\left(\text{Sin}\left[\frac{2\,a11}{\sqrt{3}}\right]\left(\text{Cos}\left[\frac{a10}{\sqrt{3}}\right]\left(\text{Cos}\left[\frac{a9}{\sqrt{3}}\right]\left(\text{Cos}\left[\frac{a8}{\sqrt{3}}\right](\text{Cos}[a7]\;(\text{Cos}[a6]\;(\text{Cos}[a2]\;\text{Cos}[a3]\;\text{Sin}[a1] + \text{Cos}[a1]\;\text{Sin}[a2]\;\text{Sin}[a3]) - \right.\right.\right.\right.\right.$

$(-\text{Cos}[a5]\;(\text{Cos}[a3]\;\text{Sin}[a1]\;\text{Sin}[a2] + \text{Cos}[a1]\;\text{Cos}[a2]\;\text{Sin}[a3])\;\text{Sin}[a4] - \text{Cos}[a4]\;(\text{Cos}[a1]\;\text{Cos}[a2]\;\text{Cos}[a3] - \text{Sin}[a1]\;\text{Sin}[a2]\;\text{Sin}[a3])$

$\text{Sin}[a5])\;\text{Sin}[a6]) - (\text{Cos}[a6]\;(-\text{Cos}[a5]\;(\text{Cos}[a1]\;\text{Cos}[a2]\;\text{Cos}[a3] - \text{Sin}[a1]\;\text{Sin}[a2]\;\text{Sin}[a3])\;\text{Sin}[a4] + \text{Cos}[a4]\;(\text{Cos}[a3]\;\text{Sin}[a1]$

$\text{Sin}[a2] + \text{Cos}[a1]\;\text{Cos}[a2]\;\text{Sin}[a3])\;\text{Sin}[a5]) - (\text{Cos}[a1]\;\text{Cos}[a3]\;\text{Sin}[a2] - \text{Cos}[a2]\;\text{Sin}[a1]\;\text{Sin}[a3])\;\text{Sin}[a6])\;\text{Sin}[a7]) -$

$(\text{Cos}[a7]\;(\text{Cos}[a6]\;(\text{Cos}[a2]\;\text{Cos}[a3]\;\text{Sin}[a1] + \text{Cos}[a1]\;\text{Sin}[a2]\;\text{Sin}[a3]) + (-\text{Cos}[a5]\;(\text{Cos}[a1]\;\text{Cos}[a2]\;\text{Cos}[a3] - \text{Sin}[a1]\;\text{Sin}[a2]$

$\text{Sin}[a3])\;\text{Sin}[a4] + \text{Cos}[a4]\;(\text{Cos}[a3]\;\text{Sin}[a1]\;\text{Sin}[a2] + \text{Cos}[a1]\;\text{Cos}[a2]\;\text{Sin}[a3])\;\text{Sin}[a5])\;\text{Sin}[a6]) -$

$(\text{Cos}[a6]\;(-\text{Cos}[a5]\;(\text{Cos}[a3]\;\text{Sin}[a1]\;\text{Sin}[a2] + \text{Cos}[a1]\;\text{Cos}[a2]\;\text{Sin}[a3])\;\text{Sin}[a4] - \text{Cos}[a4]\;(\text{Cos}[a1]\;\text{Cos}[a2]\;\text{Cos}[a3] -$

$$\quad \text{Sin}[a1]\ \text{Sin}[a2]\ \text{Sin}[a3])\ \text{Sin}[a5]) + (\text{Cos}[a2]\ \text{Cos}[a3]\ \text{Sin}[a1] + \text{Cos}[a1]\ \text{Sin}[a2]\ \text{Sin}[a3])\ \text{Sin}[a6])\ \text{Sin}[a7])\ \text{Sin}\!\left[\frac{a8}{\sqrt{3}}\right]) -$$

$$\left(\text{Cos}\!\left[\frac{a8}{\sqrt{3}}\right](\text{Cos}[a4]\ \text{Cos}[a5]\ (\text{Cos}[a1]\ \text{Cos}[a2]\ \text{Cos}[a3] - \text{Sin}[a1]\ \text{Sin}[a2]\ \text{Sin}[a3]) - (\text{Cos}[a3]\ \text{Sin}[a1]\ \text{Sin}[a2] + \text{Cos}[a1]\ \text{Cos}[a2]\ \text{Sin}[a3])\right.$$
$$\text{Sin}[a4]\ \text{Sin}[a5]) - (\text{Cos}[a4]\ \text{Cos}[a5]\ (\text{Cos}[a3]\ \text{Sin}[a1]\ \text{Sin}[a2] + \text{Cos}[a1]\ \text{Cos}[a2]\ \text{Sin}[a3]) +$$
$$\left.(\text{Cos}[a1]\ \text{Cos}[a2]\ \text{Cos}[a3] - \text{Sin}[a1]\ \text{Sin}[a2]\ \text{Sin}[a3])\ \text{Sin}[a4]\ \text{Sin}[a5])\ \text{Sin}\!\left[\frac{a8}{\sqrt{3}}\right])\ \text{Sin}\!\left[\frac{a9}{\sqrt{3}}\right]\right) +$$

$$\text{Sin}\!\left[\frac{a10}{\sqrt{3}}\right]\left(\text{Sin}\!\left[\frac{a9}{\sqrt{3}}\right]\left(\text{Cos}\!\left[\frac{a8}{\sqrt{3}}\right](\text{Cos}[a4]\ \text{Cos}[a5]\ (\text{Cos}[a3]\ \text{Sin}[a1]\ \text{Sin}[a2] + \text{Cos}[a1]\ \text{Cos}[a2]\ \text{Sin}[a3]) +\right.\right.$$
$$(\text{Cos}[a1]\ \text{Cos}[a2]\ \text{Cos}[a3] - \text{Sin}[a1]\ \text{Sin}[a2]\ \text{Sin}[a3])\ \text{Sin}[a4]\ \text{Sin}[a5]) + (\text{Cos}[a4]\ \text{Cos}[a5]$$
$$(\text{Cos}[a1]\ \text{Cos}[a2]\ \text{Cos}[a3] - \text{Sin}[a1]\ \text{Sin}[a2]\ \text{Sin}[a3]) - (\text{Cos}[a3]\ \text{Sin}[a1]\ \text{Sin}[a2] + \text{Cos}[a1]\ \text{Cos}[a2]\ \text{Sin}[a3])\ \text{Sin}[a4]\ \text{Sin}[a5])\ \text{Sin}\!\left[\frac{a8}{\sqrt{3}}\right]) -$$

$$\left(\text{Cos}\!\left[\frac{a8}{\sqrt{3}}\right](\text{Cos}[a7]\ (\text{Cos}[a6]\ (\text{Cos}[a1]\ \text{Cos}[a3]\ \text{Sin}[a2] - \text{Cos}[a2]\ \text{Sin}[a1]\ \text{Sin}[a3]) + (-\text{Cos}[a5]\ (\text{Cos}[a1]\ \text{Cos}[a2]\ \text{Cos}[a3] -\right.$$
$$\text{Sin}[a1]\ \text{Sin}[a2]\ \text{Sin}[a3])\ \text{Sin}[a4] + \text{Cos}[a4]\ (\text{Cos}[a3]\ \text{Sin}[a1]\ \text{Sin}[a2] + \text{Cos}[a1]\ \text{Cos}[a2]\ \text{Sin}[a3])\ \text{Sin}[a5])\ \text{Sin}[a6]) -$$
$$(\text{Cos}[a6]\ (-\text{Cos}[a5]\ (\text{Cos}[a3]\ \text{Sin}[a1]\ \text{Sin}[a2] + \text{Cos}[a1]\ \text{Cos}[a2]\ \text{Sin}[a3])\ \text{Sin}[a4] - \text{Cos}[a4]\ (\text{Cos}[a1]\ \text{Cos}[a2]\ \text{Cos}[a3] -$$
$$\text{Sin}[a1]\ \text{Sin}[a2]\ \text{Sin}[a3])\ \text{Sin}[a5]) + (\text{Cos}[a2]\ \text{Cos}[a3]\ \text{Sin}[a1] + \text{Cos}[a1]\ \text{Sin}[a2]\ \text{Sin}[a3])\ \text{Sin}[a6])\ \text{Sin}[a7]) +$$
$$(\text{Cos}[a7]\ (\text{Cos}[a6]\ (\text{Cos}[a2]\ \text{Cos}[a3]\ \text{Sin}[a1] + \text{Cos}[a1]\ \text{Sin}[a2]\ \text{Sin}[a3]) - (-\text{Cos}[a5]\ (\text{Cos}[a3]\ \text{Sin}[a1]\ \text{Sin}[a2] + \text{Cos}[a1]\ \text{Cos}[a2]$$
$$\text{Sin}[a3])\ \text{Sin}[a4] - \text{Cos}[a4]\ (\text{Cos}[a1]\ \text{Cos}[a2]\ \text{Cos}[a3] - \text{Sin}[a1]\ \text{Sin}[a2]\ \text{Sin}[a3])\ \text{Sin}[a5])\ \text{Sin}[a6]) -$$
$$(\text{Cos}[a6]\ (-\text{Cos}[a5]\ (\text{Cos}[a1]\ \text{Cos}[a2]\ \text{Cos}[a3] - \text{Sin}[a1]\ \text{Sin}[a2]\ \text{Sin}[a3])\ \text{Sin}[a4] + \text{Cos}[a4]\ (\text{Cos}[a3]\ \text{Sin}[a1]\ \text{Sin}[a2] + \text{Cos}[a1]$$
$$\left.\left.\left.\text{Cos}[a2]\ \text{Sin}[a3])\ \text{Sin}[a5]) - (\text{Cos}[a1]\ \text{Cos}[a3]\ \text{Sin}[a2] - \text{Cos}[a2]\ \text{Sin}[a1]\ \text{Sin}[a3])\ \text{Sin}[a6])\ \text{Sin}[a7])\ \text{Sin}\!\left[\frac{a8}{\sqrt{3}}\right])\ \text{Sin}\!\left[\frac{a9}{\sqrt{3}}\right]\right)\right) +$$

$$\text{Cos}\!\left[\frac{2\ a11}{\sqrt{3}}\right]\left(\text{Sin}\!\left[\frac{2\ a10}{\sqrt{3}}\right]\left(\text{Cos}\!\left[\frac{2\ a8}{\sqrt{3}}\right](\text{Cos}[a7]\ (\text{Cos}[a6]\ (-\text{Cos}[a5]\ (\text{Cos}[a3]\ \text{Sin}[a1]\ \text{Sin}[a2] + \text{Cos}[a1]\ \text{Cos}[a2]\ \text{Sin}[a3])\ \text{Sin}[a4] - \text{Cos}[a4]\right.\right.$$
$$(\text{Cos}[a1]\ \text{Cos}[a2]\ \text{Cos}[a3] - \text{Sin}[a1]\ \text{Sin}[a2]\ \text{Sin}[a3])\ \text{Sin}[a5]) + (\text{Cos}[a2]\ \text{Cos}[a3]\ \text{Sin}[a1] + \text{Cos}[a1]\ \text{Sin}[a2]\ \text{Sin}[a3])\ \text{Sin}[a6]) +$$
$$(\text{Cos}[a6]\ (\text{Cos}[a1]\ \text{Cos}[a3]\ \text{Sin}[a2] - \text{Cos}[a2]\ \text{Sin}[a1]\ \text{Sin}[a3]) + (-\text{Cos}[a5]\ (\text{Cos}[a1]\ \text{Cos}[a2]\ \text{Cos}[a3] - \text{Sin}[a1]\ \text{Sin}[a2]\ \text{Sin}[a3])$$
$$\text{Sin}[a4] + \text{Cos}[a4]\ (\text{Cos}[a3]\ \text{Sin}[a1]\ \text{Sin}[a2] + \text{Cos}[a1]\ \text{Cos}[a2]\ \text{Sin}[a3])\ \text{Sin}[a5])\ \text{Sin}[a6])\ \text{Sin}[a7]) -$$
$$(\text{Cos}[a7]\ (\text{Cos}[a6]\ (-\text{Cos}[a5]\ (\text{Cos}[a1]\ \text{Cos}[a2]\ \text{Cos}[a3] - \text{Sin}[a1]\ \text{Sin}[a2]\ \text{Sin}[a3])\ \text{Sin}[a4] + \text{Cos}[a4]\ (\text{Cos}[a3]\ \text{Sin}[a1]\ \text{Sin}[a2] +$$
$$\text{Cos}[a1]\ \text{Cos}[a2]\ \text{Sin}[a3])\ \text{Sin}[a5]) - (\text{Cos}[a1]\ \text{Cos}[a3]\ \text{Sin}[a2] - \text{Cos}[a2]\ \text{Sin}[a1]\ \text{Sin}[a3])\ \text{Sin}[a6]) +$$
$$(\text{Cos}[a6]\ (\text{Cos}[a2]\ \text{Cos}[a3]\ \text{Sin}[a1] + \text{Cos}[a1]\ \text{Sin}[a2]\ \text{Sin}[a3]) - (-\text{Cos}[a5]\ (\text{Cos}[a3]\ \text{Sin}[a1]\ \text{Sin}[a2] + \text{Cos}[a1]\ \text{Cos}[a2]\ \text{Sin}[a3])$$
$$\left.\text{Sin}[a4] - \text{Cos}[a4]\ (\text{Cos}[a1]\ \text{Cos}[a2]\ \text{Cos}[a3] - \text{Sin}[a1]\ \text{Sin}[a2]\ \text{Sin}[a3])\ \text{Sin}[a5])\ \text{Sin}[a6])\ \text{Sin}[a7])\ \text{Sin}\!\left[\frac{2\ a8}{\sqrt{3}}\right]) +$$

$$\text{Cos}\!\left[\frac{2\ a10}{\sqrt{3}}\right]\left(\text{Cos}\!\left[\frac{2\ a8}{\sqrt{3}}\right](\text{Cos}[a7]\ (\text{Cos}[a6]\ (-\text{Cos}[a5]\ (\text{Cos}[a1]\ \text{Cos}[a2]\ \text{Cos}[a3] - \text{Sin}[a1]\ \text{Sin}[a2]\ \text{Sin}[a3])\ \text{Sin}[a4] + \text{Cos}[a4]\right.$$
$$(\text{Cos}[a3]\ \text{Sin}[a1]\ \text{Sin}[a2] + \text{Cos}[a1]\ \text{Cos}[a2]\ \text{Sin}[a3])\ \text{Sin}[a5]) - (\text{Cos}[a1]\ \text{Cos}[a3]\ \text{Sin}[a2] - \text{Cos}[a2]\ \text{Sin}[a1]\ \text{Sin}[a3])\ \text{Sin}[a6]) +$$
$$(\text{Cos}[a6]\ (\text{Cos}[a2]\ \text{Cos}[a3]\ \text{Sin}[a1] + \text{Cos}[a1]\ \text{Sin}[a2]\ \text{Sin}[a3]) - (-\text{Cos}[a5]\ (\text{Cos}[a3]\ \text{Sin}[a1]\ \text{Sin}[a2] + \text{Cos}[a1]\ \text{Cos}[a2]\ \text{Sin}[a3])$$
$$\text{Sin}[a4] - \text{Cos}[a4]\ (\text{Cos}[a1]\ \text{Cos}[a2]\ \text{Cos}[a3] - \text{Sin}[a1]\ \text{Sin}[a2]\ \text{Sin}[a3])\ \text{Sin}[a5])\ \text{Sin}[a6])\ \text{Sin}[a7]) -$$
$$(\text{Cos}[a7]\ (\text{Cos}[a6]\ (-\text{Cos}[a5]\ (\text{Cos}[a3]\ \text{Sin}[a1]\ \text{Sin}[a2] + \text{Cos}[a1]\ \text{Cos}[a2]\ \text{Sin}[a3])\ \text{Sin}[a4] - \text{Cos}[a4]\ (\text{Cos}[a1]\ \text{Cos}[a2]\ \text{Cos}[a3] -$$
$$\text{Sin}[a1]\ \text{Sin}[a2]\ \text{Sin}[a3])\ \text{Sin}[a5]) + (\text{Cos}[a2]\ \text{Cos}[a3]\ \text{Sin}[a1] + \text{Cos}[a1]\ \text{Sin}[a2]\ \text{Sin}[a3])\ \text{Sin}[a6]) +$$
$$(\text{Cos}[a6]\ (\text{Cos}[a1]\ \text{Cos}[a3]\ \text{Sin}[a2] - \text{Cos}[a2]\ \text{Sin}[a1]\ \text{Sin}[a3]) + (-\text{Cos}[a5]\ (\text{Cos}[a1]\ \text{Cos}[a2]\ \text{Cos}[a3] - \text{Sin}[a1]\ \text{Sin}[a2]\ \text{Sin}[a3])$$
$$\left.\left.\left.\text{Sin}[a4] + \text{Cos}[a4]\ (\text{Cos}[a3]\ \text{Sin}[a1]\ \text{Sin}[a2] + \text{Cos}[a1]\ \text{Cos}[a2]\ \text{Sin}[a3])\ \text{Sin}[a5])\ \text{Sin}[a6])\ \text{Sin}[a7])\ \text{Sin}\!\left[\frac{2\ a8}{\sqrt{3}}\right])\ \text{Sin}\!\left[\frac{2\ a9}{\sqrt{3}}\right]\right)\right)\right) +$$

$$\text{Cos}\!\left[\frac{2\ a13}{\sqrt{3}}\right]\left(\text{Sin}\!\left[\frac{a12}{\sqrt{3}}\right]\left(\text{Cos}\!\left[\frac{a11}{\sqrt{3}}\right]\text{Cos}\!\left[\frac{2\ a9}{\sqrt{3}}\right]\left(\text{Cos}\!\left[\frac{2\ a8}{\sqrt{3}}\right](\text{Cos}[a7]\ (\text{Cos}[a6]\ (-\text{Cos}[a5]\ (\text{Cos}[a1]\ \text{Cos}[a2]\ \text{Cos}[a3] - \text{Sin}[a1]\ \text{Sin}[a2]\ \text{Sin}[a3])\ \text{Sin}[a4] +\right.\right.\right.$$
$$\text{Cos}[a4]\ (\text{Cos}[a3]\ \text{Sin}[a1]\ \text{Sin}[a2] + \text{Cos}[a1]\ \text{Cos}[a2]\ \text{Sin}[a3])\ \text{Sin}[a5]) - (\text{Cos}[a1]\ \text{Cos}[a3]\ \text{Sin}[a2] - \text{Cos}[a2]\ \text{Sin}[a1]\ \text{Sin}[a3])\ \text{Sin}[a6]) +$$
$$(\text{Cos}[a6]\ (\text{Cos}[a2]\ \text{Cos}[a3]\ \text{Sin}[a1] + \text{Cos}[a1]\ \text{Sin}[a2]\ \text{Sin}[a3]) - (-\text{Cos}[a5]\ (\text{Cos}[a3]\ \text{Sin}[a1]\ \text{Sin}[a2] + \text{Cos}[a1]\ \text{Cos}[a2]\ \text{Sin}[a3])\ \text{Sin}[a6]) +$$
$$\text{Cos}[a4]\ (\text{Cos}[a1]\ \text{Cos}[a2]\ \text{Cos}[a3] - \text{Sin}[a1]\ \text{Sin}[a2]\ \text{Sin}[a3])\ \text{Sin}[a5])\ \text{Sin}[a6])\ \text{Sin}[a7]) -$$
$$(\text{Cos}[a7]\ (\text{Cos}[a6]\ (-\text{Cos}[a5]\ (\text{Cos}[a3]\ \text{Sin}[a1]\ \text{Sin}[a2] + \text{Cos}[a1]\ \text{Cos}[a2]\ \text{Sin}[a3])\ \text{Sin}[a4] - \text{Cos}[a4]\ (\text{Cos}[a1]\ \text{Cos}[a2]\ \text{Cos}[a3] -$$
$$\text{Sin}[a1]\ \text{Sin}[a2]\ \text{Sin}[a3])\ \text{Sin}[a5]) + (\text{Cos}[a2]\ \text{Cos}[a3]\ \text{Sin}[a1] + \text{Cos}[a1]\ \text{Sin}[a2]\ \text{Sin}[a3])\ \text{Sin}[a6]) +$$
$$(\text{Cos}[a6]\ (\text{Cos}[a1]\ \text{Cos}[a3]\ \text{Sin}[a2] - \text{Cos}[a2]\ \text{Sin}[a1]\ \text{Sin}[a3]) + (-\text{Cos}[a5]\ (\text{Cos}[a1]\ \text{Cos}[a2]\ \text{Cos}[a3] - \text{Sin}[a1]\ \text{Sin}[a2]\ \text{Sin}[a3])\ \text{Sin}[a4] +$$
$$\left.\text{Cos}[a4]\ (\text{Cos}[a3]\ \text{Sin}[a1]\ \text{Sin}[a2] + \text{Cos}[a1]\ \text{Cos}[a2]\ \text{Sin}[a3])\ \text{Sin}[a5])\ \text{Sin}[a6])\ \text{Sin}[a7])\ \text{Sin}\!\left[\frac{2\ a8}{\sqrt{3}}\right]) +$$

$$\text{Sin}\!\left[\frac{a11}{\sqrt{3}}\right]\left(-\text{Sin}\!\left[\frac{a10}{\sqrt{3}}\right]\left(\text{Cos}\!\left[\frac{a9}{\sqrt{3}}\right]\left(\text{Cos}\!\left[\frac{a8}{\sqrt{3}}\right](\text{Cos}[a7]\ (\text{Cos}[a6]\ (\text{Cos}[a1]\ \text{Cos}[a3]\ \text{Sin}[a2] - \text{Cos}[a2]\ \text{Sin}[a1]\ \text{Sin}[a3]) + (-\text{Cos}[a5]\ (\text{Cos}[a1]\ \text{Cos}[a2]\right.\right.\right.$$

$$
\begin{aligned}
&\Big(\text{Cos}[a3] - \text{Sin}[a1]\,\text{Sin}[a2]\,\text{Sin}[a3]\Big)\,\text{Sin}[a4] + \text{Cos}[a4]\,(\text{Cos}[a3]\,\text{Sin}[a1]\,\text{Sin}[a2] + \text{Cos}[a1]\,\text{Cos}[a2]\,\text{Sin}[a3])\Big)\,\text{Sin}[a5]\Big)\,\text{Sin}[a6] - \\
&(\text{Cos}[a6]\,(-\text{Cos}[a5]\,(\text{Cos}[a3]\,\text{Sin}[a1]\,\text{Sin}[a2] + \text{Cos}[a1]\,\text{Cos}[a2]\,\text{Sin}[a3])\,\text{Sin}[a4] - \text{Cos}[a4]\,(\text{Cos}[a3] - \\
&\quad \text{Sin}[a1]\,\text{Sin}[a2]\,\text{Sin}[a3])\,\text{Sin}[a5]) + (\text{Cos}[a2]\,\text{Cos}[a3]\,\text{Sin}[a1] + \text{Cos}[a1]\,\text{Sin}[a2]\,\text{Sin}[a3])\,\text{Sin}[a7]) + \\
&(\text{Cos}[a7]\,(\text{Cos}[a6]\,(\text{Cos}[a2]\,\text{Cos}[a3]\,\text{Sin}[a1] + \text{Cos}[a1]\,\text{Sin}[a2]\,\text{Sin}[a3]) - (-\text{Cos}[a5]\,(\text{Cos}[a3]\,\text{Sin}[a1]\,\text{Sin}[a2] + \text{Cos}[a1]\,\text{Cos}[a2] \\
&\quad \text{Sin}[a3])\,\text{Sin}[a4] - \text{Cos}[a4]\,(\text{Cos}[a3] - \text{Sin}[a1]\,\text{Sin}[a2]\,\text{Sin}[a3])\,\text{Sin}[a5])\,\text{Sin}[a6]) - \\
&(\text{Cos}[a6]\,(-\text{Cos}[a5]\,(\text{Cos}[a1]\,\text{Cos}[a2]\,\text{Cos}[a3] - \text{Sin}[a1]\,\text{Sin}[a2]\,\text{Sin}[a3])\,\text{Sin}[a4] + \text{Cos}[a4]\,(\text{Cos}[a3]\,\text{Sin}[a1]\,\text{Sin}[a2] + \\
&\quad \text{Cos}[a1]\,(\text{Cos}[a2]\,\text{Sin}[a3]))\,\text{Sin}[a5]) - (\text{Cos}[a1]\,\text{Cos}[a3]\,\text{Sin}[a2] - \text{Cos}[a2]\,\text{Sin}[a1]\,\text{Sin}[a3])\,\text{Sin}[a6])\,\text{Sin}[a7])\,\frac{a8}{\sqrt{3}}\Big] \Big) + \\[4pt]
&\left(\text{Cos}\!\left[\frac{a8}{\sqrt{3}}\right] (\text{Cos}[a4]\,\text{Cos}[a5]\,(\text{Cos}[a3]\,\text{Sin}[a1]\,\text{Sin}[a2] + \text{Cos}[a1]\,\text{Cos}[a2]\,\text{Sin}[a3]) + (\text{Cos}[a1]\,\text{Cos}[a2]\,\text{Cos}[a3] - \text{Sin}[a1]\,\text{Sin}[a2]\,\text{Sin}[a3])\right. \\
&\quad \text{Sin}[a4]\,\text{Sin}[a5]) + (\text{Cos}[a4]\,\text{Cos}[a5]\,(\text{Cos}[a1]\,\text{Cos}[a2]\,\text{Cos}[a3] - \text{Sin}[a1]\,\text{Sin}[a2]\,\text{Sin}[a3]) - \\
&\quad \left. (\text{Cos}[a3]\,\text{Sin}[a1]\,\text{Sin}[a2] + \text{Cos}[a1]\,\text{Cos}[a2]\,\text{Sin}[a3])\,\text{Sin}[a4]\,\text{Sin}[a5])\,\text{Sin}\!\left[\frac{a8}{\sqrt{3}}\right]\right)\text{Sin}\!\left[\frac{a9}{\sqrt{3}}\right] + \\[4pt]
&\text{Cos}\!\left[\frac{a10}{\sqrt{3}}\right] \left(\text{Cos}\!\left[\frac{a9}{\sqrt{3}}\right] \left(\text{Cos}\!\left[\frac{a8}{\sqrt{3}}\right] (\text{Cos}[a4]\,\text{Cos}[a5]\,(\text{Cos}[a1]\,\text{Cos}[a2]\,\text{Cos}[a3] - \text{Sin}[a1]\,\text{Sin}[a2]\,\text{Sin}[a3]) - \right.\right. \\
&\quad (\text{Cos}[a3]\,\text{Sin}[a1]\,\text{Sin}[a2] + \text{Cos}[a1]\,\text{Cos}[a2]\,\text{Sin}[a3])\,\text{Sin}[a4]\,\text{Sin}[a5]) - (\text{Cos}[a4]\,\text{Cos}[a5] \\
&\quad \left.(\text{Cos}[a3]\,\text{Sin}[a1]\,\text{Sin}[a2] + \text{Cos}[a1]\,\text{Cos}[a2]\,\text{Sin}[a3]) + (\text{Cos}[a1]\,\text{Cos}[a2]\,\text{Cos}[a3] - \text{Sin}[a1]\,\text{Sin}[a2]\,\text{Sin}[a3])\,\text{Sin}[a4]\,\text{Sin}[a5])\,\text{Sin}\!\left[\frac{a8}{\sqrt{3}}\right]\right) + \\[4pt]
&\left(\text{Cos}\!\left[\frac{a8}{\sqrt{3}}\right] (\text{Cos}[a7]\,(\text{Cos}[a6]\,(\text{Cos}[a2]\,\text{Cos}[a3]\,\text{Sin}[a1] + \text{Cos}[a1]\,\text{Sin}[a2]\,\text{Sin}[a3]) - (-\text{Cos}[a5]\,(\text{Cos}[a3]\,\text{Sin}[a1]\,\text{Sin}[a2] + \right. \\
&\quad \text{Cos}[a1]\,\text{Cos}[a2]\,\text{Sin}[a3])\,\text{Sin}[a4] - \text{Cos}[a4]\,(\text{Cos}[a1]\,\text{Cos}[a2]\,\text{Cos}[a3] - \text{Sin}[a1]\,\text{Sin}[a2]\,\text{Sin}[a3])\,\text{Sin}[a5])\,\text{Sin}[a6]) - \\
&\quad (\text{Cos}[a6]\,(-\text{Cos}[a5]\,(\text{Cos}[a1]\,\text{Cos}[a2]\,\text{Cos}[a3] - \text{Sin}[a1]\,\text{Sin}[a2]\,\text{Sin}[a3])\,\text{Sin}[a4] + \text{Cos}[a4]\,(\text{Cos}[a3]\,\text{Sin}[a1]\,\text{Sin}[a2] + \\
&\quad \text{Cos}[a1]\,\text{Cos}[a2]\,\text{Sin}[a3])\,\text{Sin}[a5]) - (\text{Cos}[a1]\,\text{Cos}[a3]\,\text{Sin}[a2] - \text{Cos}[a2]\,\text{Sin}[a1]\,\text{Sin}[a3])\,\text{Sin}[a6])\,\text{Sin}[a7]) - \\
&\quad (\text{Cos}[a7]\,(\text{Cos}[a6]\,(\text{Cos}[a1]\,\text{Cos}[a3]\,\text{Sin}[a2] - \text{Cos}[a2]\,\text{Sin}[a1]\,\text{Sin}[a3]) + (-\text{Cos}[a5]\,(\text{Cos}[a1]\,\text{Cos}[a2]\,\text{Cos}[a3] - \text{Sin}[a1]\,\text{Sin}[a2] \\
&\quad \text{Sin}[a3])\,\text{Sin}[a4] + \text{Cos}[a4]\,(\text{Cos}[a3]\,\text{Sin}[a1]\,\text{Sin}[a2] + \text{Cos}[a1]\,\text{Cos}[a2]\,\text{Sin}[a3])\,\text{Sin}[a5])\,\text{Sin}[a6]) - \\
&\quad (\text{Cos}[a6]\,(-\text{Cos}[a5]\,(\text{Cos}[a3]\,\text{Sin}[a1]\,\text{Sin}[a2] + \text{Cos}[a1]\,\text{Cos}[a2]\,\text{Sin}[a3])\,\text{Sin}[a4] - \text{Cos}[a4]\,(\text{Cos}[a3] - \text{Sin}[a1] \\
&\quad \left.\left.\left. \text{Sin}[a2]\,\text{Sin}[a3])\,\text{Sin}[a5]) + (\text{Cos}[a2]\,\text{Cos}[a3]\,\text{Sin}[a1] + \text{Cos}[a1]\,\text{Sin}[a2]\,\text{Sin}[a3])\,\text{Sin}[a6])\,\text{Sin}[a7])\,\text{Sin}\!\left[\frac{a8}{\sqrt{3}}\right]\right)\text{Sin}\!\left[\frac{a9}{\sqrt{3}}\right]\right)\right) + \\[6pt]
&\text{Cos}\!\left[\frac{a12}{\sqrt{3}}\right] \left(\text{Cos}\!\left[\frac{a11}{\sqrt{3}}\right] \left(-\text{Sin}\!\left[\frac{a10}{\sqrt{3}}\right] \left(\text{Cos}\!\left[\frac{a9}{\sqrt{3}}\right] \left(\text{Cos}\!\left[\frac{a8}{\sqrt{3}}\right] (\text{Cos}[a7]\,(\text{Cos}[a6]\,(\text{Cos}[a2]\,\text{Cos}[a3]\,\text{Sin}[a1] + \text{Cos}[a1]\,\text{Sin}[a2]\,\text{Sin}[a3]) - \right.\right.\right.\right.\right. \\
&\quad (-\text{Cos}[a5]\,(\text{Cos}[a3]\,\text{Sin}[a1]\,\text{Sin}[a2] + \text{Cos}[a1]\,\text{Cos}[a2]\,\text{Sin}[a3])\,\text{Sin}[a4] - \text{Cos}[a4]\,(\text{Cos}[a1]\,\text{Cos}[a2]\,\text{Cos}[a3] - \text{Sin}[a1]\,\text{Sin}[a2]\,\text{Sin}[a3]) \\
&\quad \text{Sin}[a5])\,\text{Sin}[a6]) - (\text{Cos}[a6]\,(-\text{Cos}[a5]\,(\text{Cos}[a1]\,\text{Cos}[a2]\,\text{Cos}[a3] - \text{Sin}[a1]\,\text{Sin}[a2]\,\text{Sin}[a3])\,\text{Sin}[a4] + \text{Cos}[a4]\,(\text{Cos}[a3]\,\text{Sin}[a1] \\
&\quad \text{Sin}[a2] + \text{Cos}[a1]\,\text{Cos}[a2]\,\text{Sin}[a3])\,\text{Sin}[a5]) - (\text{Cos}[a1]\,\text{Cos}[a3]\,\text{Sin}[a2] - \text{Cos}[a2]\,\text{Sin}[a1]\,\text{Sin}[a3])\,\text{Sin}[a6])\,\text{Sin}[a7]) - \\
&\quad (\text{Cos}[a7]\,(\text{Cos}[a6]\,(\text{Cos}[a1]\,\text{Cos}[a3]\,\text{Sin}[a2] - \text{Cos}[a2]\,\text{Sin}[a1]\,\text{Sin}[a3]) + (-\text{Cos}[a5]\,(\text{Cos}[a1]\,\text{Cos}[a2]\,\text{Cos}[a3] - \text{Sin}[a1]\,\text{Sin}[a2] \\
&\quad \text{Sin}[a3])\,\text{Sin}[a4] + \text{Cos}[a4]\,(\text{Cos}[a3]\,\text{Sin}[a1]\,\text{Sin}[a2] + \text{Cos}[a1]\,\text{Cos}[a2]\,\text{Sin}[a3])\,\text{Sin}[a5])\,\text{Sin}[a6]) - \\
&\quad (\text{Cos}[a6]\,(-\text{Cos}[a5]\,(\text{Cos}[a3]\,\text{Sin}[a1]\,\text{Sin}[a2] + \text{Cos}[a1]\,\text{Cos}[a2]\,\text{Sin}[a3])\,\text{Sin}[a4] - \text{Cos}[a4]\,(\text{Cos}[a3] - \\
&\quad \left.\text{Sin}[a1]\,\text{Sin}[a2]\,\text{Sin}[a3])\,\text{Sin}[a5]) + (\text{Cos}[a2]\,\text{Cos}[a3]\,\text{Sin}[a1] + \text{Cos}[a1]\,\text{Sin}[a2]\,\text{Sin}[a3])\,\text{Sin}[a6])\,\text{Sin}[a7])\,\text{Sin}\!\left[\frac{a8}{\sqrt{3}}\right]\right) - \\[4pt]
&\quad \left(\text{Cos}\!\left[\frac{a8}{\sqrt{3}}\right] (\text{Cos}[a4]\,\text{Cos}[a5]\,(\text{Cos}[a1]\,\text{Cos}[a2]\,\text{Cos}[a3] - \text{Sin}[a1]\,\text{Sin}[a2]\,\text{Sin}[a3]) - (\text{Cos}[a3]\,\text{Sin}[a1]\,\text{Sin}[a2] + \text{Cos}[a1]\,\text{Cos}[a2]\,\text{Sin}[a3])\right. \\
&\quad \text{Sin}[a4]\,\text{Sin}[a5]) - (\text{Cos}[a4]\,\text{Cos}[a5]\,(\text{Cos}[a3]\,\text{Sin}[a1]\,\text{Sin}[a2] + \text{Cos}[a1]\,\text{Cos}[a2]\,\text{Sin}[a3]) + \\
&\quad \left. (\text{Cos}[a1]\,\text{Cos}[a2]\,\text{Cos}[a3] - \text{Sin}[a1]\,\text{Sin}[a2]\,\text{Sin}[a3])\,\text{Sin}[a4]\,\text{Sin}[a5])\,\text{Sin}\!\left[\frac{a8}{\sqrt{3}}\right]\right)\text{Sin}\!\left[\frac{a9}{\sqrt{3}}\right]\right) + \\[4pt]
&\quad \text{Cos}\!\left[\frac{a10}{\sqrt{3}}\right] \left(\text{Cos}\!\left[\frac{a9}{\sqrt{3}}\right] \left(\text{Cos}\!\left[\frac{a8}{\sqrt{3}}\right] (\text{Cos}[a4]\,\text{Cos}[a5]\,(\text{Cos}[a3]\,\text{Sin}[a1]\,\text{Sin}[a2] + \text{Cos}[a1]\,\text{Cos}[a2]\,\text{Sin}[a3]) + \right.\right. \\
&\quad (\text{Cos}[a1]\,\text{Cos}[a2]\,\text{Cos}[a3] - \text{Sin}[a1]\,\text{Sin}[a2]\,\text{Sin}[a3])\,\text{Sin}[a4]\,\text{Sin}[a5]) + (\text{Cos}[a4]\,\text{Cos}[a5] \\
&\quad \left.(\text{Cos}[a1]\,\text{Cos}[a2]\,\text{Cos}[a3] - \text{Sin}[a1]\,\text{Sin}[a2]\,\text{Sin}[a3]) - (\text{Cos}[a3]\,\text{Sin}[a1]\,\text{Sin}[a2] + \text{Cos}[a1]\,\text{Cos}[a2]\,\text{Sin}[a3])\,\text{Sin}[a4]\,\text{Sin}[a5])\,\text{Sin}\!\left[\frac{a8}{\sqrt{3}}\right]\right) - \\[4pt]
&\quad \left(\text{Cos}\!\left[\frac{a8}{\sqrt{3}}\right] (\text{Cos}[a7]\,(\text{Cos}[a6]\,(\text{Cos}[a1]\,\text{Cos}[a3]\,\text{Sin}[a2] - \text{Cos}[a2]\,\text{Sin}[a1]\,\text{Sin}[a3]) + (-\text{Cos}[a5]\,(\text{Cos}[a1]\,\text{Cos}[a2]\,\text{Cos}[a3] - \right. \\
&\quad \text{Sin}[a1]\,\text{Sin}[a2]\,\text{Sin}[a3])\,\text{Sin}[a4] + \text{Cos}[a4]\,(\text{Cos}[a3]\,\text{Sin}[a1]\,\text{Sin}[a2] + \text{Cos}[a1]\,\text{Cos}[a2]\,\text{Sin}[a3])\,\text{Sin}[a5])\,\text{Sin}[a6]) - \\
&\quad (\text{Cos}[a6]\,(-\text{Cos}[a5]\,(\text{Cos}[a3]\,\text{Sin}[a1]\,\text{Sin}[a2] + \text{Cos}[a1]\,\text{Cos}[a2]\,\text{Sin}[a3])\,\text{Sin}[a4] - \text{Cos}[a4]\,(\text{Cos}[a3] - \\
&\quad \text{Sin}[a1]\,\text{Sin}[a2]\,\text{Sin}[a3])\,\text{Sin}[a5]) + (\text{Cos}[a2]\,\text{Cos}[a3]\,\text{Sin}[a1] + \text{Cos}[a1]\,\text{Sin}[a2]\,\text{Sin}[a3])\,\text{Sin}[a6])\,\text{Sin}[a7]) + 
\end{aligned}
$$

$$(\mathrm{Cos}[a7]\ (\mathrm{Cos}[a6]\ (\mathrm{Cos}[a2]\ \mathrm{Cos}[a3]\ \mathrm{Sin}[a1] + \mathrm{Cos}[a1]\ \mathrm{Sin}[a2]\ \mathrm{Sin}[a3]) - (\mathrm{Cos}[a5]\ (\mathrm{Cos}[a3]\ \mathrm{Sin}[a1]\ \mathrm{Sin}[a2] + \mathrm{Cos}[a1]\ \mathrm{Cos}[a2]$$
$$\mathrm{Sin}[a3])\ \mathrm{Sin}[a4] - \mathrm{Cos}[a4]\ (\mathrm{Cos}[a1]\ \mathrm{Cos}[a2]\ \mathrm{Cos}[a3] - \mathrm{Sin}[a1]\ \mathrm{Sin}[a2]\ \mathrm{Sin}[a3])\ \mathrm{Sin}[a5])\ \mathrm{Sin}[a6]) -$$
$$(\mathrm{Cos}[a6]\ (-\mathrm{Cos}[a5]\ (\mathrm{Cos}[a1]\ \mathrm{Cos}[a2]\ \mathrm{Cos}[a3] - \mathrm{Sin}[a1]\ \mathrm{Sin}[a2]\ \mathrm{Sin}[a3])\ \mathrm{Sin}[a4] + \mathrm{Cos}[a4]\ (\mathrm{Cos}[a3]\ \mathrm{Sin}[a1]\ \mathrm{Sin}[a2] + \mathrm{Cos}[a1]$$
$$\mathrm{Cos}[a2]\ \mathrm{Sin}[a3])\ \mathrm{Sin}[a5]) - (\mathrm{Cos}[a1]\ \mathrm{Cos}[a3]\ \mathrm{Sin}[a2] - \mathrm{Cos}[a2]\ \mathrm{Sin}[a1]\ \mathrm{Sin}[a3])\ \mathrm{Sin}[a6])\ \mathrm{Sin}[a7])\ \mathrm{Sin}\!\left[\tfrac{a8}{\sqrt{3}}\right]\!\Big)\ \mathrm{Sin}\!\left[\tfrac{a9}{\sqrt{3}}\right]\Big)\Big) +$$

$$\mathrm{Sin}\!\left[\tfrac{a11}{\sqrt{3}}\right]\!\left(\mathrm{Cos}\!\left[\tfrac{2\,a10}{\sqrt{3}}\right]\!\left(\mathrm{Cos}\!\left[\tfrac{2\,a8}{\sqrt{3}}\right]\ (\mathrm{Cos}[a7]\ (\mathrm{Cos}[a6]\ (-\mathrm{Cos}[a5]\ (\mathrm{Cos}[a3]\ \mathrm{Sin}[a1]\ \mathrm{Sin}[a2] + \mathrm{Cos}[a1]\ \mathrm{Cos}[a2]\ \mathrm{Sin}[a3])\ \mathrm{Sin}[a4] - \mathrm{Cos}[a4]\right.\right.$$
$$(\mathrm{Cos}[a1]\ \mathrm{Cos}[a2]\ \mathrm{Cos}[a3] - \mathrm{Sin}[a1]\ \mathrm{Sin}[a2]\ \mathrm{Sin}[a3])\ \mathrm{Sin}[a5]) + (\mathrm{Cos}[a2]\ \mathrm{Cos}[a3]\ \mathrm{Sin}[a1] + \mathrm{Cos}[a1]\ \mathrm{Cos}[a2]\ \mathrm{Sin}[a3])\ \mathrm{Sin}[a6]) +$$
$$(\mathrm{Cos}[a6]\ (\mathrm{Cos}[a1]\ \mathrm{Cos}[a2]\ \mathrm{Cos}[a3] - \mathrm{Sin}[a1]\ \mathrm{Sin}[a2]\ \mathrm{Sin}[a3]) + (-\mathrm{Cos}[a1]\ \mathrm{Cos}[a2]\ \mathrm{Cos}[a3] - \mathrm{Sin}[a1]\ \mathrm{Sin}[a2]\ \mathrm{Sin}[a3])$$
$$\mathrm{Sin}[a4] + \mathrm{Cos}[a4]\ (\mathrm{Cos}[a3]\ \mathrm{Sin}[a1]\ \mathrm{Sin}[a2] + \mathrm{Cos}[a1]\ \mathrm{Cos}[a2]\ \mathrm{Sin}[a3])\ \mathrm{Sin}[a5])\ \mathrm{Sin}[a6])\ \mathrm{Sin}[a7]) -$$
$$(\mathrm{Cos}[a7]\ (\mathrm{Cos}[a6]\ (\mathrm{Cos}[a1]\ \mathrm{Cos}[a2]\ \mathrm{Cos}[a3] - \mathrm{Sin}[a1]\ \mathrm{Sin}[a2]\ \mathrm{Sin}[a3])\ \mathrm{Sin}[a5]) - (\mathrm{Cos}[a1]\ \mathrm{Cos}[a3]\ \mathrm{Sin}[a2] - \mathrm{Cos}[a2]\ \mathrm{Sin}[a1]\ \mathrm{Sin}[a3])\ \mathrm{Sin}[a6]) +$$
$$(\mathrm{Cos}[a6]\ (\mathrm{Cos}[a2]\ \mathrm{Cos}[a3]\ \mathrm{Sin}[a1] + \mathrm{Cos}[a1]\ \mathrm{Cos}[a2]\ \mathrm{Sin}[a3]) - (-\mathrm{Cos}[a5]\ (\mathrm{Cos}[a3]\ \mathrm{Sin}[a1]\ \mathrm{Sin}[a2] + \mathrm{Cos}[a1]\ \mathrm{Cos}[a2]\ \mathrm{Sin}[a3])$$
$$\mathrm{Sin}[a4] - \mathrm{Cos}[a4]\ (\mathrm{Cos}[a1]\ \mathrm{Cos}[a2]\ \mathrm{Cos}[a3] - \mathrm{Sin}[a1]\ \mathrm{Sin}[a2]\ \mathrm{Sin}[a3])\ \mathrm{Sin}[a5])\ \mathrm{Sin}[a6])\ \mathrm{Sin}[a7])\ \mathrm{Sin}\!\left[\tfrac{2\,a8}{\sqrt{3}}\right]\Big) -$$

$$\mathrm{Sin}\!\left[\tfrac{2\,a10}{\sqrt{3}}\right]\!\left(\mathrm{Cos}\!\left[\tfrac{2\,a8}{\sqrt{3}}\right]\ (\mathrm{Cos}[a7]\ (\mathrm{Cos}[a6]\ (-\mathrm{Cos}[a5]\ (\mathrm{Cos}[a1]\ \mathrm{Cos}[a2]\ \mathrm{Cos}[a3] - \mathrm{Sin}[a1]\ \mathrm{Sin}[a2]\ \mathrm{Sin}[a3])\ \mathrm{Sin}[a4] + \mathrm{Cos}[a4]\right.$$
$$(\mathrm{Cos}[a3]\ \mathrm{Sin}[a1]\ \mathrm{Sin}[a2] + \mathrm{Cos}[a1]\ \mathrm{Cos}[a2]\ \mathrm{Sin}[a3])\ \mathrm{Sin}[a5]) - (\mathrm{Cos}[a1]\ \mathrm{Cos}[a3]\ \mathrm{Sin}[a2] - \mathrm{Cos}[a2]\ \mathrm{Sin}[a1]\ \mathrm{Sin}[a3])\ \mathrm{Sin}[a6]) +$$
$$(\mathrm{Cos}[a6]\ (\mathrm{Cos}[a2]\ \mathrm{Cos}[a3]\ \mathrm{Sin}[a1] + \mathrm{Cos}[a1]\ \mathrm{Cos}[a2]\ \mathrm{Sin}[a3]) - (-\mathrm{Cos}[a5]\ (\mathrm{Cos}[a3]\ \mathrm{Sin}[a1]\ \mathrm{Sin}[a2] + \mathrm{Cos}[a1]\ \mathrm{Cos}[a2]\ \mathrm{Sin}[a3])$$
$$\mathrm{Sin}[a4] - \mathrm{Cos}[a4]\ (\mathrm{Cos}[a1]\ \mathrm{Cos}[a2]\ \mathrm{Cos}[a3] - \mathrm{Sin}[a1]\ \mathrm{Sin}[a2]\ \mathrm{Sin}[a3])\ \mathrm{Sin}[a5])\ \mathrm{Sin}[a6])\ \mathrm{Sin}[a7]) -$$
$$(\mathrm{Cos}[a7]\ (\mathrm{Cos}[a6]\ (-\mathrm{Cos}[a5]\ (\mathrm{Cos}[a3]\ \mathrm{Sin}[a1]\ \mathrm{Sin}[a2] + \mathrm{Cos}[a1]\ \mathrm{Cos}[a2]\ \mathrm{Sin}[a3])\ \mathrm{Sin}[a4] - \mathrm{Cos}[a4]\ (\mathrm{Cos}[a1]\ \mathrm{Cos}[a2]\ \mathrm{Cos}[a3] -$$
$$\mathrm{Sin}[a1]\ \mathrm{Sin}[a2]\ \mathrm{Sin}[a3])\ \mathrm{Sin}[a5]) + (\mathrm{Cos}[a2]\ \mathrm{Cos}[a3]\ \mathrm{Sin}[a1] + \mathrm{Cos}[a1]\ \mathrm{Sin}[a2]\ \mathrm{Sin}[a3])\ \mathrm{Sin}[a6]) +$$
$$(\mathrm{Cos}[a6]\ (\mathrm{Cos}[a1]\ \mathrm{Cos}[a3]\ \mathrm{Sin}[a2] - \mathrm{Cos}[a2]\ \mathrm{Sin}[a1]\ \mathrm{Sin}[a3]) + (-\mathrm{Cos}[a5]\ (\mathrm{Cos}[a3]\ \mathrm{Sin}[a1]\ \mathrm{Sin}[a2] + \mathrm{Cos}[a1]\ \mathrm{Sin}[a2]\ \mathrm{Sin}[a3])$$
$$\mathrm{Sin}[a4] + \mathrm{Cos}[a4]\ (\mathrm{Cos}[a3]\ \mathrm{Sin}[a1]\ \mathrm{Sin}[a2] + \mathrm{Cos}[a1]\ \mathrm{Cos}[a2]\ \mathrm{Sin}[a3])\ \mathrm{Sin}[a5])\ \mathrm{Sin}[a6])\ \mathrm{Sin}[a7])\ \mathrm{Sin}\!\left[\tfrac{2\,a8}{\sqrt{3}}\right]\!\Big)\ \mathrm{Sin}\!\left[\tfrac{2\,a9}{\sqrt{3}}\right]\Big)\Big)\Big)$$

Similar experiment reveals similar 'coiled' behavior [3]:

```
a = {a1, a2, a3, a4, a5, a6, a7, a8, a9, a10, a11, a12, a13, a14};
gexp = IdentityMatrix[7];
Table[gexp = gexp.MatrixExp[(a[[i]] * gbasis2[[3]][[i]])], {i, 1, 14}];

gexp[[7]][[6]]
```

$-\mathrm{Sin}[a11]\ \mathrm{Sin}[a12]$
$\quad(\mathrm{Cos}[a10]\ (\mathrm{Cos}[a9]\ (\mathrm{Cos}[a8]\ (\mathrm{Cos}[a3]\ \mathrm{Cos}[a5]\ \mathrm{Cos}[a6] + \mathrm{Sin}[a3]\ \mathrm{Sin}[a5]\ \mathrm{Sin}[a6]) +$
$\qquad (-\mathrm{Cos}[a6]\ \mathrm{Sin}[a3]\ \mathrm{Sin}[a5] + \mathrm{Cos}[a3]\ \mathrm{Cos}[a5]\ \mathrm{Sin}[a6])\ \mathrm{Sin}[a7])\ \mathrm{Sin}[a8]) -$
$\qquad (\mathrm{Cos}[a8]\ (-\mathrm{Cos}[a3]\ \mathrm{Cos}[a6]\ \mathrm{Sin}[a5] - \mathrm{Cos}[a5]\ \mathrm{Sin}[a3]\ \mathrm{Sin}[a6])\ \mathrm{Sin}[a7] +$
$\qquad (\mathrm{Cos}[a5]\ \mathrm{Cos}[a6]\ \mathrm{Sin}[a3] - \mathrm{Cos}[a3]\ \mathrm{Sin}[a5]\ \mathrm{Sin}[a6])\ \mathrm{Sin}[a8])\ \mathrm{Sin}[a9]) +$
$\quad \mathrm{Sin}[a10]\ (\mathrm{Cos}[a9]\ (\mathrm{Cos}[a8]\ (\mathrm{Cos}[a5]\ \mathrm{Cos}[a6]\ \mathrm{Sin}[a3] - \mathrm{Cos}[a3]\ \mathrm{Sin}[a5]\ \mathrm{Sin}[a6]) -$
$\qquad (-\mathrm{Cos}[a3]\ \mathrm{Cos}[a6]\ \mathrm{Sin}[a5] - \mathrm{Cos}[a5]\ \mathrm{Sin}[a3]\ \mathrm{Sin}[a6])\ \mathrm{Sin}[a7]\ \mathrm{Sin}[a8]) -$
$\qquad (\mathrm{Cos}[a8]\ (-\mathrm{Cos}[a6]\ \mathrm{Sin}[a3]\ \mathrm{Sin}[a5] + \mathrm{Cos}[a3]\ \mathrm{Cos}[a5]\ \mathrm{Sin}[a6])\ \mathrm{Sin}[a7] -$
$\qquad (\mathrm{Cos}[a3]\ \mathrm{Cos}[a5]\ \mathrm{Cos}[a6] + \mathrm{Sin}[a3]\ \mathrm{Sin}[a5]\ \mathrm{Sin}[a6])\ \mathrm{Sin}[a8])\ \mathrm{Sin}[a9])) +$
$\quad \mathrm{Cos}[a12]\ (-\mathrm{Cos}[a7]\ \mathrm{Sin}[a11]\ (-\mathrm{Cos}[a3]\ \mathrm{Cos}[a6]\ \mathrm{Sin}[a5] - \mathrm{Cos}[a5]\ \mathrm{Sin}[a3]\ \mathrm{Sin}[a6]) +$
$\qquad \mathrm{Cos}[a11]\ (\mathrm{Sin}[a10]\ (\mathrm{Cos}[a9]\ (\mathrm{Cos}[a8]\ (-\mathrm{Cos}[a3]\ \mathrm{Cos}[a6]\ \mathrm{Sin}[a5] - \mathrm{Cos}[a5]\ \mathrm{Sin}[a3]\ \mathrm{Sin}[a6])$
$\qquad\quad \mathrm{Sin}[a7] + (\mathrm{Cos}[a5]\ \mathrm{Cos}[a6]\ \mathrm{Sin}[a3] - \mathrm{Cos}[a3]\ \mathrm{Sin}[a5]\ \mathrm{Sin}[a6])\ \mathrm{Sin}[a8]) +$
$\qquad\quad (\mathrm{Cos}[a8]\ (\mathrm{Cos}[a3]\ \mathrm{Cos}[a5]\ \mathrm{Cos}[a6] + \mathrm{Sin}[a3]\ \mathrm{Sin}[a5]\ \mathrm{Sin}[a6]) +$
$\qquad\quad (-\mathrm{Cos}[a6]\ \mathrm{Sin}[a3]\ \mathrm{Sin}[a5] + \mathrm{Cos}[a3]\ \mathrm{Cos}[a5]\ \mathrm{Sin}[a6])\ \mathrm{Sin}[a7]\ \mathrm{Sin}[a8])\ \mathrm{Sin}[a9]) +$
$\qquad \mathrm{Cos}[a10]\ (\mathrm{Cos}[a9]\ (\mathrm{Cos}[a8]\ (-\mathrm{Cos}[a6]\ \mathrm{Sin}[a3]\ \mathrm{Sin}[a5] + \mathrm{Cos}[a3]\ \mathrm{Cos}[a5]\ \mathrm{Sin}[a6])$
$\qquad\quad \mathrm{Sin}[a7] - (\mathrm{Cos}[a3]\ \mathrm{Cos}[a5]\ \mathrm{Cos}[a6] + \mathrm{Sin}[a3]\ \mathrm{Sin}[a5]\ \mathrm{Sin}[a6])\ \mathrm{Sin}[a8]) +$
$\qquad\quad (\mathrm{Cos}[a8]\ (\mathrm{Cos}[a5]\ \mathrm{Cos}[a6]\ \mathrm{Sin}[a3] - \mathrm{Cos}[a3]\ \mathrm{Sin}[a5]\ \mathrm{Sin}[a6]) - (-\mathrm{Cos}[a3]$
$\qquad\quad\quad \mathrm{Cos}[a6]\ \mathrm{Sin}[a5] - \mathrm{Cos}[a5]\ \mathrm{Sin}[a3]\ \mathrm{Sin}[a6])\ \mathrm{Sin}[a7]\ \mathrm{Sin}[a8])\ \mathrm{Sin}[a9])))$

# 4.1 BCH Formula for Exp: $\mathfrak{g}_2 \longrightarrow G_2$

```
(* Choose any vectors from g₂, for the sake of clarity choose C1 and C14 *)
A = C1;
B = C14;
(* Notice C1C14 - C14C1 ≠ 0 i.e. they do not commute *)
N[MatrixCommutator[A, B]] // MatrixForm
```

$$\begin{pmatrix}
0. & 0. & 0. & 1.1547 & 0. & 0. & 0. \\
0. & 0. & 0. & 0. & 0. & 0. & 0.57735 \\
0. & 0. & 0. & 0. & 0. & -0.57735 & \\
-1.1547 & 0. & 0. & 0. & 0. & 0. & 0. \\
0. & 0. & 0. & 0. & 0. & 0. & 0. \\
0. & 0. & 0.57735 & 0. & 0. & 0. & 0. \\
0. & -0.57735 & 0. & 0. & 0. & 0. & 0.
\end{pmatrix}$$

Let's compute the exponentiations and verify $e^{C1} e^{C14} \neq e^{C1+C14}$. Notice exponentiations are multiplied in $\mathsf{G}_2$ i.e. matrix multiplied.

```
A = C1;
B = C14;
(* additional call to N necessary to assure Real printout *)
N[MatrixExp[N[A + B]]] // MatrixForm
N[MatrixExp[B].MatrixExp[A]] // MatrixForm
```

$$\begin{pmatrix}
0.452726 & 0.0971483 & 0. & -0.45928 & 0. & 0. & -0.758066 \\
-0.0971483 & 0.850473 & 0. & 0.463166 & 0. & 0. & -0.22964 \\
0. & 0. & 0.851048 & 0. & -0.457337 & 0.257992 & 0. \\
-0.45928 & -0.463166 & 0. & 0.452726 & 0. & 0. & -0.60793 \\
0. & 0. & 0.457337 & 0. & 0.404192 & -0.792131 & 0. \\
0. & 0. & 0.257992 & 0. & 0.792131 & 0.553144 & 0. \\
0.758066 & -0.22964 & 0. & 0.60793 & 0. & 0. & 0.0549779
\end{pmatrix}$$

$$\begin{pmatrix}
0.404192 & 0. & 0. & -0.769672 & 0. & 0. & -0.4942 \\
0. & 0.837912 & 0. & 0.2949 & 0. & 0. & -0.45928 \\
0. & 0. & 0.837912 & 0. & -0.2949 & 0.45928 & 0. \\
0. & -0.545806 & 0. & 0.452726 & 0. & 0. & -0.705078 \\
0. & 0. & 0.545806 & 0. & 0.452726 & -0.705078 & 0. \\
0. & 0. & 0. & 0. & 0.841471 & 0.540302 & 0. \\
0.914674 & 0. & 0. & 0.340116 & 0. & 0. & 0.218386
\end{pmatrix}$$

Let's compute the exponentiations and verify $e^{C1+C14} \neq e^{C14+C1}$ i.e. the exponentiations in $\mathsf{G}_2$ do not commute although the exponents do i.e. C1 + C14 = C14 + C1

```
A = C1;
B = C14;
N[MatrixExp[A].MatrixExp[B]] // MatrixForm
N[MatrixExp[B].MatrixExp[A]] // MatrixForm
```

$$
\begin{pmatrix}
0.404192 & 0. & 0. & 0. & 0. & 0. & -0.914674 \\
0. & 0.837912 & 0. & 0.545806 & 0. & 0. & 0. \\
0. & 0. & 0.837912 & 0. & -0.545806 & 0. & 0. \\
-0.769672 & -0.2949 & 0. & 0.452726 & 0. & 0. & -0.340116 \\
0. & 0. & 0.2949 & 0. & 0.452726 & -0.841471 & 0. \\
0. & 0. & 0.45928 & 0. & 0.705078 & 0.540302 & 0. \\
0.4942 & -0.45928 & 0. & 0.705078 & 0. & 0. & 0.218386
\end{pmatrix}
$$

$$
\begin{pmatrix}
0.404192 & 0. & 0. & -0.769672 & 0. & 0. & -0.4942 \\
0. & 0.837912 & 0. & 0.2949 & 0. & 0. & -0.45928 \\
0. & 0. & 0.837912 & 0. & -0.2949 & 0.45928 & 0. \\
0. & -0.545806 & 0. & 0.452726 & 0. & 0. & -0.705078 \\
0. & 0. & 0.545806 & 0. & 0.452726 & -0.705078 & 0. \\
0. & 0. & 0. & 0. & 0.841471 & 0.540302 & 0. \\
0.914674 & 0. & 0. & 0.340116 & 0. & 0. & 0.218386
\end{pmatrix}
$$

## 4.2 $\displaystyle\lim_{n\to\infty} e^{A+B+\text{BCH}[A,B,n]} = e^A\, e^B$

Now calculate the BCH say up to second order terms i.e. make the third argument to BCH[A, B, 2] number 2:

```
A = C1;
B = C14;
bch = BCH[A, B, 2];
bch // MatrixForm
```

$$
\begin{pmatrix}
0 & 0 & 0 & \frac{1}{\sqrt{3}} & 0 & 0 & 0 \\
0 & 0 & 0 & 0 & 0 & 0 & \frac{1}{2\sqrt{3}} \\
0 & 0 & 0 & 0 & 0 & -\frac{1}{2\sqrt{3}} & 0 \\
-\frac{1}{\sqrt{3}} & 0 & 0 & 0 & 0 & 0 & 0 \\
0 & 0 & 0 & 0 & 0 & 0 & 0 \\
0 & 0 & \frac{1}{2\sqrt{3}} & 0 & 0 & 0 & 0 \\
0 & -\frac{1}{2\sqrt{3}} & 0 & 0 & 0 & 0 & 0
\end{pmatrix}
$$

See if there are any improvements:

```
A = C1;
B = C14;

(* Calculate eᴬeᴮ *)
expAB = N[MatrixExp[A].MatrixExp[B]];
(* calculate  eᴬ⁺ᴮ *)
expAplusB = N[MatrixExp[N[A + B]]];

(* calculate the distance between eᴬeᴮ and eᴬ⁺ᴮ , this would be the case with no BCH *)
Norm[expAB - expAplusB ]

(* calculate the BCH series up to two second degree terms *)
bch = BCH[A, B, 2];
(* add the BCH to the exponent i.e. A+B+BCH *)
expAplusBplusBCH = N[MatrixExp[N[A + B + bch]]];

(* calculate the distance between eᴬeᴮ and eᴬ⁺ᴮ⁺ᴮᶜᴴ,
the distance should be less than the case without BCH *)
Norm[expAB - expAplusBplusBCH ]
```

```
0.52272

0.176082
```

Now calculate the BCH up to 4th degree terms, the distance between $e^A\,e^B$ and $e^{A+B+\text{BCH}[A,B,4]}$ lessens :

```
A = C1;
B = C14;

(* Calculate eᴬeᴮ *)
expAB = N[MatrixExp[A].MatrixExp[B]];
(* calculate  eᴬ⁺ᴮ *)
expAplusB = N[MatrixExp[N[A + B]]];

(* calculate the distance between eᴬeᴮ and eᴬ⁺ᴮ , this would be the case with no BCH *)
Norm[expAB - expAplusB ]

(* calculate the BCH series up to two second degree terms *)
bch = BCH[A, B, 4];
(* add the BCH to the exponent i.e. A+B+BCH *)
expAplusBplusBCH = N[MatrixExp[N[A + B + bch]]];

(* calculate the distance between eᴬeᴮ and eᴬ⁺ᴮ⁺ᴮᶜᴴ,
the distance should be less than the case without BCH *)
Norm[expAB - expAplusBplusBCH ]
```

Now calculate the BCH up to 8th degree terms, the distance between $e^A\,e^B$ and $e^{A+B+\text{BCH}[A,B,8]}$ drastically drops :

```
A = C1;
B = C14;

(* Calculate e^A e^B *)
expAB = N[MatrixExp[A].MatrixExp[B]];
(* calculate  e^{A+B} *)
expAplusB = N[MatrixExp[N[A + B]]];

(* calculate the distance between e^A e^B and e^{A+B} , this would be the case with no BCH *)
Norm[expAB - expAplusB ]

(* calculate the BCH series up to two second degree terms *)
bch = BCH[A, B, 8];
(* add the BCH to the exponent i.e. A+B+BCH *)
expAplusBplusBCH = N[MatrixExp[N[A + B + bch]]];

(* calculate the distance between e^A e^B and e^{A+B+BCH},
the distance should be less than the case without BCH *)
Norm[expAB - expAplusBplusBCH ]
```

0.52272

0.000159433

In order to understand the metric applied to the matrices in $G_2$ let's output the computations, first matrix is $e^{A+B}$, second $e^A e^B$ and the third $e^{A+B+\text{BCH}[A,B,8]}$ and as you can see the last two matrices quite close in terms of the numbers in corresponding entries:

```
A = C1;
B = C14;

(* calculate  e^{A+B} *)
expAplusB = N[MatrixExp[N[A + B]]];

expAplusB // MatrixForm

(* Calculate e^A e^B *)
expAB = N[MatrixExp[A].MatrixExp[B]];

expAB // MatrixForm

(* calculate the BCH series up to two second degree terms *)
bch = BCH[A, B, 8];
(* add the BCH to the exponent i.e. A+B+BCH *)
expAplusBplusBCH = N[MatrixExp[N[A + B + bch]]];

expAplusBplusBCH // MatrixForm
```

$$
\begin{pmatrix}
0.452726 & 0.0971483 & 0. & -0.45928 & 0. & 0. & -0.758066 \\
-0.0971483 & 0.850473 & 0. & 0.463166 & 0. & 0. & -0.22964 \\
0. & 0. & 0.851048 & 0. & -0.457337 & 0.257992 & 0. \\
-0.45928 & -0.463166 & 0. & 0.452726 & 0. & 0. & -0.60793 \\
0. & 0. & 0.457337 & 0. & 0.404192 & -0.792131 & 0. \\
0. & 0. & 0.257992 & 0. & 0.792131 & 0.553144 & 0. \\
0.758066 & -0.22964 & 0. & 0.60793 & 0. & 0. & 0.0549779
\end{pmatrix}
$$

$$
\begin{pmatrix}
0.404192 & 0. & 0. & 0. & 0. & 0. & -0.914674 \\
0. & 0.837912 & 0. & 0.545806 & 0. & 0. & 0. \\
0. & 0. & 0.837912 & 0. & -0.545806 & 0. & 0. \\
-0.769672 & -0.2949 & 0. & 0.452726 & 0. & 0. & -0.340116 \\
0. & 0. & 0.2949 & 0. & 0.452726 & -0.841471 & 0. \\
0. & 0. & 0.45928 & 0. & 0.705078 & 0.540302 & 0. \\
0.4942 & -0.45928 & 0. & 0.705078 & 0. & 0. & 0.218386
\end{pmatrix}
$$

$$
\begin{pmatrix}
0.404211 & -0.000134605 & 0. & 0.0000819156 & 0. & 0. & -0.914666 \\
0.000156294 & 0.837922 & 0. & 0.54579 & 0. & 0. & -5.36129 \times 10^{-6} \\
0. & 0. & 0.837913 & 0. & -0.545804 & -1.31531 \times 10^{-6} & 0. \\
-0.769662 & -0.294824 & 0. & 0.452844 & 0. & 0. & -0.340047 \\
0. & 0. & 0.294899 & 0. & 0.452727 & -0.841471 & 0. \\
0. & 0. & 0.459279 & 0. & 0.705079 & 0.540303 & 0. \\
0.4942 & -0.45931 & 0. & 0.705015 & 0. & 0. & 0.218529
\end{pmatrix}
$$

# 5. BCH Kolmogorov Complexity

As it turned out:

$$\underset{n\to\infty}{\text{Lim}}\ e^{A+B+BCH[A,B,n]} = e^A\,e^B \qquad\qquad 5.1$$

Let's look at the algorithm that computes BCH[A, B, n], the core of the algorithm is the function p which generates the series [6]:

```
(* Alogorithm by Matthias W. Reinsch [6] *)
p[n_] := p[n] = (F = Table[1 / (j - i) !, {i, n + 1}, {j, n + 1}];
    G = Table[1 / (j - i) ! Product[s[k], {k, i, j - 1}], {i, n + 1}, {j, n + 1}];
    qthpower = IdentityMatrix[n + 1]; FGm1 = F.G - qthpower;
    Expand[-Sum[qthpower = qthpower.FGm1; (-1) ^q / q qthpower, {q, n}]][[1, n + 1]])

(* This code just plugs in the symbols for matrices *)
translated[n_] :=
  (temp = Expand[Product[s[k]^2, {k, n}] p[n]]; Sum[term = Apply[List, temp[[i]]];
    term[[1]] Apply[Dot, Take[term, -n] /. {s[i_]^2 -> Symbol["x"], s[i_]^3 -> Symbol["y"]}],
    {i, Length[temp]}])
```

For n = 2 the commutator is obtained from the algorithm:

```
translated[2] /. {Symbol["x"] -> A, Symbol["y"] -> B}
```

$$\frac{A.B}{2} - \frac{B.A}{2}$$

n = 3:

```
translated[3] /. {Symbol["x"] -> A, Symbol["y"] -> B}
```

$$\frac{A.A.B}{12} - \frac{A.B.A}{6} + \frac{A.B.B}{12} + \frac{B.A.A}{12} - \frac{B.A.B}{6} + \frac{B.B.A}{12}$$

n = 6:

```
translated[6] /. {Symbol["x"] -> A, Symbol["y"] -> B}
```

$$-\frac{A.A.A.B.B}{1440} + \frac{1}{360}A.A.A.B.A.B + \frac{1}{360}A.A.A.B.B.B - \frac{1}{240}A.A.B.A.A.B - \frac{1}{240}A.A.B.A.B.B -$$

$$\frac{1}{240}A.A.B.B.A.B - \frac{A.A.B.B.B.B}{1440} + \frac{1}{360}A.B.A.A.A.B - \frac{1}{240}A.B.A.A.B.B + \frac{1}{60}A.B.A.B.A.B +$$

$$\frac{1}{360}A.B.A.B.B.B - \frac{1}{240}A.B.B.A.A.B - \frac{1}{240}A.B.B.A.B.B + \frac{1}{360}A.B.B.B.A.B -$$

$$\frac{1}{360}B.A.A.B.A.B + \frac{1}{240}B.A.A.B.A.A + \frac{1}{240}B.A.A.B.B.A - \frac{1}{360}B.A.B.A.A.A -$$

$$\frac{1}{60}B.A.B.A.B.A + \frac{1}{240}B.A.B.B.A.A - \frac{1}{360}B.A.B.B.B.A + \frac{B.B.A.A.A.A}{1440} + \frac{1}{240}B.B.A.A.B.A +$$

$$\frac{1}{240}B.B.A.B.A.A + \frac{1}{240}B.B.B.A.A - \frac{1}{360}B.B.B.A.A.A - \frac{1}{360}B.B.B.A.B.A + \frac{B.B.B.B.A.A}{1440}$$

It can easily be seen that the algorithm is actually comprised of series of loops, while the loop-counters are functions of n, and even though the time and output length grows linearly, the actual code for p is constant in size. In other words no matter what precision BCH series needed the same code can generate the series. And we can most probably assume that such p is the shortest such program thanks to its simplicity.

Therefore Kolmogorov Complexity of BCH[A, B, n] is a known constant value.

Moreover it is easy to verify that the extension of this algorithm for more than 2 arguments (parameters) i.e. BCH[A, B, C... ] is just p function with another additional loop added. As an example:

$$\left(e^A e^B\right) e^C = e^{A+B+\text{BCH}[A,B]} e^C = e^{A+B+C+\text{BCH}[A,B]+\text{BCH}[A+B+\text{BCH}[A,B],C]} \qquad 5.2$$

Then p must generated the nested expressions e.g.  BCH[A+B+BCH[A, B], C] via the mentioned additional loop. The corresponding loop-counter is the number of arguments or parameters.

**Remark**: *Assume n is fixed and we omit n from the argument list of BCH for short-hand.*

Therefore by Theorem B.4 :

**Theorem 5**:

$$K\left(\left(\left(e^A e^B\right) e^C \, ...\right)\right) \leq K\left(e^{A+B+C+\cdots}\right) + K(\text{BCH}) \quad . \qquad 5.3$$

**Interpretation 1**: *Complexity of non-commutative non-associative algebraic expression is at most the Complexity of corresponding commutative associative algebra plus K(BCH).*

**Interpretation 2**: *Complexity of the **Separated-Variables** is at most the Complexity of the **Unseparated-Variables** plus K(BCH).*

**Remark 5.1**: $\left(\left(e^A e^B\right) e^C \, ...\right)$ *is separated-variables,* $e^{A+B+C+\cdots}$ *is unseparated variables.*

Example of separated variables: http://hyperphysics.phy-astr.gsu.edu/hbase/quantum/hydsch.html .

Assuming observable variables are the separated variables (or else their eigenvalues not computable), then we can conclude:

**Thesis 5**: *Observation of a variable increases the overall Complexity. Or the observable universe is in perpetual monotonic increase of complexity.*

# 6. One-Parameter Subgroup

One-Parameter Subgroup is rendered as:

1. Transparent strip of matrices each rendered as color-blocks. Buy using the matrix norm the blocks are distanced from each other accordingly. Each matrix is an exponentiation of $e^{tx}$ which x is a member of the Lie Algebra and for positive t, in other words $e^{tx}$ belongs to the corresponding Lie Group approaching the identity element
2. Color scheme is used to indicate the approach to 0 from left (shades of red) and from right (shades of green)
3. The transparent strip has a tip to indicate an arrow or vector as member of the tangent space of a matrix group or in Lie Algebra
4. Cylinder is a member of the Lie Group, in specific the identity member
5. Diagonal line of green/black colors is the identity matrix
6. Subscript for the matrices/expressions indicates which group they belong to e.g. $x_{G_2}$ means x belongs to $G_2$, $x_{\mathfrak{g}_2}$ means x belongs to $\mathfrak{g}_2$

This is a 7x7 Reals matrix color coded for entries:

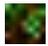     FIG 6.1

This is the Identity matrix color coded:

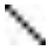     FIG 6.2

This is 7x7 matrix quite close to Identity matrix:

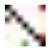     FIG 6.3

The following is a group, indeed an Abelian subgroup of Lie Group G :

$$e : \mathfrak{g} \longrightarrow G \qquad 6.1$$

$$one - param - group = \{e^{tx} \mid \forall\, t \in \mathbb{R},\ \text{fixed}\, x \in \mathfrak{g}\} \qquad 6.2$$

Since BCH $[s\,x,\ t\,x] = 0_{\mathfrak{g}}$ then $e^{sx} * e^{tx} = e^{(s+t).x}$ and again by the same inference $e^{tx} * e^{-tx} = e^{(t-t).x} = 1_G$

```
dim = 7;
range = 2;
cylrad = 1.7;
a = 1;
n = 200;

expnames =
    {{"e^tC3", "d/dt e^tC3|_t=0 = C3"}, {"e^tC5", "d/dt e^tC5|_t=0 = C5"}, {"e^tC14", "d/dt e^tC14|_t=0 = C14"}};
matrices = {C3, C5, C14};

(* arrowheadsize, arrowheadoffset,
arrowheadtextoffset, separation, edge text factor *)
arrowconfig = {0.05, 4.5, 5.8, 400, 2.4, "g2", "G2"};

oneparam = Table[oneparamgroupmaker[matrices[[i]], i * Pi / 10,
    expnames[[i]], dim, a, cylrad, n, range, arrowconfig], {i, 1, 3, 1}];

Show[oneparam[[1]], oneparam[[2]], oneparam[[3]], ImageSize → 500]
```

FIG 6.4

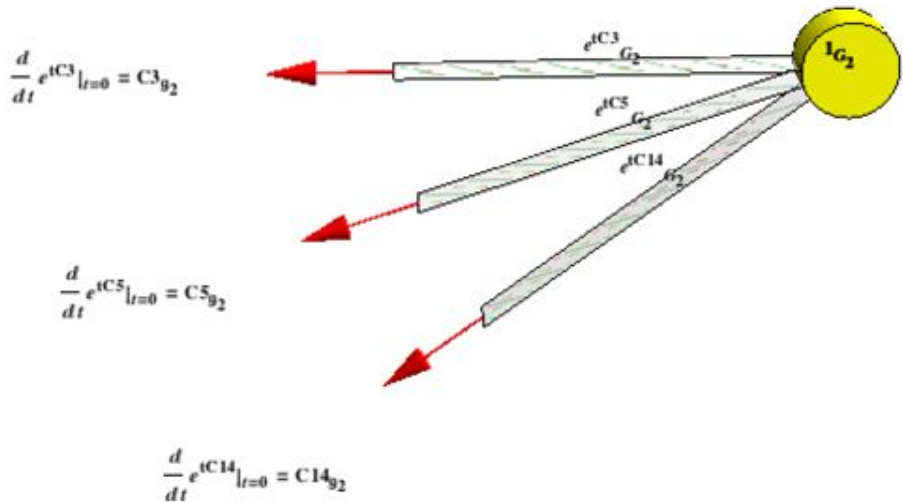

## Matrix Norm

Definitions                   and                   properties:

$$\|A\| = \sup_{x \neq 0} \frac{\|Ax\|}{\|x\|}, \quad \|AB\| \leq \|A\| \, \|B\|, \quad \|A+B\| \leq \|A\| + \|B\|, \quad \text{for convergence } \|A_n - A\| \to 0$$

6.3

```
t = 1;
x = C14;
Norm[t * x]
```

$$\frac{2}{\sqrt{3}}$$

```
t = 0.01;
x = C14;
Norm[t * x]
```

```
0.011547
```

```
t = 0.0001;
x = C14;
Norm[t * x]
```

```
0.00011547
```

## Matrix Sequence Convergence

```
t = 1;
x = C14;
(* Notice ‖e^tC14‖ = 1 i.e. it is similar to the unit norm of  e^iz *)
Norm[MatrixExp[t * x]]
N[Norm[MatrixExp[t * x] - MatrixExp[0 * x]]]
```

```
1
```

```
1.09161
```

```
t = 0.01;
x = C14;
(* Notice ‖e^tC14‖ = 1 i.e. it is similar to the unit norm of  e^iz *)
Norm[MatrixExp[t * x]]
Norm[MatrixExp[t * x] - MatrixExp[0 * x]]
```

```
1.
```

```
0.0115469
```

```
t = 0.0001;
x = C14;
(* Notice ‖e^tC14‖ = 1 i.e. it is similar to the unit norm of  e^iz *)
Norm[MatrixExp[t * x]]
Norm[MatrixExp[t * x] - MatrixExp[0 * x]]
```

```
1.
```

```
0.00011547
```

$$x = \frac{d}{dt} e^{tx} \Big|_{t=0}$$

Definition:

$$\frac{d}{dt} e^{tx} = \mathrm{Lim}_{\Delta t \to 0} \frac{e^{(t+\Delta t)x} - e^{tx}}{\Delta t}$$

$$\frac{d}{dt} \, e^{tx} = \mathrm{Lim}_{\Delta t \to 0} \, \frac{e^{(t+\Delta t)x} - e^{tx}}{\Delta t} \qquad\qquad 6.4$$

Set t = 0 and $\Delta t = 0.01$ :

```
t = 0.01;
x = C14;
derEXP[x_, t_] := ((MatrixExp[t * x] - MatrixExp[(0 * x)]) / t);

Norm[x - derEXP[x, t]]
```

0.00666664

Set t = 0 and $\Delta t = 0.0001$, see that the derivative at t = 0 becomes closer and closer to x :

```
t = 0.0001;
x = C14;
derEXP[x_, t_] := ((MatrixExp[t * x] - MatrixExp[(0 * x)]) / t);

Norm[x - derEXP[x, t]]
```

0.0000666667

## Ribbon Rendition

```
x = .;
y = .;
flow = .;
flow2 = .;
t = .;
n = .;
dim = 7;
range = 1;

x = RandomReal[{-range, range}, {dim, dim}];

n = 750;
(* e^tx e^sx = e^(t+s)x which is called the one-parameter subgroup *)
flow = Table[ArrayPlot[MatrixExp[t * x], PixelConstrained → 3, ColorRules →
      {y_ /; y < (-0.005 * range) → Darker[Red, Abs[y / range]], y_ /; y > (0.005 * range) →
        Darker[Green, Abs[y / range]]}, Frame → False], {t, 1, 0, -1 / n}];

(* Most of the middle part of the one-
 parameter groups is cutout for graphics purposes *)
flow2 = Drop[flow, {10, n - 15}];

Column[{
  Text["x a member of Lie Algebra g" ],
  ArrayPlot[x, PixelConstrained → 3,
   ColorRules → {y_ /; y < (-0.005 * range) → Darker[Red, Abs[y / range]],
      y_ /; y > (0.005 * range) → Darker[Green, Abs[y / range]]}, Frame → False],

  Text["\ne^x= ∑_{n=0}^{∞} x^n/n! a member of Lie Group G"],

  ArrayPlot[MatrixExp[1 * x], PixelConstrained → 3,
   ColorRules → {y_ /; y < (-0.005 * range) → Darker[Red, Abs[y / range]],
      y_ /; y > (0.005 * range) → Darker[Green, Abs[y / range]]}, Frame → False],

  (* d/dt e^tx= x *)

  Text["\n d/dt e^tx |_{t=0}"],
  ArrayPlot[(MatrixExp[(1 / n) * x] - MatrixExp[(0 * x)]) / (1 / n), PixelConstrained → 3,
   ColorRules → {y_ /; y < (-0.005 * range) → Darker[Red, Abs[y / range]],
      y_ /; y > (0.005 * range) → Darker[Green, Abs[y / range]]}, Frame → False],

  (* Norm of the x-d/dt e^tx → 0 *)

  Text["\n‖x- d/dt e^tx‖"],

  (* ‖A‖ = sup_{x≠0} ‖Ax‖/‖x‖ , ‖AB‖≤‖A‖‖B‖, ‖A+B‖≤‖A‖+‖B‖, for convergence ‖A_n-A‖→0 *)
  Norm[x - ((MatrixExp[(1 / n) * x] - MatrixExp[(0 * x)]) / (1 / n))],

  (* for 1 row/column use GraphicsRow/Column *)

  Text["\none-parameter subgroup (middle part removed)"],

  GraphicsRow[flow2, Spacings → 0 ]}]
```

x a member of Lie Algebra g

$e^x = \sum_{n=0}^{\infty} \frac{x^n}{n!}$ a member of Lie Group G

$\frac{d}{dt} e^{tx} \rvert_{t=0}$

$\|x - \frac{d}{dt} e^{tx}\|$

0.00393603

one−parameter subgroup (middle part removed)

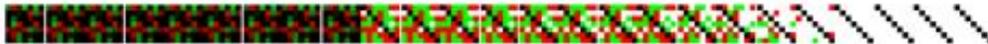

## 6.1 SO(4)

```
θ =.;
so4basis = SO4Basis["θ"];
Table[so4basis[[i]] // MatrixForm, {i, 1, 6}]
Table[MatrixExp[so4basis[[i]]] // MatrixForm, {i, 1, 6}]
```

$$\left\{ \begin{pmatrix} 0 & \theta & 0 & 0 \\ -\theta & 0 & 0 & 0 \\ 0 & 0 & 0 & \theta \\ 0 & 0 & -\theta & 0 \end{pmatrix}, \begin{pmatrix} 0 & 0 & 0 & \theta \\ 0 & 0 & -\theta & 0 \\ 0 & \theta & 0 & 0 \\ -\theta & 0 & 0 & 0 \end{pmatrix}, \begin{pmatrix} 0 & 0 & \theta & 0 \\ 0 & 0 & 0 & -\theta \\ -\theta & 0 & 0 & 0 \\ 0 & \theta & 0 & 0 \end{pmatrix}, \right.$$

$$\begin{pmatrix} 0 & 0 & -\theta & 0 \\ 0 & 0 & 0 & -\theta \\ \theta & 0 & 0 & 0 \\ 0 & \theta & 0 & 0 \end{pmatrix}, \begin{pmatrix} 0 & 0 & 0 & \theta \\ 0 & 0 & -\theta & 0 \\ 0 & \theta & 0 & 0 \\ -\theta & 0 & 0 & 0 \end{pmatrix}, \left. \begin{pmatrix} 0 & -\theta & 0 & 0 \\ \theta & 0 & 0 & 0 \\ 0 & 0 & 0 & \theta \\ 0 & 0 & -\theta & 0 \end{pmatrix} \right\}$$

$$\left\{ \begin{pmatrix} \cos[\theta] & \sin[\theta] & 0 & 0 \\ -\sin[\theta] & \cos[\theta] & 0 & 0 \\ 0 & 0 & \cos[\theta] & \sin[\theta] \\ 0 & 0 & -\sin[\theta] & \cos[\theta] \end{pmatrix}, \begin{pmatrix} \cos[\theta] & 0 & 0 & \sin[\theta] \\ 0 & \cos[\theta] & -\sin[\theta] & 0 \\ 0 & \sin[\theta] & \cos[\theta] & 0 \\ -\sin[\theta] & 0 & 0 & \cos[\theta] \end{pmatrix}, \right.$$

$$\begin{pmatrix} \cos[\theta] & 0 & \sin[\theta] & 0 \\ 0 & \cos[\theta] & 0 & -\sin[\theta] \\ -\sin[\theta] & 0 & \cos[\theta] & 0 \\ 0 & \sin[\theta] & 0 & \cos[\theta] \end{pmatrix}, \begin{pmatrix} \cos[\theta] & 0 & -\sin[\theta] & 0 \\ 0 & \cos[\theta] & 0 & -\sin[\theta] \\ \sin[\theta] & 0 & \cos[\theta] & 0 \\ 0 & \sin[\theta] & 0 & \cos[\theta] \end{pmatrix}, \right.$$

$$\begin{pmatrix} \cos[\theta] & 0 & 0 & \sin[\theta] \\ 0 & \cos[\theta] & -\sin[\theta] & 0 \\ 0 & \sin[\theta] & \cos[\theta] & 0 \\ -\sin[\theta] & 0 & 0 & \cos[\theta] \end{pmatrix}, \left. \begin{pmatrix} \cos[\theta] & -\sin[\theta] & 0 & 0 \\ \sin[\theta] & \cos[\theta] & 0 & 0 \\ 0 & 0 & \cos[\theta] & \sin[\theta] \\ 0 & 0 & -\sin[\theta] & \cos[\theta] \end{pmatrix} \right\}$$

```
dim = 4;
range = Pi / 3;
cylrad = 1.7;
a = 1;
n = 500;

θ =.;
liealgebramatrices = so4basis /. {θ → Pi / 4};
lieagroupmatrices = Table[MatrixExp[liealgebramatrices[[i]]], {i, 1, 6}];

expnames = Table[{lieagroupmatrices[[i]] // MatrixForm,
    liealgebramatrices[[i]] // MatrixForm}, {i, 1, 6}];

(* arrowheadsize, arrowheadoffset, arrowheadtextoffset, separation, edge factor *)
arrowconfig = {0.05, 3.5, 2.8, 400, 4.5, "\!\(\*
StyleBox[\"s \",\nFontSize->14]\)(4)", "SO(4)"};

oneparam = Table[oneparamgroupmaker[liealgebramatrices[[i]], i * 2 * Pi / 3,
    expnames[[i]], dim, a, cylrad, n, range, arrowconfig], {i, 1, 3}];

Show[oneparam[[1]], oneparam[[2]], oneparam[[3]], ImageSize → 500]
```

FIG 6.1.1

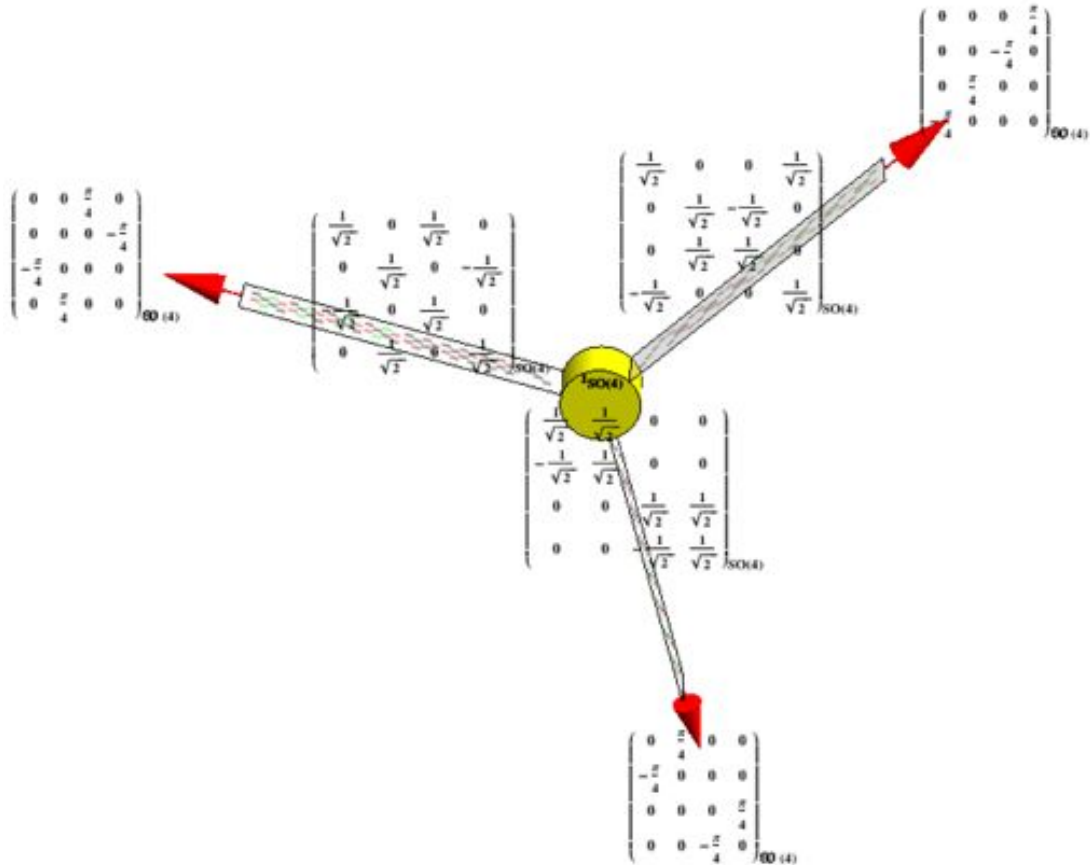

# 7. Maximal Torus Generated by C5 And C11

It happens that $\mathbf{g}_2$ basis vectors C5 and C11 commute [1]:

```
Clear[x, y];
A = C5;
B = C11;
N[MatrixCommutator[A, B]] // MatrixForm
```

$$
\begin{pmatrix}
0. & 0. & 0. & 0. & 0. & 0. & 0. \\
0. & 0. & 0. & 0. & 0. & 0. & 0. \\
0. & 0. & 0. & 0. & 0. & 0. & 0. \\
0. & 0. & 0. & 0. & 0. & 0. & 0. \\
0. & 0. & 0. & 0. & 0. & 0. & 0. \\
0. & 0. & 0. & 0. & 0. & 0. & 0. \\
0. & 0. & 0. & 0. & 0. & 0. & 0.
\end{pmatrix}
$$

Looks like their exponentiations are rotation matrices:

```
MatrixExp[z * A] // MatrixForm
MatrixExp[z * B] // MatrixForm
```

$$
\begin{pmatrix}
1 & 0 & 0 & 0 & 0 & 0 & 0 \\
0 & \cos[z] & 0 & 0 & 0 & -\sin[z] & 0 \\
0 & 0 & \cos[z] & 0 & 0 & 0 & \sin[z] \\
0 & 0 & 0 & 1 & 0 & 0 & 0 \\
0 & 0 & 0 & 0 & 1 & 0 & 0 \\
0 & \sin[z] & 0 & 0 & 0 & \cos[z] & 0 \\
0 & 0 & -\sin[z] & 0 & 0 & 0 & \cos[z]
\end{pmatrix}
$$

$$
\begin{pmatrix}
\cos\left[\frac{2z}{\sqrt{3}}\right] & 0 & 0 & -\sin\left[\frac{2z}{\sqrt{3}}\right] & 0 & 0 & 0 \\
0 & \cos\left[\frac{z}{\sqrt{3}}\right] & 0 & 0 & 0 & 0 & -\sin\left[\frac{z}{\sqrt{3}}\right] \\
0 & 0 & \cos\left[\frac{z}{\sqrt{3}}\right] & 0 & 0 & \sin\left[\frac{z}{\sqrt{3}}\right] & 0 \\
\sin\left[\frac{2z}{\sqrt{3}}\right] & 0 & 0 & \cos\left[\frac{2z}{\sqrt{3}}\right] & 0 & 0 & 0 \\
0 & 0 & 0 & 0 & 1 & 0 & 0 \\
0 & 0 & -\sin\left[\frac{z}{\sqrt{3}}\right] & 0 & 0 & \cos\left[\frac{z}{\sqrt{3}}\right] & 0 \\
0 & \sin\left[\frac{z}{\sqrt{3}}\right] & 0 & 0 & 0 & 0 & \cos\left[\frac{z}{\sqrt{3}}\right]
\end{pmatrix}
$$

The exponents of C5 and C11 multiplied by two vars e.g. x, y definitely commute since the BCH[xA, yB] = 0 :

```
g1 = FullSimplify[MatrixExp[x * A].MatrixExp[y * B]];
g2 = FullSimplify[MatrixExp[y * B].MatrixExp[x * A]];

g1 // MatrixForm
g2 // MatrixForm

g1 === g2
```

$$\begin{pmatrix}
\cos\left[\frac{2y}{\sqrt{3}}\right] & 0 & 0 & -\sin\left[\frac{2y}{\sqrt{3}}\right] & 0 & 0 & 0 \\
0 & \cos[x]\cos\left[\frac{y}{\sqrt{3}}\right] & \sin[x]\sin\left[\frac{y}{\sqrt{3}}\right] & 0 & 0 & -\cos\left[\frac{y}{\sqrt{3}}\right]\sin[x] & -\cos[x]\sin\left[\frac{y}{\sqrt{3}}\right] \\
0 & \sin[x]\sin\left[\frac{y}{\sqrt{3}}\right] & \cos[x]\cos\left[\frac{y}{\sqrt{3}}\right] & 0 & 0 & \cos[x]\sin\left[\frac{y}{\sqrt{3}}\right] & \cos\left[\frac{y}{\sqrt{3}}\right]\sin[x] \\
\sin\left[\frac{2y}{\sqrt{3}}\right] & 0 & 0 & \cos\left[\frac{2y}{\sqrt{3}}\right] & 0 & 0 & 0 \\
0 & 0 & 0 & 0 & 1 & 0 & 0 \\
0 & \cos\left[\frac{y}{\sqrt{3}}\right]\sin[x] & -\cos[x]\sin\left[\frac{y}{\sqrt{3}}\right] & 0 & 0 & \cos[x]\cos\left[\frac{y}{\sqrt{3}}\right] & -\sin[x]\sin\left[\frac{y}{\sqrt{3}}\right] \\
0 & \cos[x]\sin\left[\frac{y}{\sqrt{3}}\right] & -\cos\left[\frac{y}{\sqrt{3}}\right]\sin[x] & 0 & 0 & -\sin[x]\sin\left[\frac{y}{\sqrt{3}}\right] & \cos[x]\cos\left[\frac{y}{\sqrt{3}}\right]
\end{pmatrix}$$

$$\begin{pmatrix}
\cos\left[\frac{2y}{\sqrt{3}}\right] & 0 & 0 & -\sin\left[\frac{2y}{\sqrt{3}}\right] & 0 & 0 & 0 \\
0 & \cos[x]\cos\left[\frac{y}{\sqrt{3}}\right] & \sin[x]\sin\left[\frac{y}{\sqrt{3}}\right] & 0 & 0 & -\cos\left[\frac{y}{\sqrt{3}}\right]\sin[x] & -\cos[x]\sin\left[\frac{y}{\sqrt{3}}\right] \\
0 & \sin[x]\sin\left[\frac{y}{\sqrt{3}}\right] & \cos[x]\cos\left[\frac{y}{\sqrt{3}}\right] & 0 & 0 & \cos[x]\sin\left[\frac{y}{\sqrt{3}}\right] & \cos\left[\frac{y}{\sqrt{3}}\right]\sin[x] \\
\sin\left[\frac{2y}{\sqrt{3}}\right] & 0 & 0 & \cos\left[\frac{2y}{\sqrt{3}}\right] & 0 & 0 & 0 \\
0 & 0 & 0 & 0 & 1 & 0 & 0 \\
0 & \cos\left[\frac{y}{\sqrt{3}}\right]\sin[x] & -\cos[x]\sin\left[\frac{y}{\sqrt{3}}\right] & 0 & 0 & \cos[x]\cos\left[\frac{y}{\sqrt{3}}\right] & -\sin[x]\sin\left[\frac{y}{\sqrt{3}}\right] \\
0 & \cos[x]\sin\left[\frac{y}{\sqrt{3}}\right] & -\cos\left[\frac{y}{\sqrt{3}}\right]\sin[x] & 0 & 0 & -\sin[x]\sin\left[\frac{y}{\sqrt{3}}\right] & \cos[x]\cos\left[\frac{y}{\sqrt{3}}\right]
\end{pmatrix}$$

```
True
```

Ok then where is the Torus????????

Let's take the action of the above $G_2$ element on some fixed Octonion i.e. rotate the Im part of the Octonion which will roam around a Torus:

```
Clear[x, y]
w = {1, 1, 2.3, 1, 1, 1, 1, 1};
ParametricPlot3D[
 {actionG2[g1 /. {x → u, y → v}, w][[3]], actionG2[g1 /. {x → u, y → v}, w][[5]],
  actionG2[g1 /. {x → u, y → v}, w][[7]]}, {u, 0, 2 * Pi}, {v, 0, 2 * Pi * (√3 / 2)}]
```

FIG 7.1

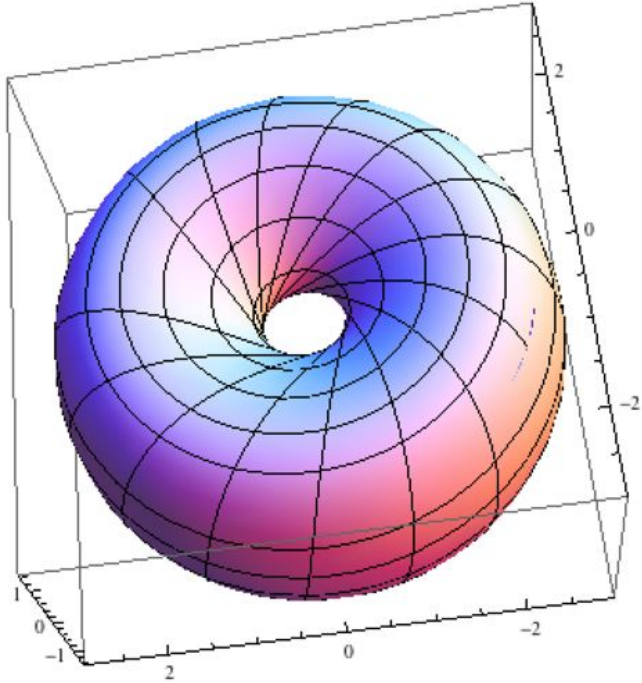

A transparent high-resolution plot to assure the inside is an empty hall:

```
Clear[x, y]
w = {1, 1, 2.3, 1, 1, 1, 1, 1};
ParametricPlot3D[{actionG2[g1 /. {x → u, y → v}, w][[3]],
  actionG2[g1 /. {x → u, y → v}, w][[5]], actionG2[g1 /. {x → u, y → v}, w][[7]]},
 {u, 0, 2 * Pi}, {v, 0, 2 * Pi * (√3 / 2)}, PlotPoints → 80,
 Mesh → None, ExclusionsStyle → {None, Red},
 PlotStyle → Directive[Cyan, Opacity[0.5], Specularity[White, 20]]]
```

FIG 7.2

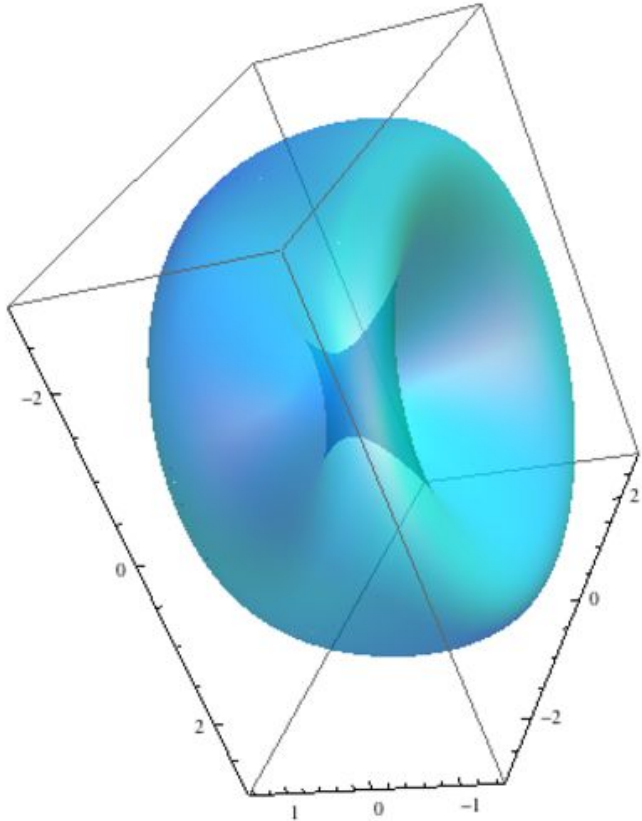

**TODO**: *This section needs to be reviewed by an expert*.

As we can see both g1 and g2 rendered the same Torus!

Since no one should trust any mathematics, let's see if we can get the Torus by changing to non-commutative elements of $G_2$:

```
Clear[x, y];
A = C3;
B = C13;
(* Non-zero commutation *)
N[MatrixCommutator[A, B]] // MatrixForm
```

$$
\begin{pmatrix}
0. & 0. & 0. & 0. & 0. & 0. & -1.1547 \\
0. & 0. & 0. & 0.57735 & 0. & 0. & 0. \\
0. & 0. & 0. & 0. & -0.57735 & 0. & 0. \\
0. & -0.57735 & 0. & 0. & 0. & 0. & 0. \\
0. & 0. & 0.57735 & 0. & 0. & 0. & 0. \\
0. & 0. & 0. & 0. & 0. & 0. & 0. \\
1.1547 & 0. & 0. & 0. & 0. & 0. & 0.
\end{pmatrix}
$$

```
MatrixExp[z * A] // MatrixForm
MatrixExp[z * B] // MatrixForm
```

$$
\begin{pmatrix}
1 & 0 & 0 & 0 & 0 & 0 & 0 \\
0 & 1 & 0 & 0 & 0 & 0 & 0 \\
0 & 0 & 1 & 0 & 0 & 0 & 0 \\
0 & 0 & 0 & \text{Cos}[z] & -\text{Sin}[z] & 0 & 0 \\
0 & 0 & 0 & \text{Sin}[z] & \text{Cos}[z] & 0 & 0 \\
0 & 0 & 0 & 0 & 0 & \text{Cos}[z] & -\text{Sin}[z] \\
0 & 0 & 0 & 0 & 0 & \text{Sin}[z] & \text{Cos}[z]
\end{pmatrix}
$$

$$
\begin{pmatrix}
\text{Cos}\left[\frac{2z}{\sqrt{3}}\right] & 0 & 0 & 0 & 0 & -\text{Sin}\left[\frac{2z}{\sqrt{3}}\right] & 0 \\
0 & \text{Cos}\left[\frac{z}{\sqrt{3}}\right] & 0 & 0 & -\text{Sin}\left[\frac{z}{\sqrt{3}}\right] & 0 & 0 \\
0 & 0 & \text{Cos}\left[\frac{z}{\sqrt{3}}\right] & -\text{Sin}\left[\frac{z}{\sqrt{3}}\right] & 0 & 0 & 0 \\
0 & 0 & \text{Sin}\left[\frac{z}{\sqrt{3}}\right] & \text{Cos}\left[\frac{z}{\sqrt{3}}\right] & 0 & 0 & 0 \\
0 & \text{Sin}\left[\frac{z}{\sqrt{3}}\right] & 0 & 0 & \text{Cos}\left[\frac{z}{\sqrt{3}}\right] & 0 & 0 \\
\text{Sin}\left[\frac{2z}{\sqrt{3}}\right] & 0 & 0 & 0 & 0 & \text{Cos}\left[\frac{2z}{\sqrt{3}}\right] & 0 \\
0 & 0 & 0 & 0 & 0 & 0 & 1
\end{pmatrix}
$$

Exponents fail to commute:

```
g1 = FullSimplify[MatrixExp[x * A].MatrixExp[y * B]];
g2 = FullSimplify[MatrixExp[y * B].MatrixExp[x * A]];

Style[g1 // MatrixForm, FontSize → 10]
Style[g2 // MatrixForm, FontSize → 10]

g1 === g2
```

$$
\begin{pmatrix}
\cos\left[\frac{2y}{\sqrt{3}}\right] & 0 & 0 & 0 & 0 & -\sin\left[\frac{2y}{\sqrt{3}}\right] & 0 \\
0 & \cos\left[\frac{y}{\sqrt{3}}\right] & 0 & 0 & -\sin\left[\frac{y}{\sqrt{3}}\right] & 0 & 0 \\
0 & 0 & \cos\left[\frac{y}{\sqrt{3}}\right] & -\sin\left[\frac{y}{\sqrt{3}}\right] & 0 & 0 & 0 \\
0 & -\sin[x]\sin\left[\frac{y}{\sqrt{3}}\right] & \cos[x]\sin\left[\frac{y}{\sqrt{3}}\right] & \cos[x]\cos\left[\frac{y}{\sqrt{3}}\right] & -\cos\left[\frac{y}{\sqrt{3}}\right]\sin[x] & 0 & 0 \\
0 & \cos[x]\sin\left[\frac{y}{\sqrt{3}}\right] & \sin[x]\sin\left[\frac{y}{\sqrt{3}}\right] & \cos\left[\frac{y}{\sqrt{3}}\right]\sin[x] & \cos[x]\cos\left[\frac{y}{\sqrt{3}}\right] & 0 & 0 \\
\cos[x]\sin\left[\frac{2y}{\sqrt{3}}\right] & 0 & 0 & 0 & 0 & \cos[x]\cos\left[\frac{2y}{\sqrt{3}}\right] & -\sin[x] \\
\sin[x]\sin\left[\frac{2y}{\sqrt{3}}\right] & 0 & 0 & 0 & 0 & \cos\left[\frac{2y}{\sqrt{3}}\right]\sin[x] & \cos[x]
\end{pmatrix}
$$

$$
\begin{pmatrix}
\cos\left[\frac{2y}{\sqrt{3}}\right] & 0 & 0 & 0 & 0 & -\cos[x]\sin\left[\frac{2y}{\sqrt{3}}\right] & \sin[x]\sin\left[\frac{2y}{\sqrt{3}}\right] \\
0 & \cos\left[\frac{y}{\sqrt{3}}\right] & 0 & -\sin[x]\sin\left[\frac{y}{\sqrt{3}}\right] & -\cos[x]\sin\left[\frac{y}{\sqrt{3}}\right] & 0 & 0 \\
0 & 0 & \cos\left[\frac{y}{\sqrt{3}}\right] & -\cos[x]\sin\left[\frac{y}{\sqrt{3}}\right] & \sin[x]\sin\left[\frac{y}{\sqrt{3}}\right] & 0 & 0 \\
0 & 0 & \sin\left[\frac{y}{\sqrt{3}}\right] & \cos[x]\cos\left[\frac{y}{\sqrt{3}}\right] & -\cos\left[\frac{y}{\sqrt{3}}\right]\sin[x] & 0 & 0 \\
0 & \sin\left[\frac{y}{\sqrt{3}}\right] & 0 & \cos\left[\frac{y}{\sqrt{3}}\right]\sin[x] & \cos[x]\cos\left[\frac{y}{\sqrt{3}}\right] & 0 & 0 \\
\sin\left[\frac{2y}{\sqrt{3}}\right] & 0 & 0 & 0 & 0 & \cos[x]\cos\left[\frac{2y}{\sqrt{3}}\right] & -\cos\left[\frac{2y}{\sqrt{3}}\right]\sin[x] \\
0 & 0 & 0 & 0 & 0 & \sin[x] & \cos[x]
\end{pmatrix}
$$

```
False
```

Huh? So we got one Torus looking shape by g1 and another non-Torus convoluted shape (perhaps a Mobius Band) for g2!!!!!!

Relax since g1 ≠ g2 it actually means that when we rotate the fixed Octonion with g1 and then with g2 we move around one surface which is a Torus and then we move in reverse i.e. first rotate with g2 then with g1 the Octonion moves around totally different surface!

So the two surfaces unioned is no where close to a Torus!

```
Clear[x, y]
w = {1, 1, 2.3, 1, 1, 1, 1, 1};

i = 2;
j = 5;
k = 6;
ParametricPlot3D[{actionG2[g1 /. {x → u, y → v}, w][[i]],
  actionG2[g1 /. {x → u, y → v}, w][[j]], actionG2[g1 /. {x → u, y → v}, w][[k]]},
 {u, 0, 2 * Pi}, {v, 0, 2 * Pi * (√3 / 2)}, Mesh → None, PlotPoints → 80,
 Mesh → None, ExclusionsStyle → {None, Red},
 PlotStyle → Directive[Cyan, Opacity[0.5], Specularity[White, 20]]]

ParametricPlot3D[{actionG2[g2 /. {x → u, y → v}, w][[i]],
  actionG2[g2 /. {x → u, y → v}, w][[j]], actionG2[g2 /. {x → u, y → v}, w][[k]]},
 {u, 0, 2 * Pi}, {v, 0, 2 * Pi * (√3 / 2)}, Mesh → None, PlotPoints → 80,
 Mesh → None, ExclusionsStyle → {None, Red},
 PlotStyle → Directive[Cyan, Opacity[0.5], Specularity[White, 20]]]
```

FIG 7.3

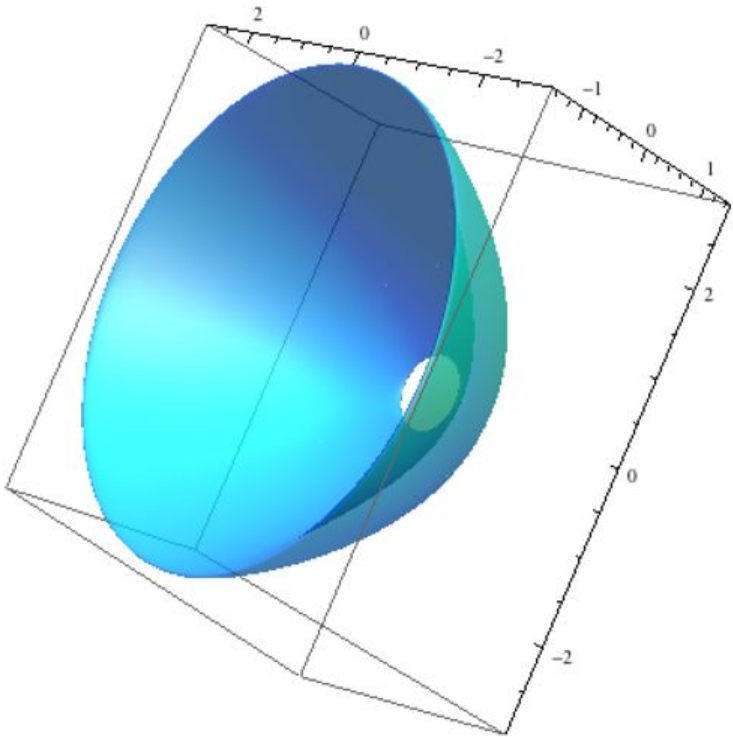

FIG 7.4

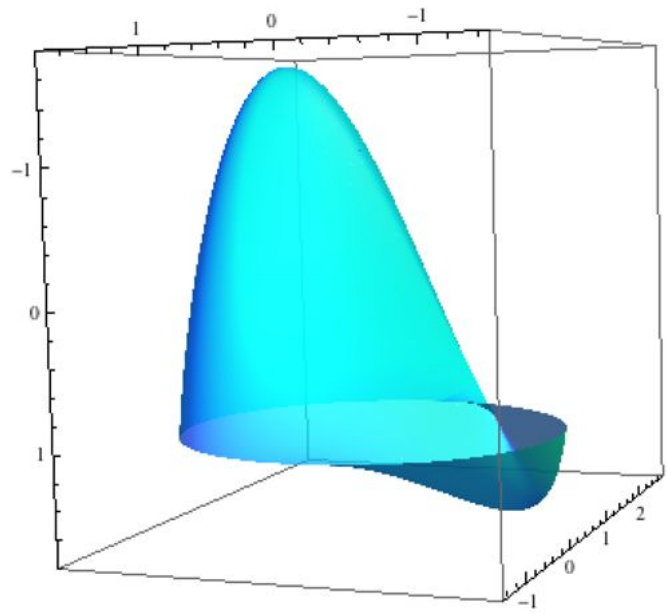

# 8. Parametrization

From [1, 2], $G_2$ is parametrized by 14 Reals or in other words $G_2$, itself as the automorphism of Octonions or as the holonomy of a 7-dimensional manifold, is a 14-dimensional manifold as well:

$$g = \sigma(a_1, a_2, a_3)\, s(a_4, a_5, a_6)\, e^{\sqrt{3}\ C11}\, e^{a_8\, C5}\, u(a_9, a_{10}, a_{11}, a_{12}, a_{13}, a_{14}) \qquad 8.1$$

where

$$s(x, y, z) = e^{xC3}\, e^{yC9}\, e^{zC8} \qquad\qquad 8.2$$

$$\sigma(x', y', z') = e^{\sqrt{3}\ x'C8}\, e^{\sqrt{3}\ y'C9}\, e^{\sqrt{3}\ z'C8} \qquad 8.3$$

$$u(x, y, z, x', y', z') = s(x, y, z)\, \sigma(x', y', z') \qquad 8.4$$

Let's print some parts of the parametrization, note they will not fit the page, but to get an idea of complexities of the equations:

```
expr = sG2[x, y, z];
Style[expr // MatrixForm, FontSize → 7]
```

$$
\begin{pmatrix}
1 & 0 & 0 & 0 & 0 & 0 & 0 \\
0 & 1 & 0 & 0 & 0 & 0 & 0 \\
0 & 0 & 1 & 0 & 0 & 0 & 0 \\
0 & 0 & 0 & \text{Cos}[x]\,\text{Cos}[y]\,\text{Cos}[z] & -\text{Cos}[y]\,\text{Sin}[x]\,\text{Sin}[z] & -\text{Cos}[y]\,\text{Cos}[z]\,\text{Sin}[x]-\text{Cos}[x]\,\text{Cos}[y]\,\text{Sin}[z] & \text{Cos}[x]\,\text{Cos}[z]\,\text{Sin}[y]+\text{Sin}[x]\,\text{Sin}[y]\,\text{Sin}[z] & \text{Cos}[z]\,\text{Sin}[x]\,\text{Sin}[y]- \\
0 & 0 & 0 & \text{Cos}[y]\,\text{Cos}[z]\,\text{Sin}[x]+\text{Cos}[x]\,\text{Cos}[y]\,\text{Sin}[z] & \text{Cos}[x]\,\text{Cos}[y]\,\text{Cos}[z]-\text{Cos}[y]\,\text{Sin}[x]\,\text{Sin}[z] & \text{Cos}[z]\,\text{Sin}[x]\,\text{Sin}[y]-\text{Cos}[x]\,\text{Sin}[y]\,\text{Sin}[z] & -\text{Cos}[x]\,\text{Cos}[z]\,\text{Sin}[y]- \\
0 & 0 & 0 & -\text{Cos}[x]\,\text{Cos}[z]\,\text{Sin}[y]-\text{Sin}[x]\,\text{Sin}[y]\,\text{Sin}[z] & -\text{Cos}[z]\,\text{Sin}[x]\,\text{Sin}[y]-\text{Cos}[x]\,\text{Sin}[y]\,\text{Sin}[z] & \text{Cos}[x]\,\text{Sin}[x]\,\text{Sin}[z] & -\text{Cos}[y]\,\text{Cos}[z]\,\text{Sin}[x]- \\
0 & 0 & 0 & \text{Cos}[z]\,\text{Sin}[x]\,\text{Sin}[y]+\text{Cos}[x]\,\text{Sin}[y]\,\text{Sin}[z] & \text{Cos}[x]\,\text{Cos}[z]\,\text{Sin}[y]+\text{Sin}[x]\,\text{Sin}[y]\,\text{Sin}[z] & \text{Cos}[y]\,\text{Cos}[z]\,\text{Sin}[x]+\text{Cos}[x]\,\text{Cos}[y]\,\text{Sin}[z] & \text{Cos}[x]\,\text{Cos}[y]\,\text{Cos}[z]-
\end{pmatrix}
$$

```
expr = σG2[x', y', z'];
Style[expr // MatrixForm, FontSize → 7]
```

$$
\begin{pmatrix}
\text{Cos}[2\,y'] & -\text{Cos}[2\,z']\,\text{Sin}[2\,y'] & \text{Sin}[2\,y']\,\text{Sin}[2\,z'] & 0 \\
\text{Cos}[2\,x']\,\text{Sin}[2\,y'] & \text{Cos}[2\,x']\,\text{Cos}[2\,y']\,\text{Cos}[2\,z']-\text{Sin}[2\,x']\,\text{Sin}[2\,z'] & -\text{Cos}[2\,z']\,\text{Sin}[2\,x']-\text{Cos}[2\,x']\,\text{Cos}[2\,y']\,\text{Sin}[2\,z'] & 0 \\
\text{Sin}[2\,x']\,\text{Sin}[2\,y'] & \text{Cos}[2\,y']\,\text{Cos}[2\,z']\,\text{Sin}[2\,x']+\text{Cos}[2\,x']\,\text{Sin}[2\,z'] & \text{Cos}[2\,x']\,\text{Cos}[2\,z']-\text{Cos}[2\,y']\,\text{Sin}[2\,x']\,\text{Sin}[2\,z'] & 0 \\
0 & 0 & 0 & \text{Cos}[x']\,\text{Cos}[y']\,\text{Cos}[z']-\text{Cos}[y']\,\text{Sin}[x']\,\text{Sin}[z'] & C \\
0 & 0 & 0 & -\text{Cos}[y']\,\text{Cos}[z']\,\text{Sin}[x']-\text{Cos}[x']\,\text{Cos}[y']\,\text{Sin}[z'] & C \\
0 & 0 & 0 & \text{Cos}[z']\,\text{Sin}[x']\,\text{Sin}[y']-\text{Cos}[x']\,\text{Sin}[y']\,\text{Sin}[z'] & C \\
0 & 0 & 0 & -\text{Cos}[x']\,\text{Cos}[z']\,\text{Sin}[y']-\text{Sin}[x']\,\text{Sin}[y']\,\text{Sin}[z'] & C
\end{pmatrix}
$$

Parametrization for SO(4)

```
expr = uSO4[x, y, z, x', y', z'];
Style[expr // MatrixForm, FontSize → 7]
```

$$
\begin{pmatrix}
\text{Cos}[2\,y'] & -\text{Cos}[2\,z']\,\text{Sin}[2\,y'] & \text{Sin}[2\,y']\,\text{Sin}[2\,z'] & \\
\text{Cos}[2\,x']\,\text{Sin}[2\,y'] & \text{Cos}[2\,x']\,\text{Cos}[2\,y']\,\text{Cos}[2\,z']-\text{Sin}[2\,x']\,\text{Sin}[2\,z'] & -\text{Cos}[2\,z']\,\text{Sin}[2\,x']-\text{Cos}[2\,x']\,\text{Cos}[2\,y']\,\text{Sin}[2\,z'] & \\
\text{Sin}[2\,x']\,\text{Sin}[2\,y'] & \text{Cos}[2\,y']\,\text{Cos}[2\,z']\,\text{Sin}[2\,x']+\text{Cos}[2\,x']\,\text{Sin}[2\,z'] & \text{Cos}[2\,x']\,\text{Cos}[2\,z']-\text{Cos}[2\,y']\,\text{Sin}[2\,x']\,\text{Sin}[2\,z'] & \\
0 & 0 & 0 & (-\text{Cos}[y]\,\text{Cos}[z]\,\text{Sin}[x]-\text{Cos}[x]\,\text{Cos}[y]\,\text{Sin}[z])\ (- \\
0 & 0 & 0 & (\text{Cos}[x]\,\text{Cos}[y]\,\text{Cos}[z]-\text{Cos}[y]\,\text{Sin}[x]\,\text{Sin}[z])\ (-C \\
0 & 0 & 0 & (-\text{Cos}[z]\,\text{Sin}[x]\,\text{Sin}[y]+\text{Cos}[x]\,\text{Sin}[y]\,\text{Sin}[z])\ (-C \\
0 & 0 & 0 & (\text{Cos}[x]\,\text{Cos}[z]\,\text{Sin}[y]+\text{Sin}[x]\,\text{Sin}[y]\,\text{Sin}[z])\ (-C
\end{pmatrix}
$$

Numerical examples:

```
sG2[Pi * 1 / 2, Pi * 1 / 3, Pi * 1 / 2] // MatrixForm
oG2[Pi * 1 / 2, Pi * 1 / 3, Pi * 1 / 2] // MatrixForm
uSO4[Pi * 1 / 2, Pi * 1 / 3, Pi * 1 / 2, Pi * 1 / 3, Pi * 1 / 2, Pi * 1 / 3] // MatrixForm

(* parametrization is an array of 14 reals with certain bounds *)
r = {1 / 2, 1 / 3, 1 / 2, 1 / 3, 1 / 2, 1 / 3,
     -1 / 2, -1 / 3, -1 / 2, -1 / 3, -1 / 2, -1 / 3, 1 / 5, 1 / 5} * Pi;

(* gG2 gives the parametrization of the Italian paper *)
gG2[r] // MatrixForm
```

$$\begin{pmatrix}
1 & 0 & 0 & 0 & 0 & 0 & 0 \\
0 & 1 & 0 & 0 & 0 & 0 & 0 \\
0 & 0 & 1 & 0 & 0 & 0 & 0 \\
0 & 0 & 0 & -\frac{1}{2} & 0 & \frac{\sqrt{3}}{2} & 0 \\
0 & 0 & 0 & 0 & -\frac{1}{2} & 0 & -\frac{\sqrt{3}}{2} \\
0 & 0 & 0 & -\frac{\sqrt{3}}{2} & 0 & -\frac{1}{2} & 0 \\
0 & 0 & 0 & 0 & \frac{\sqrt{3}}{2} & 0 & -\frac{1}{2}
\end{pmatrix}$$

$$\begin{pmatrix}
-\frac{1}{2} & \frac{\sqrt{3}}{2} & 0 & 0 & 0 & 0 & 0 \\
-\frac{\sqrt{3}}{2} & -\frac{1}{2} & 0 & 0 & 0 & 0 & 0 \\
0 & 0 & 1 & 0 & 0 & 0 & 0 \\
0 & 0 & 0 & -\frac{1}{2} & 0 & 0 & \frac{\sqrt{3}}{2} \\
0 & 0 & 0 & 0 & -\frac{1}{2} & -\frac{\sqrt{3}}{2} & 0 \\
0 & 0 & 0 & 0 & \frac{\sqrt{3}}{2} & -\frac{1}{2} & 0 \\
0 & 0 & 0 & -\frac{\sqrt{3}}{2} & 0 & 0 & -\frac{1}{2}
\end{pmatrix}$$

$$\begin{pmatrix}
-1 & 0 & 0 & 0 & 0 & 0 & 0 \\
0 & -1 & 0 & 0 & 0 & 0 & 0 \\
0 & 0 & 1 & 0 & 0 & 0 & 0 \\
0 & 0 & 0 & 0 & \frac{\sqrt{3}}{2} & 0 & -\frac{1}{2} \\
0 & 0 & 0 & \frac{\sqrt{3}}{2} & 0 & \frac{1}{2} & 0 \\
0 & 0 & 0 & 0 & -\frac{1}{2} & 0 & -\frac{\sqrt{3}}{2} \\
0 & 0 & 0 & \frac{1}{2} & 0 & -\frac{\sqrt{3}}{2} & 0
\end{pmatrix}$$

$$\begin{pmatrix}
-0.463221 & -0.565616 & 0.527262 & -0.136699 & -0.150591 & 0.0482028 & -0.379226 \\
0.624263 & -0.0127994 & 0.740022 & 0.0789229 & 0.0869435 & -0.0278299 & 0.218946 \\
-0.41182 & 0.671943 & 0.359022 & -0.46617 & 0.127126 & -0.0683558 & 0.108869 \\
0.20591 & 0.139557 & -0.0250025 & -0.233104 & -0.909043 & -0.225194 & -0.078045 \\
0.118882 & -0.193973 & -0.103641 & -0.52211 & -0.0784719 & 0.748232 & 0.314471 \\
-0.20591 & 0.335972 & 0.179511 & 0.630899 & -0.299557 & 0.568703 & -0.0361772 \\
0.356646 & 0.241719 & -0.0433056 & -0.181109 & 0.177062 & 0.241336 & -0.830726
\end{pmatrix}$$

# 9. Shape of $G_2$ Action

A randomly distributed set of points in $G_2$ is generated via its randomly distributed parametrization. The **Im** part of an Octonion (w) is fixed and the action of said random points in $G_2$ are computed.

Then combinations of 3 coordinates are selected from 7 coordinates to visualize in 3D and similarly in 2D.

For example below from the 1-7 coordinates $\{2, 5, 7\}$ is selected. All possible combinations were looked at and the shape is a similar globular fuzz.

```
(* Im part of some random but fixed Octonion *)
w = RandomReal [{-1, 1}, 7];
w

{-0.864919, 0.346332, 0.654304, 0.745847, -0.616471, -0.0923945, -0.806799}

rlist =.;
v = w;
coverG2 = {{0, 1}, {0, 1/2}, {0, 1/2}, {0, 2}, {0, 1/4}, {0, 1}, {0, 1/6},
     {0, 1/2}, {0, 2}, {0, 1/2}, {0, 1}, {0, 1}, {0, 1/2}, {0, 1}} * Pi;

(* generate random grid, but not all points are valid *)
m = 4000;
pts = randomGrid[coverG2, m];

(* Only allow for the points which satisfy the (4.15) in [1] *)
pts2 = {};
Table[Which[paramBooleG2[pts[[i]]] == 1, pts2 = AppendTo[pts2, pts[[i]]]],
  {i, 1, Length[pts]}];

glist = Table[gG2[pts2[[i]]], {i, 1, Length[pts2]}];

rlist = Table[glist[[i]].v, {i, 1, Length[glist]}];

Graphics3D[Point[rlist[[All, {2, 5, 7}]]]]
(*comb3=Subsets[Range[7], {3}];
Table[Graphics3D[Point[rlist[[All,comb3[[i]]]]]],{i, 1, Length[comb3]}]*)
```

FIG 9.1

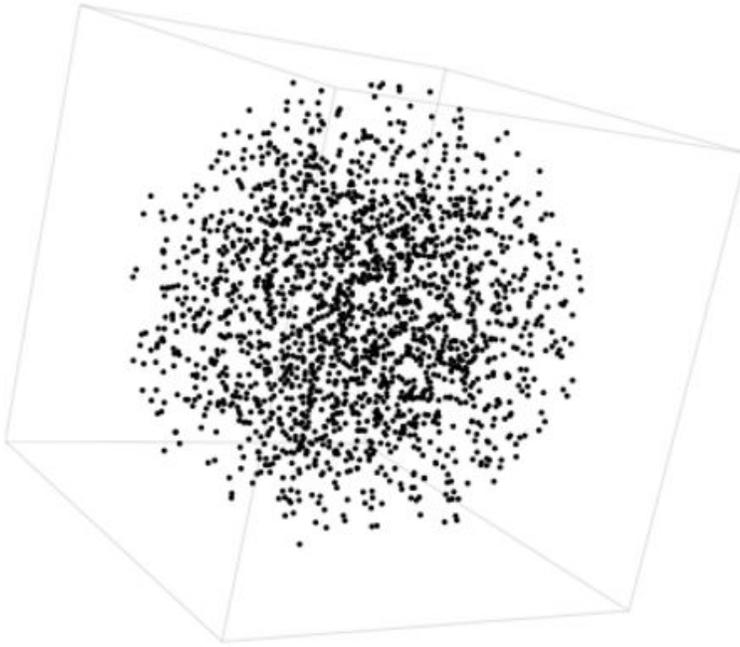

```
Graphics[Point[rlist[[All, {2, 7}]]]]]
(*comb2=Subsets[Range[7], {2}];
Table[Graphics[Point[rlist[[All,comb2[[i]]]]]]], {i, 1, Length[comb2]}]*)
```

FIG 9.2

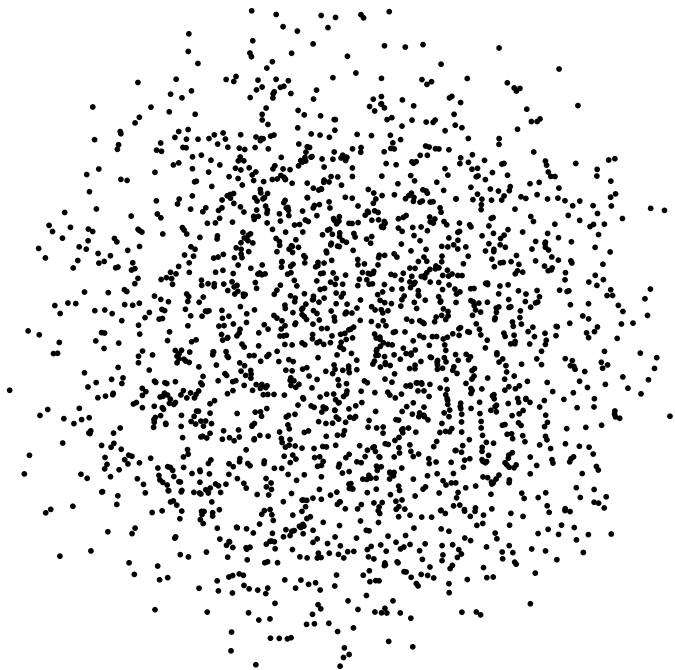

Powers of One Member

Get one member of $G_2$ e.g. **glist[500]** in below code and compute its first n powers i.e. $g^1$, $g^2$, $g^3$, .... $g^n$ and compute their action on the fixed Im part of an Octonion and then plot all possible combination of 3 coordinates out of 7 and plot correspondingly:

**Remark 9.1**: *The changing of this single member drastically changes the shape of the Action*

```
n = 1700;
rlist = Table[MatrixPower [glist[[900]], i].v, {i, 1, n}];

Graphics3D[Point[rlist[[All, {2, 5, 7}]]]]

(* calculate the combination of 3 indices out of 7 and make a list out of them
   comb3=Subsets[Range[7], {3}];
 make a list of all the above combinations and plot in 3d
 Table[Graphics3D[Point[rlist[[All,comb3[[i]]]]]]],{i, 1, Length[comb3]}] *)
```

FIG 9.2

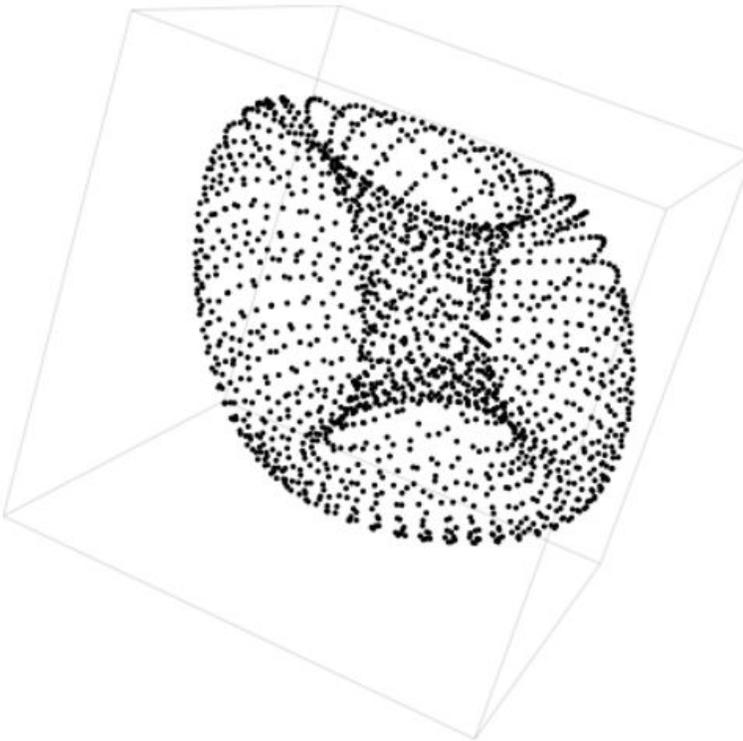

Accumulative Action of Line Across the HyperCube

```
v = w;
(* Generate 2000 points via the parametrization, this case the line going from

   {0,0,0,0,0,0,0,0,0,0,0,0,0,0} → {1,1/2,1/2,2,1/4,1,1/6,1/2,2,1/2,1,1,1/2,1}*Pi

   when t goes from 0 to 1.

   This line is in the parametrization space which is a hypercube
*)
pts = Table[
   t * {1, 1 / 2, 1 / 2, 2, 1 / 4, 1, 1 / 6, 1 / 2, 2, 1 / 2, 1, 1, 1 / 2, 1} * Pi, {t, 0, 1, 1 / 2000}];

(* Only allow for the points which satisfy the (4.15) in [1] *)
pts2 = {};
Table[Which[paramBooleG2[pts[[i]]] == 1, pts2 = AppendTo[pts2, pts[[i]]]],
   {i, 1, Length[pts]}];

glist = Table[gG2[pts2[[i]]], {i, 1, Length[pts2]}];

(* Normalize v and accumulate the action i.e. g₁, g₂g₁, g₃g₂g₁ ...  *)
rlist = Table[v = glist[[i]].v / Norm[v], {i, 1, Length[glist]}];
Graphics3D[Line[Accumulate[rlist[[All, {2, 5, 7}]]]]]]
```

FIG 9.4

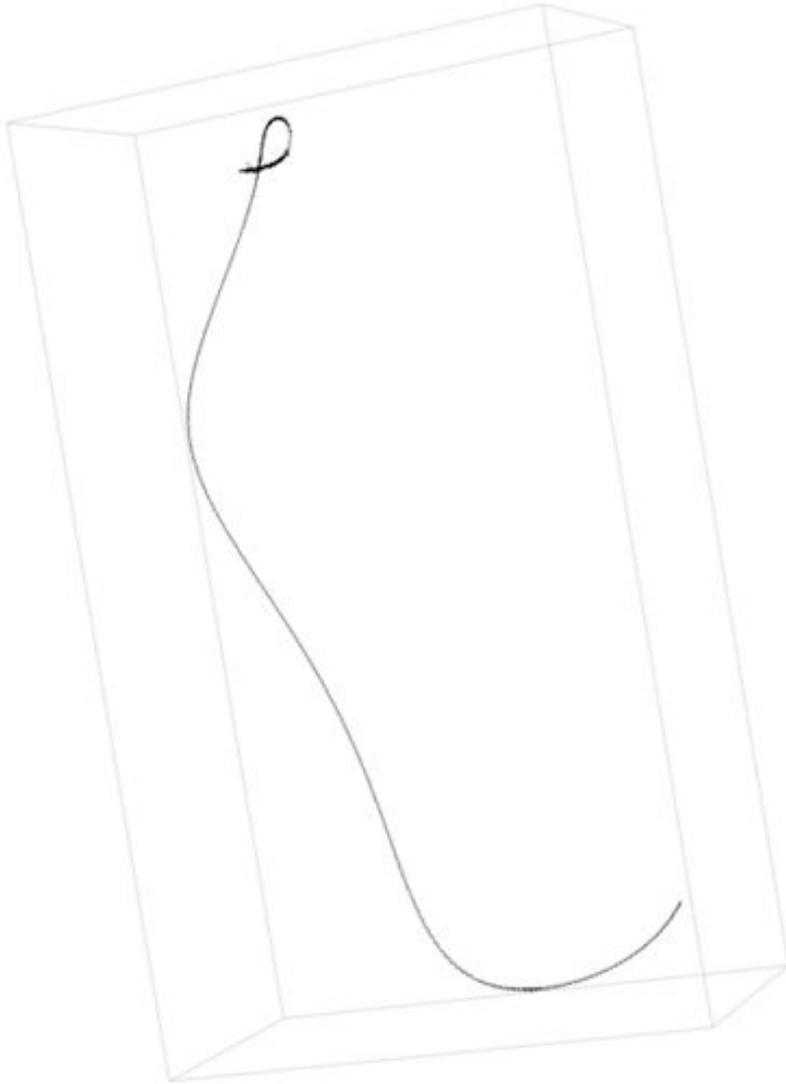

Top part of the curve magnified:

FIG 9.5

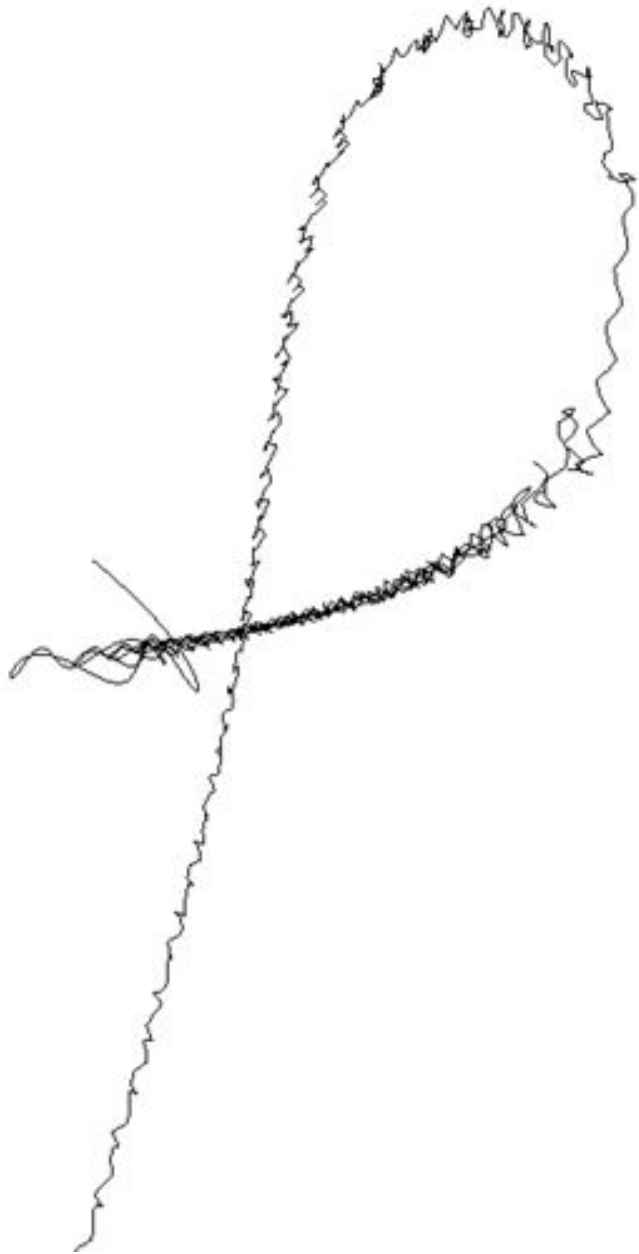

## Accumulative Action of Line A<sub>vector</sub> SO(4) Parameter Boundary

All first 8 parameters set to 0 (or any other fixed number) and a line in hypercube parameter space is parametrized in the boundary of SO(4) see (2.8) [1]:

```
v = w;
pts = Table[
    t * {0, 0 / 2, 0 / 2, 0, 0 / 4, 0, 0 / 6, 0 / 2, 2, 1 / 2, 1, 1, 1 / 2, 1} * Pi, {t, 0, 1, 1 / 2000}];

pts2 = {};
 Table[Which[paramBooleG2[pts[[i]]] == 1, pts2 = AppendTo[pts2, pts[[i]]]],
  {i, 1, Length[pts]}];

glist = Table[gG2[pts2[[i]]], {i, 1, Length[pts2]}];

rlist = Table[v = glist[[i]].v / Norm[v], {i, 1, Length[glist]}];

(*comb3=Subsets[Range[7], {3}];
Table[Graphics3D[Line[Accumulate[rlist[[All,comb3[[i]]]]]]],{i, 1, Length[comb3]}]*)
Graphics3D[Line[Accumulate[rlist[[All, {2, 5, 7}]]]]]
```

FIG 9.6

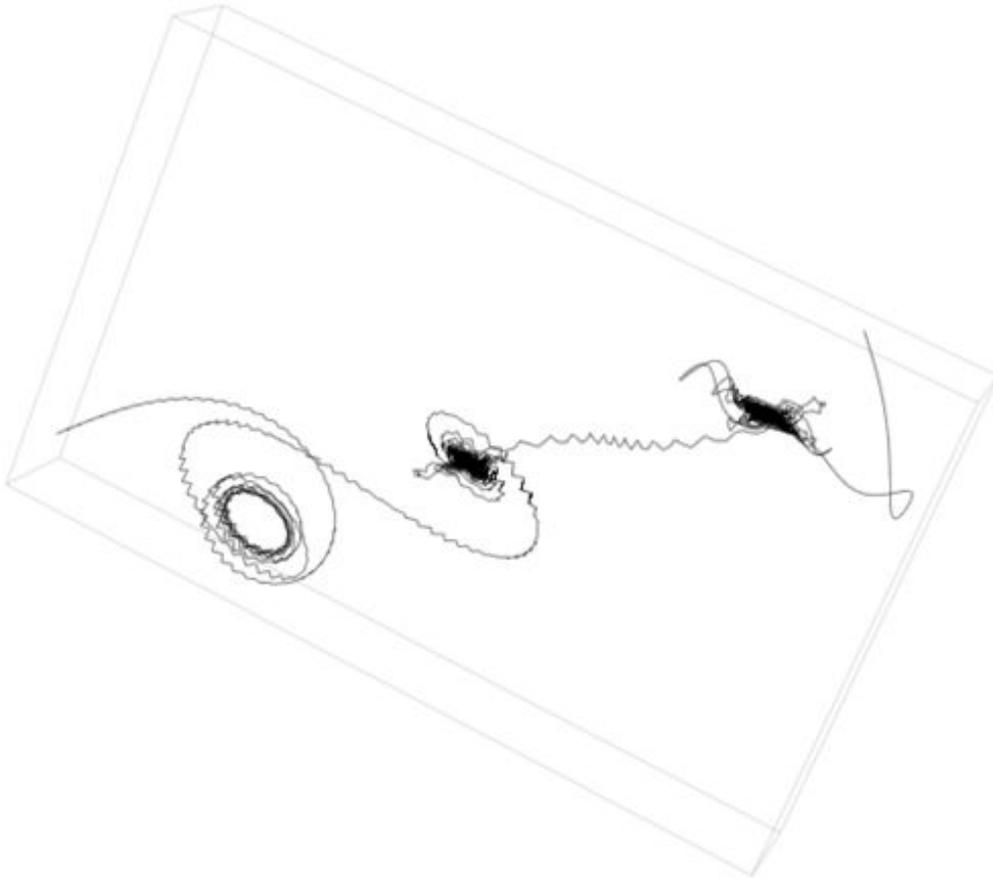

Accumulative Action of Line Across S SubSpace Spanned by {C1, C2, C3}

```
v = w;
pts = Table[
    t * {1, 1 / 2, 1 / 2, 0, 0 / 4, 0, 0 / 6, 0 / 6, 0, 0 / 2, 0, 0, 0 / 2, 0} * Pi, {t, 0, 1, 1 / 2000}];

pts2 = {};
Table[Which[paramBooleG2[pts[[i]]] == 1, pts2 = AppendTo[pts2, pts[[i]]]],
  {i, 1, Length[pts]}];

glist = Table[gG2[pts2[[i]]], {i, 1, Length[pts2]}];

rlist = Table[v = glist[[i]].v / Norm[v], {i, 1, Length[glist]}];
(*comb3=Subsets[Range[7], {3}];
Table[Graphics3D[Line[Accumulate[rlist[[All,comb3[[i]]]]]]],{i, 1, Length[comb3]}]*)
Graphics3D[Line[Accumulate[rlist[[All, {2, 5, 7}]]]]]
```

FIG 9.7

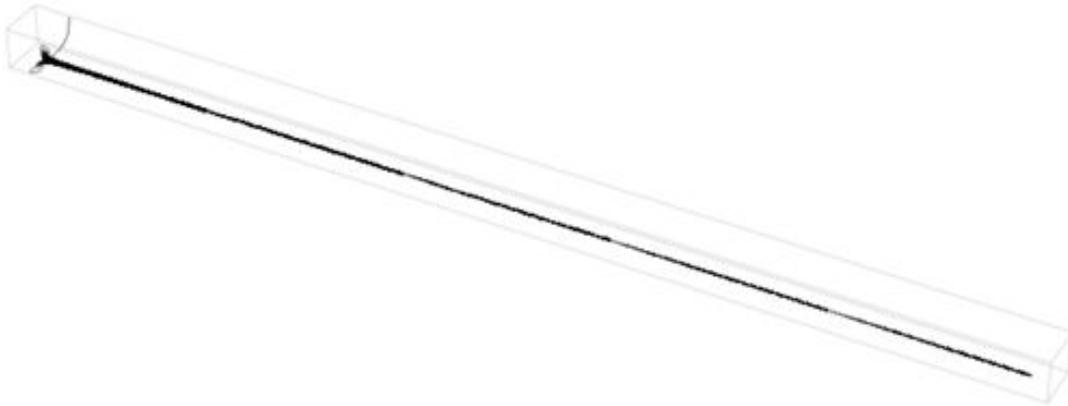

Left-most top part magnified:

FIG 9.8

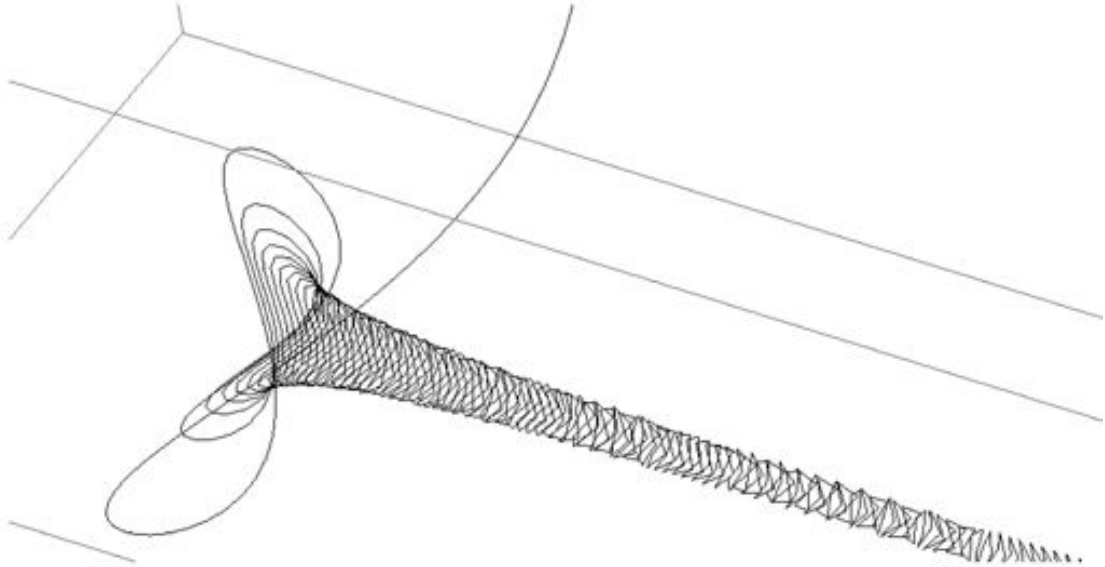

Accumulative Action of Line Across $\sigma$ SubSpace Spanned by {C8, C9, C10}

```
v = w;
pts = Table[
    t * {0, 0 / 2, 0 / 2, 2, 1 / 4, 1, 0 / 6, 0 / 6, 0, 0 / 2, 0, 0, 0 / 2, 0} * Pi, {t, 0, 1, 1 / 2000}];

pts2 = {};
 Table[Which[paramBooleG2[pts[[i]]] == 1, pts2 = AppendTo[pts2, pts[[i]]]],
   {i, 1, Length[pts]}];

glist = Table[gG2[pts2[[i]]], {i, 1, Length[pts2]}];

rlist = Table[v = glist[[i]].v / Norm[v], {i, 1, Length[glist]}];
(*comb3=Subsets[Range[7], {3}];
Table[Graphics3D[Line[Accumulate[rlist[[All,comb3[[i]]]]]]],{i, 1, Length[comb3]}]*)
Graphics3D[Line[Accumulate[rlist[[All, {5, 6, 7}]]]]]
```

FIG 9.9

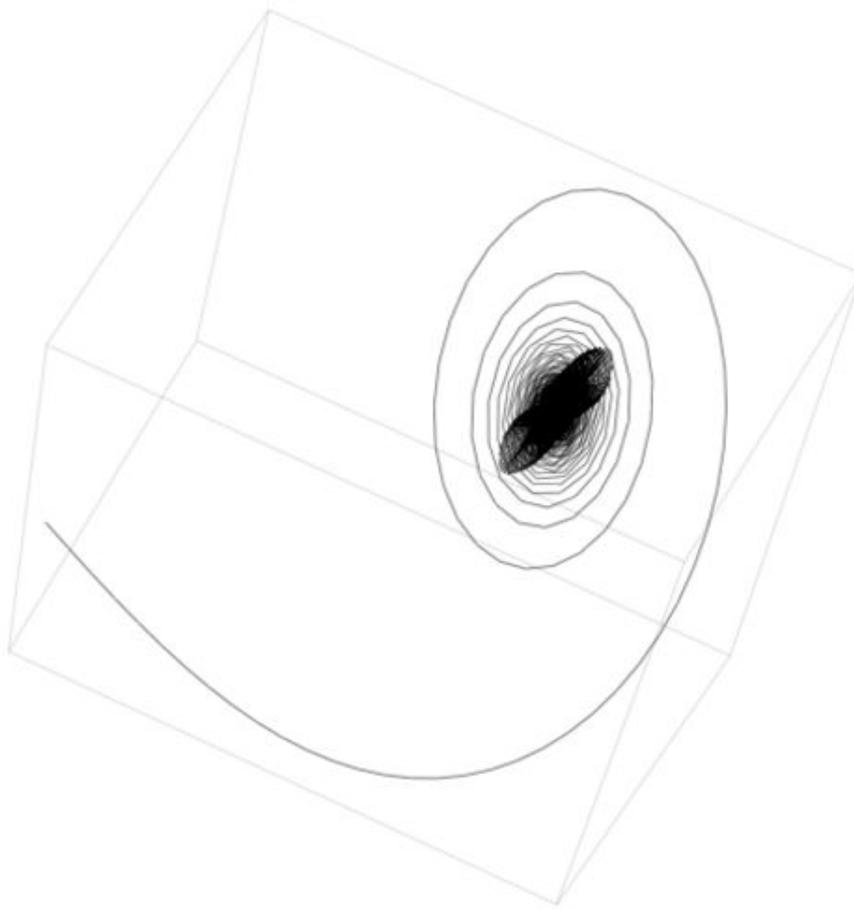

Core magnified:

FIG 9.10

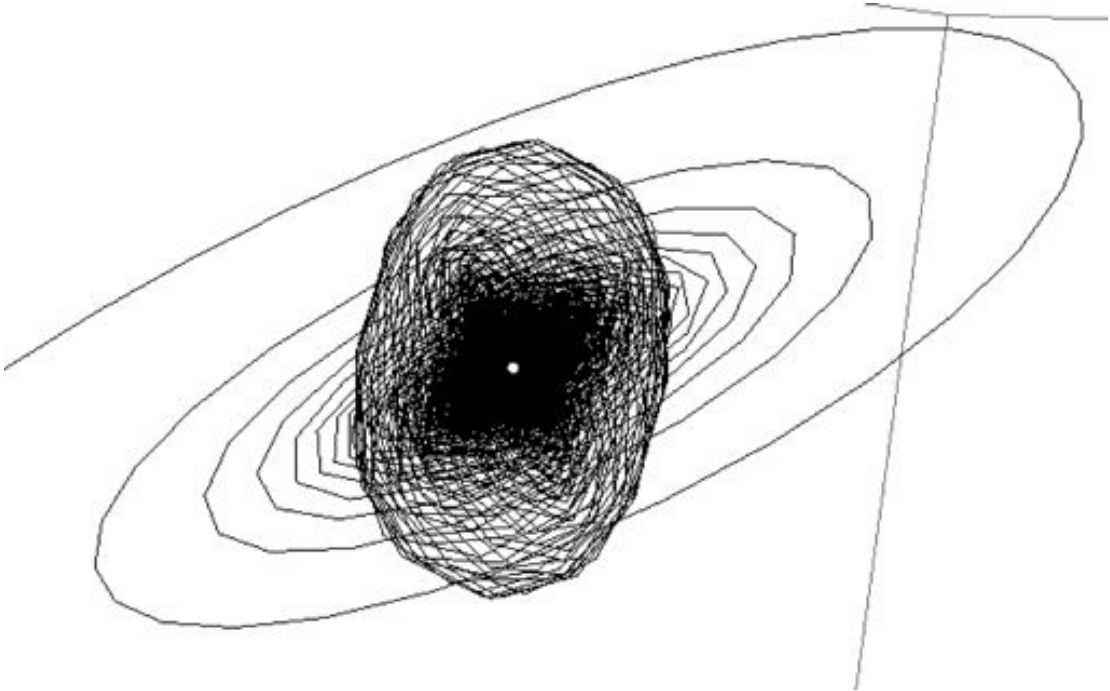

Accumulative Action of Line Across SubSpace Spanned by {C5, C11}

```
v = w;
pts = Table[t * {0, 0, 0, 0, 0, 0, 1 / 6, 1 / 2, 0, 0, 0, 0, 0, 0} * Pi, {t, 0, 1, 1 / 2000}];

pts2 = {};
 Table[Which[paramBooleG2[pts[[i]]] == 1, pts2 = AppendTo[pts2, pts[[i]]]],
   {i, 1, Length[pts]}];
glist = Table[gG2[pts2[[i]]], {i, 1, Length[pts2]}];

rlist = Table[v = glist[[i]].v / Norm[v], {i, 1, Length[glist]}];
(*comb3=Subsets[Range[7], {3}];
Table[Graphics3D[Line[Accumulate[rlist[[All,comb3[[i]]]]]]],{i, 1, Length[comb3]}]*)
Graphics3D[Line[Accumulate[rlist[[All, {5, 6, 7}]]]]]
```

FIG 9.11

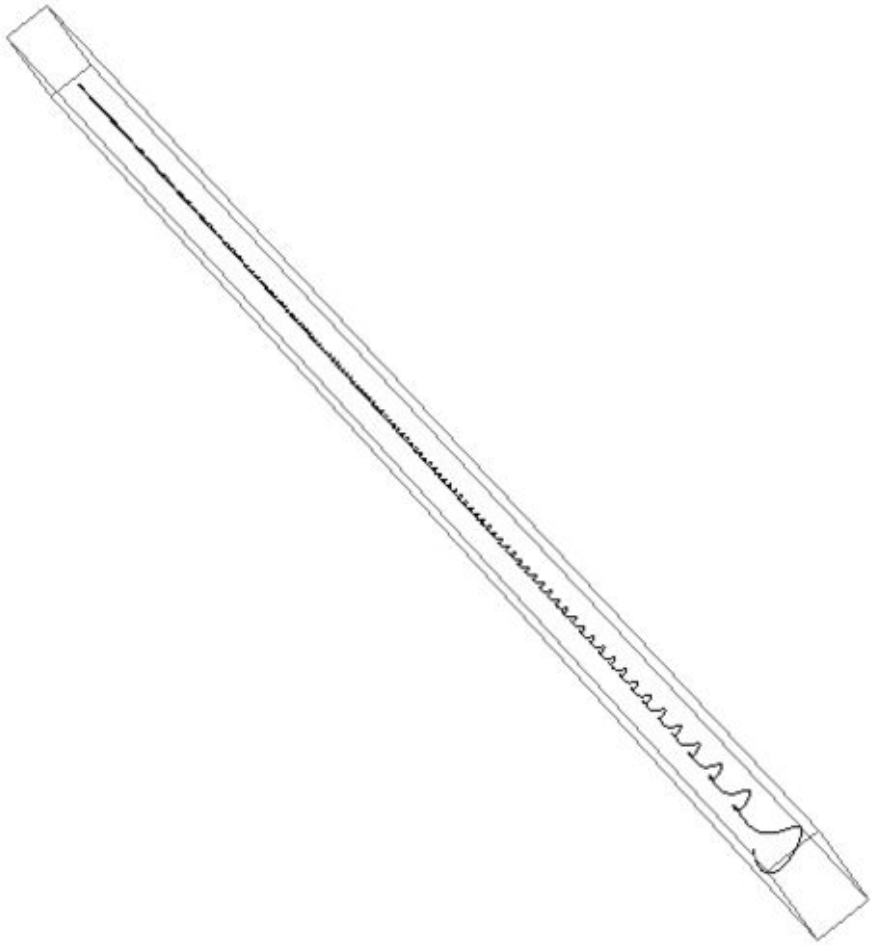

Alternative view magnified:

FIG 9.12

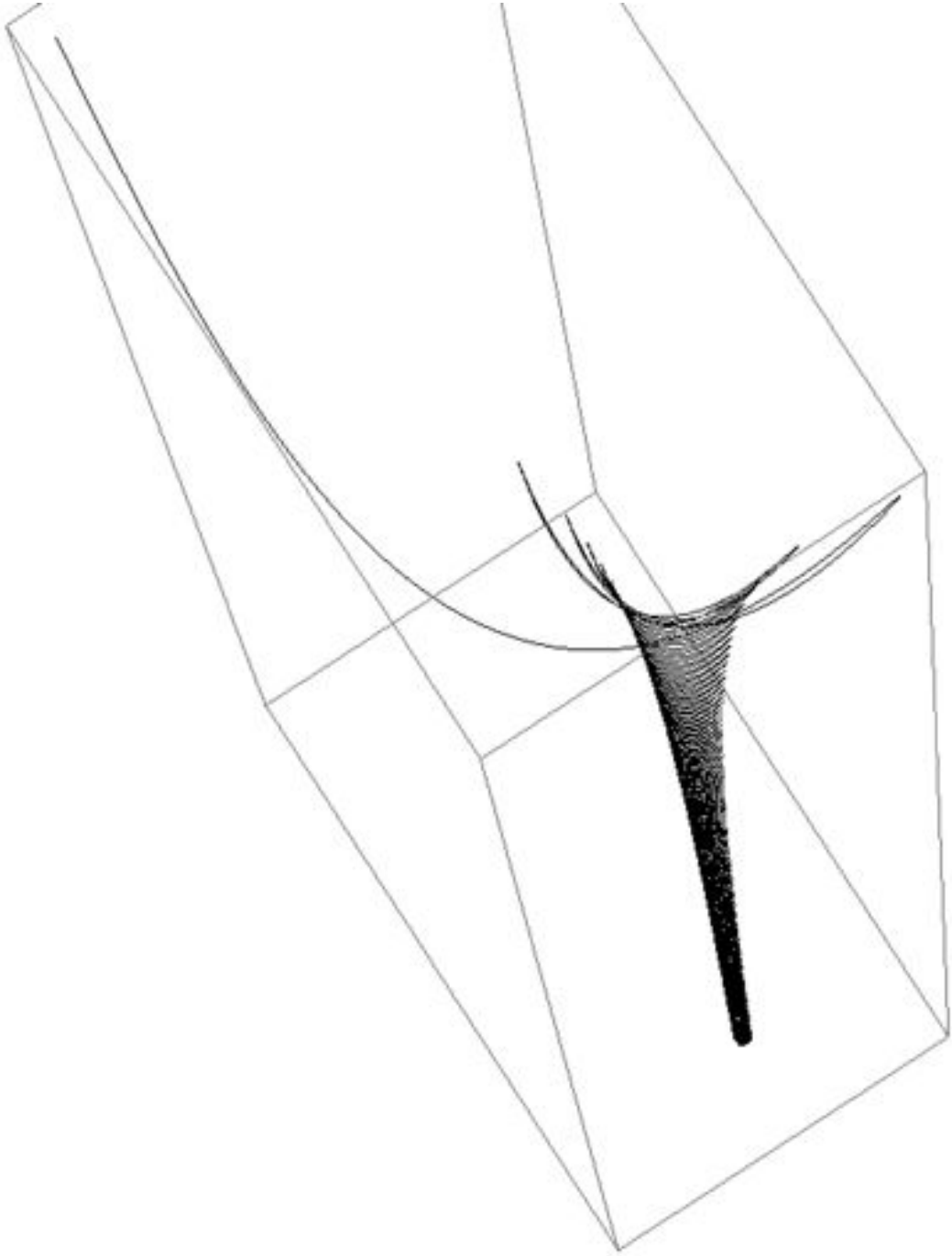

Accumulative Action of Random Sampling Across the HyperCube

```
v = w;
coverG2 = {{0, 1}, {0, 1 / 2}, {0, 1 / 2}, {0, 2}, {0, 1 / 4}, {0, 1}, {0, 1 / 6},
    {0, 1 / 2}, {0, 2}, {0, 1 / 2}, {0, 1}, {0, 1}, {0, 1 / 2}, {0, 1}} * Pi;

m = 4000;
pts = randomGrid[coverG2, m];

pts2 = {};
Table[Which[paramBooleG2[pts[[i]]] == 1, pts2 = AppendTo[pts2, pts[[i]]]],
    {i, 1, Length[pts]}];
glist = Table[gG2[pts2[[i]]], {i, 1, Length[pts2]}];

rlist = Table[v = glist[[i]].v / Norm[v], {i, 1, Length[glist]}];
Graphics3D[Line[Accumulate[rlist[[All, {2, 5, 7}]]]]]
```

FIG 9.13

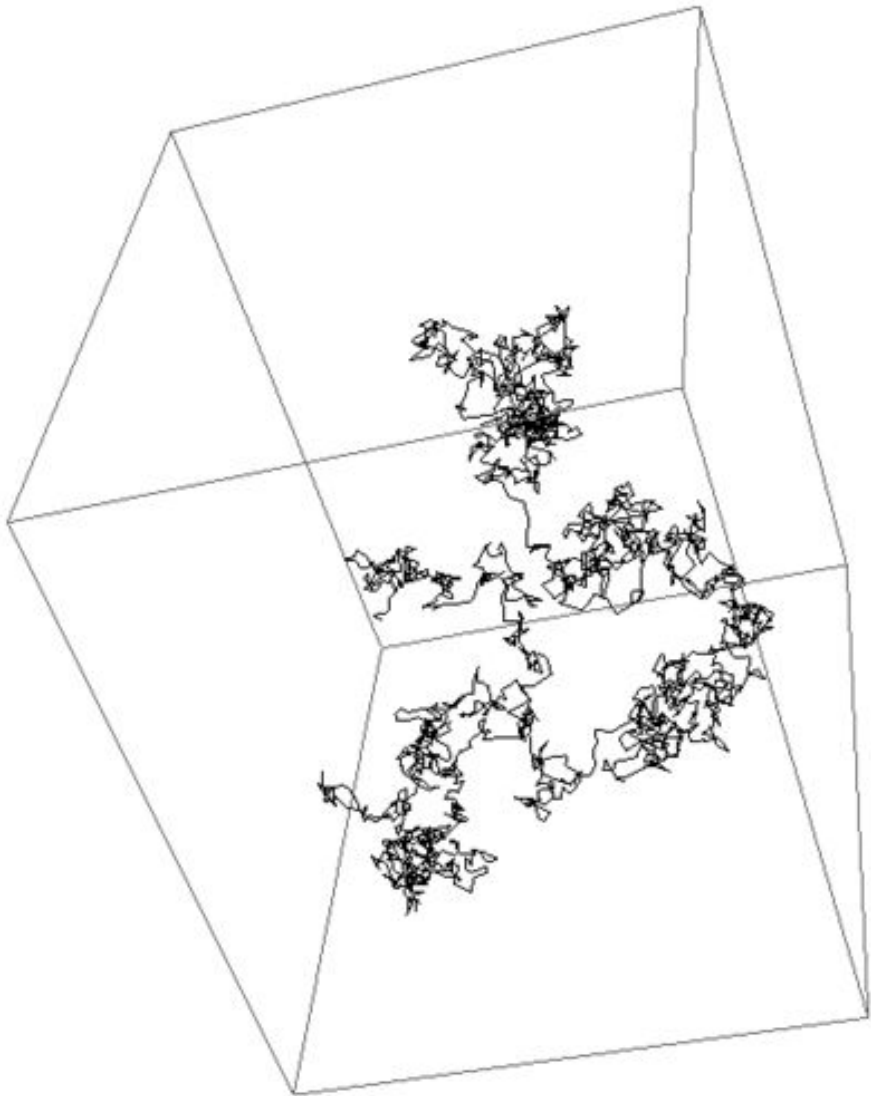

# Appendix A

```
(* Definition of vector Product *)
x = {x1, x2, x3, x4, x5, x6, x7, x8};
y = {y1, y2, y3, y4, y5, y6, y7, y8};

Simplify[OctCommutator[x, y] / 2]

{0, -x4 y3 + x3 y4 - x6 y5 + x5 y6 + x8 y7 - x7 y8,
 x4 y2 - x2 y4 - x7 y5 - x8 y6 + x5 y7 + x6 y8, -x3 y2 + x2 y3 - x8 y5 + x7 y6 - x6 y7 + x5 y8,
 x6 y2 + x7 y3 + x8 y4 - x2 y6 - x3 y7 - x4 y8, -x5 y2 + x8 y3 - x7 y4 + x2 y5 + x4 y7 - x3 y8,
 -x8 y2 - x5 y3 + x6 y4 + x3 y5 - x4 y6 + x2 y8, x7 y2 - x6 y3 - x5 y4 + x4 y5 + x3 y6 - x2 y7}
```

Let's symbolically check if all the Exp of all basis vectors in $\mathfrak{g}_2$ preserve the vector product, the output of 14 True(s) indicates full proof of the fact by actual computation:

```
Table[
 gexp = MatrixExp[z * gbasis[[i]]];

 (* calculate the action on the vector product of two Octonions *)
 cross1 = FullSimplify[actionG2[gexp, OctCommutator[x, y] / 2]];

 (* Apply the action two each Octonion and then compute the vector product *)
 cross2 = FullSimplify[OctCommutator[actionG2[gexp, x], actionG2[gexp, y]] / 2];

 cross1 === cross2,

 {i, 1, 14}]

{True, True, True, True, True, True, True, True, True, True, True, True, True, True}
```

And the above was checked for [1] with the following basis for $\mathfrak{g}_2$ :

```
Table[ gbasis[[i]] // MatrixForm, {i, 1, 14}]
```

$$\left\{ \begin{pmatrix} 0 & 0 & 0 & 0 & 0 & 0 & 0 \\ 0 & 0 & 0 & 0 & 0 & 0 & 0 \\ 0 & 0 & 0 & 0 & 0 & 0 & 0 \\ 0 & 0 & 0 & 0 & 0 & 0 & -1 \\ 0 & 0 & 0 & 0 & 0 & -1 & 0 \\ 0 & 0 & 0 & 0 & 1 & 0 & 0 \\ 0 & 0 & 0 & 1 & 0 & 0 & 0 \end{pmatrix}, \begin{pmatrix} 0 & 0 & 0 & 0 & 0 & 0 & 0 \\ 0 & 0 & 0 & 0 & 0 & 0 & 0 \\ 0 & 0 & 0 & 0 & 0 & 0 & 0 \\ 0 & 0 & 0 & 0 & 0 & 1 & 0 \\ 0 & 0 & 0 & 0 & 0 & 0 & -1 \\ 0 & 0 & 0 & -1 & 0 & 0 & 0 \\ 0 & 0 & 0 & 0 & 1 & 0 & 0 \end{pmatrix}, \begin{pmatrix} 0 & 0 & 0 & 0 & 0 & 0 & 0 \\ 0 & 0 & 0 & 0 & 0 & 0 & 0 \\ 0 & 0 & 0 & 0 & 0 & 0 & 0 \\ 0 & 0 & 0 & 0 & -1 & 0 & 0 \\ 0 & 0 & 0 & 1 & 0 & 0 & 0 \\ 0 & 0 & 0 & 0 & 0 & 0 & -1 \\ 0 & 0 & 0 & 0 & 0 & 1 & 0 \end{pmatrix}, \right.$$

$$\left. \begin{pmatrix} 0 & 0 & 0 & 0 & 0 & 0 & 0 \\ 0 & 0 & 0 & 0 & 0 & 0 & 1 \\ 0 & 0 & 0 & 0 & 0 & 1 & 0 \\ 0 & 0 & 0 & 0 & 0 & 0 & 0 \\ 0 & 0 & 0 & 0 & 0 & 0 & 0 \\ 0 & 0 & -1 & 0 & 0 & 0 & 0 \\ 0 & -1 & 0 & 0 & 0 & 0 & 0 \end{pmatrix}, \begin{pmatrix} 0 & 0 & 0 & 0 & 0 & 0 & 0 \\ 0 & 0 & 0 & 0 & 0 & -1 & 0 \\ 0 & 0 & 0 & 0 & 0 & 0 & 1 \\ 0 & 0 & 0 & 0 & 0 & 0 & 0 \\ 0 & 0 & 0 & 0 & 0 & 0 & 0 \\ 0 & 1 & 0 & 0 & 0 & 0 & 0 \\ 0 & 0 & -1 & 0 & 0 & 0 & 0 \end{pmatrix}, \begin{pmatrix} 0 & 0 & 0 & 0 & 0 & 0 & 0 \\ 0 & 0 & 0 & 0 & 1 & 0 & 0 \\ 0 & 0 & 0 & -1 & 0 & 0 & 0 \\ 0 & 0 & 1 & 0 & 0 & 0 & 0 \\ 0 & -1 & 0 & 0 & 0 & 0 & 0 \\ 0 & 0 & 0 & 0 & 0 & 0 & 0 \\ 0 & 0 & 0 & 0 & 0 & 0 & 0 \end{pmatrix}, \right.$$

$$\begin{pmatrix}
0&0&0&0&0&0&0\\
0&0&0&-1&0&0&0\\
0&0&0&0&-1&0&0\\
0&1&0&0&0&0&0\\
0&0&1&0&0&0&0\\
0&0&0&0&0&0&0\\
0&0&0&0&0&0&0
\end{pmatrix},\quad
\begin{pmatrix}
0&0&0&0&0&0&0\\
0&0&-\frac{2}{\sqrt3}&0&0&0&0\\
0&\frac{2}{\sqrt3}&0&0&0&0&0\\
0&0&0&0&\frac{1}{\sqrt3}&0&0\\
0&0&0&-\frac{1}{\sqrt3}&0&0&0\\
0&0&0&0&0&0&-\frac{1}{\sqrt3}\\
0&0&0&0&0&\frac{1}{\sqrt3}&0
\end{pmatrix},$$

$$\begin{pmatrix}
0&-\frac{2}{\sqrt3}&0&0&0&0&0\\
\frac{2}{\sqrt3}&0&0&0&0&0&0\\
0&0&0&0&0&0&0\\
0&0&0&0&0&0&\frac{1}{\sqrt3}\\
0&0&0&0&0&-\frac{1}{\sqrt3}&0\\
0&0&0&0&\frac{1}{\sqrt3}&0&0\\
0&0&0&-\frac{1}{\sqrt3}&0&0&0
\end{pmatrix},\quad
\begin{pmatrix}
0&0&-\frac{2}{\sqrt3}&0&0&0&0\\
0&0&0&0&0&0&0\\
\frac{2}{\sqrt3}&0&0&0&0&0&0\\
0&0&0&0&0&-\frac{1}{\sqrt3}&0\\
0&0&0&0&0&0&-\frac{1}{\sqrt3}\\
0&0&0&\frac{1}{\sqrt3}&0&0&0\\
0&0&0&0&\frac{1}{\sqrt3}&0&0
\end{pmatrix},$$

$$\begin{pmatrix}
0&0&0&-\frac{2}{\sqrt3}&0&0&0\\
0&0&0&0&0&0&-\frac{1}{\sqrt3}\\
0&0&0&0&0&\frac{1}{\sqrt3}&0\\
\frac{2}{\sqrt3}&0&0&0&0&0&0\\
0&0&0&0&0&0&0\\
0&0&-\frac{1}{\sqrt3}&0&0&0&0\\
0&\frac{1}{\sqrt3}&0&0&0&0&0
\end{pmatrix},\quad
\begin{pmatrix}
0&0&0&0&-\frac{2}{\sqrt3}&0&0\\
0&0&0&0&0&0&\frac{1}{\sqrt3}\\
0&0&0&0&0&\frac{1}{\sqrt3}&0\\
0&0&0&0&0&0&0\\
\frac{2}{\sqrt3}&0&0&0&0&0&0\\
0&-\frac{1}{\sqrt3}&0&0&0&0&0\\
0&0&-\frac{1}{\sqrt3}&0&0&0&0
\end{pmatrix},$$

$$\begin{pmatrix}
0&0&0&0&0&-\frac{2}{\sqrt3}&0\\
0&0&0&0&-\frac{1}{\sqrt3}&0&0\\
0&0&0&-\frac{1}{\sqrt3}&0&0&0\\
0&0&\frac{1}{\sqrt3}&0&0&0&0\\
0&\frac{1}{\sqrt3}&0&0&0&0&0\\
\frac{2}{\sqrt3}&0&0&0&0&0&0\\
0&0&0&0&0&0&0
\end{pmatrix},\quad
\begin{pmatrix}
0&0&0&0&0&0&-\frac{2}{\sqrt3}\\
0&0&0&\frac{1}{\sqrt3}&0&0&0\\
0&0&0&0&-\frac{1}{\sqrt3}&0&0\\
0&-\frac{1}{\sqrt3}&0&0&0&0&0\\
0&0&\frac{1}{\sqrt3}&0&0&0&0\\
0&0&0&0&0&0&0\\
\frac{2}{\sqrt3}&0&0&0&0&0&0
\end{pmatrix}\Bigg\}$$

These are the Exp for every basis vector of the Lie Algebra $\mathfrak{g}_2$ from [1]:

```
Table[MatrixExp[z * gbasis[[i]]] // MatrixForm, {i, 1, 14}]
```

$$\left\{ \begin{pmatrix} 1 & 0 & 0 & 0 & 0 & 0 & 0 \\ 0 & 1 & 0 & 0 & 0 & 0 & 0 \\ 0 & 0 & 1 & 0 & 0 & 0 & 0 \\ 0 & 0 & 0 & \cos[z] & 0 & 0 & -\sin[z] \\ 0 & 0 & 0 & 0 & \cos[z] & -\sin[z] & 0 \\ 0 & 0 & 0 & 0 & \sin[z] & \cos[z] & 0 \\ 0 & 0 & 0 & \sin[z] & 0 & 0 & \cos[z] \end{pmatrix}, \right.$$

$$\begin{pmatrix} 1 & 0 & 0 & 0 & 0 & 0 & 0 \\ 0 & 1 & 0 & 0 & 0 & 0 & 0 \\ 0 & 0 & 1 & 0 & 0 & 0 & 0 \\ 0 & 0 & 0 & \cos[z] & 0 & \sin[z] & 0 \\ 0 & 0 & 0 & 0 & \cos[z] & 0 & -\sin[z] \\ 0 & 0 & 0 & -\sin[z] & 0 & \cos[z] & 0 \\ 0 & 0 & 0 & 0 & \sin[z] & 0 & \cos[z] \end{pmatrix},\ \begin{pmatrix} 1 & 0 & 0 & 0 & 0 & 0 & 0 \\ 0 & 1 & 0 & 0 & 0 & 0 & 0 \\ 0 & 0 & 1 & 0 & 0 & 0 & 0 \\ 0 & 0 & 0 & \cos[z] & -\sin[z] & 0 & 0 \\ 0 & 0 & 0 & \sin[z] & \cos[z] & 0 & 0 \\ 0 & 0 & 0 & 0 & 0 & \cos[z] & -\sin[z] \\ 0 & 0 & 0 & 0 & 0 & \sin[z] & \cos[z] \end{pmatrix},$$

$$\begin{pmatrix} 1 & 0 & 0 & 0 & 0 & 0 & 0 \\ 0 & \cos[z] & 0 & 0 & 0 & 0 & \sin[z] \\ 0 & 0 & \cos[z] & 0 & 0 & \sin[z] & 0 \\ 0 & 0 & 0 & 1 & 0 & 0 & 0 \\ 0 & 0 & 0 & 0 & 1 & 0 & 0 \\ 0 & 0 & -\sin[z] & 0 & 0 & \cos[z] & 0 \\ 0 & -\sin[z] & 0 & 0 & 0 & 0 & \cos[z] \end{pmatrix},\ \begin{pmatrix} 1 & 0 & 0 & 0 & 0 & 0 & 0 \\ 0 & \cos[z] & 0 & 0 & 0 & -\sin[z] & 0 \\ 0 & 0 & \cos[z] & 0 & 0 & 0 & \sin[z] \\ 0 & 0 & 0 & 1 & 0 & 0 & 0 \\ 0 & 0 & 0 & 0 & 1 & 0 & 0 \\ 0 & \sin[z] & 0 & 0 & 0 & \cos[z] & 0 \\ 0 & 0 & -\sin[z] & 0 & 0 & 0 & \cos[z] \end{pmatrix},$$

$$\begin{pmatrix} 1 & 0 & 0 & 0 & 0 & 0 & 0 \\ 0 & \cos[z] & 0 & 0 & \sin[z] & 0 & 0 \\ 0 & 0 & \cos[z] & -\sin[z] & 0 & 0 & 0 \\ 0 & 0 & \sin[z] & \cos[z] & 0 & 0 & 0 \\ 0 & -\sin[z] & 0 & 0 & \cos[z] & 0 & 0 \\ 0 & 0 & 0 & 0 & 0 & 1 & 0 \\ 0 & 0 & 0 & 0 & 0 & 0 & 1 \end{pmatrix},\ \begin{pmatrix} 1 & 0 & 0 & 0 & 0 & 0 & 0 \\ 0 & \cos[z] & 0 & -\sin[z] & 0 & 0 & 0 \\ 0 & 0 & \cos[z] & 0 & -\sin[z] & 0 & 0 \\ 0 & \sin[z] & 0 & \cos[z] & 0 & 0 & 0 \\ 0 & 0 & \sin[z] & 0 & \cos[z] & 0 & 0 \\ 0 & 0 & 0 & 0 & 0 & 1 & 0 \\ 0 & 0 & 0 & 0 & 0 & 0 & 1 \end{pmatrix},$$

$$\begin{pmatrix} 1 & 0 & 0 & 0 & 0 & 0 & 0 \\ 0 & \cos\left[\frac{2z}{\sqrt{3}}\right] & -\sin\left[\frac{2z}{\sqrt{3}}\right] & 0 & 0 & 0 & 0 \\ 0 & \sin\left[\frac{2z}{\sqrt{3}}\right] & \cos\left[\frac{2z}{\sqrt{3}}\right] & 0 & 0 & 0 & 0 \\ 0 & 0 & 0 & \cos\left[\frac{z}{\sqrt{3}}\right] & \sin\left[\frac{z}{\sqrt{3}}\right] & 0 & 0 \\ 0 & 0 & 0 & -\sin\left[\frac{z}{\sqrt{3}}\right] & \cos\left[\frac{z}{\sqrt{3}}\right] & 0 & 0 \\ 0 & 0 & 0 & 0 & 0 & \cos\left[\frac{z}{\sqrt{3}}\right] & -\sin\left[\frac{z}{\sqrt{3}}\right] \\ 0 & 0 & 0 & 0 & 0 & \sin\left[\frac{z}{\sqrt{3}}\right] & \cos\left[\frac{z}{\sqrt{3}}\right] \end{pmatrix},$$

$$\begin{pmatrix}
\cos\left[\frac{2z}{\sqrt{3}}\right] & -\sin\left[\frac{2z}{\sqrt{3}}\right] & 0 & 0 & 0 & 0 & 0 \\
\sin\left[\frac{2z}{\sqrt{3}}\right] & \cos\left[\frac{2z}{\sqrt{3}}\right] & 0 & 0 & 0 & 0 & 0 \\
0 & 0 & 1 & 0 & 0 & 0 & 0 \\
0 & 0 & 0 & \cos\left[\frac{z}{\sqrt{3}}\right] & 0 & 0 & \sin\left[\frac{z}{\sqrt{3}}\right] \\
0 & 0 & 0 & 0 & \cos\left[\frac{z}{\sqrt{3}}\right] & -\sin\left[\frac{z}{\sqrt{3}}\right] & 0 \\
0 & 0 & 0 & 0 & \sin\left[\frac{z}{\sqrt{3}}\right] & \cos\left[\frac{z}{\sqrt{3}}\right] & 0 \\
0 & 0 & 0 & -\sin\left[\frac{z}{\sqrt{3}}\right] & 0 & 0 & \cos\left[\frac{z}{\sqrt{3}}\right]
\end{pmatrix},$$

$$\begin{pmatrix}
\cos\left[\frac{2z}{\sqrt{3}}\right] & 0 & -\sin\left[\frac{2z}{\sqrt{3}}\right] & 0 & 0 & 0 & 0 \\
0 & 1 & 0 & 0 & 0 & 0 & 0 \\
\sin\left[\frac{2z}{\sqrt{3}}\right] & 0 & \cos\left[\frac{2z}{\sqrt{3}}\right] & 0 & 0 & 0 & 0 \\
0 & 0 & 0 & \cos\left[\frac{z}{\sqrt{3}}\right] & 0 & -\sin\left[\frac{z}{\sqrt{3}}\right] & 0 \\
0 & 0 & 0 & 0 & \cos\left[\frac{z}{\sqrt{3}}\right] & 0 & -\sin\left[\frac{z}{\sqrt{3}}\right] \\
0 & 0 & 0 & \sin\left[\frac{z}{\sqrt{3}}\right] & 0 & \cos\left[\frac{z}{\sqrt{3}}\right] & 0 \\
0 & 0 & 0 & 0 & \sin\left[\frac{z}{\sqrt{3}}\right] & 0 & \cos\left[\frac{z}{\sqrt{3}}\right]
\end{pmatrix},$$

$$\begin{pmatrix}
\cos\left[\frac{2z}{\sqrt{3}}\right] & 0 & 0 & -\sin\left[\frac{2z}{\sqrt{3}}\right] & 0 & 0 & 0 \\
0 & \cos\left[\frac{z}{\sqrt{3}}\right] & 0 & 0 & 0 & 0 & -\sin\left[\frac{z}{\sqrt{3}}\right] \\
0 & 0 & \cos\left[\frac{z}{\sqrt{3}}\right] & 0 & 0 & \sin\left[\frac{z}{\sqrt{3}}\right] & 0 \\
\sin\left[\frac{2z}{\sqrt{3}}\right] & 0 & 0 & \cos\left[\frac{2z}{\sqrt{3}}\right] & 0 & 0 & 0 \\
0 & 0 & 0 & 0 & 1 & 0 & 0 \\
0 & 0 & -\sin\left[\frac{z}{\sqrt{3}}\right] & 0 & 0 & \cos\left[\frac{z}{\sqrt{3}}\right] & 0 \\
0 & \sin\left[\frac{z}{\sqrt{3}}\right] & 0 & 0 & 0 & 0 & \cos\left[\frac{z}{\sqrt{3}}\right]
\end{pmatrix},$$

$$\begin{pmatrix}
\cos\left[\frac{2z}{\sqrt{3}}\right] & 0 & 0 & 0 & -\sin\left[\frac{2z}{\sqrt{3}}\right] & 0 & 0 \\
0 & \cos\left[\frac{z}{\sqrt{3}}\right] & 0 & 0 & 0 & \sin\left[\frac{z}{\sqrt{3}}\right] & 0 \\
0 & 0 & \cos\left[\frac{z}{\sqrt{3}}\right] & 0 & 0 & 0 & \sin\left[\frac{z}{\sqrt{3}}\right] \\
0 & 0 & 0 & 1 & 0 & 0 & 0 \\
\sin\left[\frac{2z}{\sqrt{3}}\right] & 0 & 0 & 0 & \cos\left[\frac{2z}{\sqrt{3}}\right] & 0 & 0 \\
0 & -\sin\left[\frac{z}{\sqrt{3}}\right] & 0 & 0 & 0 & \cos\left[\frac{z}{\sqrt{3}}\right] & 0 \\
0 & 0 & -\sin\left[\frac{z}{\sqrt{3}}\right] & 0 & 0 & 0 & \cos\left[\frac{z}{\sqrt{3}}\right]
\end{pmatrix},$$

$$\begin{pmatrix} \cos\left[\frac{2z}{\sqrt{3}}\right] & 0 & 0 & 0 & 0 & -\sin\left[\frac{2z}{\sqrt{3}}\right] & 0 \\ 0 & \cos\left[\frac{z}{\sqrt{3}}\right] & 0 & 0 & -\sin\left[\frac{z}{\sqrt{3}}\right] & 0 & 0 \\ 0 & 0 & \cos\left[\frac{z}{\sqrt{3}}\right] & -\sin\left[\frac{z}{\sqrt{3}}\right] & 0 & 0 & 0 \\ 0 & 0 & \sin\left[\frac{z}{\sqrt{3}}\right] & \cos\left[\frac{z}{\sqrt{3}}\right] & 0 & 0 & 0 \\ 0 & \sin\left[\frac{z}{\sqrt{3}}\right] & 0 & 0 & \cos\left[\frac{z}{\sqrt{3}}\right] & 0 & 0 \\ \sin\left[\frac{2z}{\sqrt{3}}\right] & 0 & 0 & 0 & 0 & \cos\left[\frac{2z}{\sqrt{3}}\right] & 0 \\ 0 & 0 & 0 & 0 & 0 & 0 & 1 \end{pmatrix},$$

$$\begin{pmatrix} \cos\left[\frac{2z}{\sqrt{3}}\right] & 0 & 0 & 0 & 0 & 0 & -\sin\left[\frac{2z}{\sqrt{3}}\right] \\ 0 & \cos\left[\frac{z}{\sqrt{3}}\right] & 0 & \sin\left[\frac{z}{\sqrt{3}}\right] & 0 & 0 & 0 \\ 0 & 0 & \cos\left[\frac{z}{\sqrt{3}}\right] & 0 & -\sin\left[\frac{z}{\sqrt{3}}\right] & 0 & 0 \\ 0 & -\sin\left[\frac{z}{\sqrt{3}}\right] & 0 & \cos\left[\frac{z}{\sqrt{3}}\right] & 0 & 0 & 0 \\ 0 & 0 & \sin\left[\frac{z}{\sqrt{3}}\right] & 0 & \cos\left[\frac{z}{\sqrt{3}}\right] & 0 & 0 \\ 0 & 0 & 0 & 0 & 0 & 1 & 0 \\ \sin\left[\frac{2z}{\sqrt{3}}\right] & 0 & 0 & 0 & 0 & 0 & \cos\left[\frac{2z}{\sqrt{3}}\right] \end{pmatrix} \}$$

Let's do exactly the same for Rarenas PhD [3] Lie Algebra $\mathfrak{g}_2$ :

```
Table[
 gexp = MatrixExp[z * gbasis2[[1]][[i]]];

 (* calculate the action on the vector product of two Octonions *)
 cross1 = FullSimplify[actionG2[gexp, OctCommutator[x, y] / 2]];

 (* Apply the action two each Octonion and then compute the vector product *)
 cross2 = FullSimplify[OctCommutator[actionG2[gexp, x], actionG2[gexp, y]] / 2];

 cross1 === cross2,

 {i, 1, 7}]
```

{True, True, True, True, True, True, True}

```
Table[
 gexp = MatrixExp[z * gbasis2[[2]][[i]]];

 (* calculate the action on the vector product of two Octonions *)
 cross1 = FullSimplify[actionG2[gexp, OctCommutator[x, y] / 2]];

 (* Apply the action two each Octonion and then compute the vector product *)
 cross2 = FullSimplify[OctCommutator[actionG2[gexp, x], actionG2[gexp, y]] / 2];

 cross1 === cross2,

 {i, 1, 7}]
```

{True, True, True, True, True, True, True}

$\mathfrak{g}_2$

And the following are the basis vectors for $\mathfrak{g}_2$ in Rarenas PhD [3]:

```
Table[ gbasis2[[1]][[i]] // MatrixForm, {i, 1, 7}]
Table[ gbasis2[[2]][[i]] // MatrixForm, {i, 1, 7}]
```

$$\left\{
\begin{pmatrix}
0 & 0 & 0 & 0 & 0 & 0 & 0\\
0 & 0 & 1 & 0 & 0 & 0 & 0\\
0 & -1 & 0 & 0 & 0 & 0 & 0\\
0 & 0 & 0 & 0 & -1 & 0 & 0\\
0 & 0 & 0 & 1 & 0 & 0 & 0\\
0 & 0 & 0 & 0 & 0 & 0 & 0\\
0 & 0 & 0 & 0 & 0 & 0 & 0
\end{pmatrix},
\begin{pmatrix}
0 & 0 & -1 & 0 & 0 & 0 & 0\\
0 & 0 & 0 & 0 & 0 & 0 & 0\\
1 & 0 & 0 & 0 & 0 & 0 & 0\\
0 & 0 & 0 & 0 & 0 & -1 & 0\\
0 & 0 & 0 & 0 & 0 & 0 & 0\\
0 & 0 & 1 & 0 & 0 & 0 & 0\\
0 & 0 & 0 & 0 & 0 & 0 & 0
\end{pmatrix},
\begin{pmatrix}
0 & 1 & 0 & 0 & 0 & 0 & 0\\
-1 & 0 & 0 & 0 & 0 & 0 & 0\\
0 & 0 & 0 & 0 & 0 & 0 & 0\\
0 & 0 & 0 & 0 & 0 & 0 & -1\\
0 & 0 & 0 & 0 & 0 & 0 & 0\\
0 & 0 & 0 & 0 & 0 & 0 & 0\\
0 & 0 & 0 & 1 & 0 & 0 & 0
\end{pmatrix},
\begin{pmatrix}
0 & 0 & 0 & 0 & -1 & 0 & 0\\
0 & 0 & 0 & 0 & 0 & 1 & 0\\
0 & 0 & 0 & 0 & 0 & 0 & 0\\
0 & 0 & 0 & 0 & 0 & 0 & 0\\
1 & 0 & 0 & 0 & 0 & 0 & 0\\
0 & -1 & 0 & 0 & 0 & 0 & 0\\
0 & 0 & 0 & 0 & 0 & 0 & 0
\end{pmatrix},
\right.$$

$$\begin{pmatrix}
0 & 0 & 0 & 1 & 0 & 0 & 0\\
0 & 0 & 0 & 0 & 0 & 0 & 1\\
0 & 0 & 0 & 0 & 0 & 0 & 0\\
-1 & 0 & 0 & 0 & 0 & 0 & 0\\
0 & 0 & 0 & 0 & 0 & 0 & 0\\
0 & 0 & 0 & 0 & 0 & 0 & 0\\
0 & -1 & 0 & 0 & 0 & 0 & 0
\end{pmatrix},
\begin{pmatrix}
0 & 0 & 0 & 0 & 0 & 0 & -1\\
0 & 0 & 0 & 1 & 0 & 0 & 0\\
0 & 0 & 0 & 0 & 0 & 0 & 0\\
0 & -1 & 0 & 0 & 0 & 0 & 0\\
0 & 0 & 0 & 0 & 0 & 0 & 0\\
0 & 0 & 0 & 0 & 0 & 0 & 0\\
1 & 0 & 0 & 0 & 0 & 0 & 0
\end{pmatrix},
\begin{pmatrix}
0 & 0 & 0 & 0 & 0 & 1 & 0\\
0 & 0 & 0 & 0 & 1 & 0 & 0\\
0 & 0 & 0 & 0 & 0 & 0 & 0\\
0 & 0 & 0 & 0 & 0 & 0 & 0\\
0 & -1 & 0 & 0 & 0 & 0 & 0\\
-1 & 0 & 0 & 0 & 0 & 0 & 0\\
0 & 0 & 0 & 0 & 0 & 0 & 0
\end{pmatrix}
\left.\vphantom{\begin{pmatrix}0\\0\\0\\0\\0\\0\\0\end{pmatrix}}\right\}$$

$$\left\{
\begin{pmatrix}
0 & 0 & 0 & 0 & 0 & 0 & 0\\
0 & 0 & 0 & 0 & 0 & 0 & 0\\
0 & 0 & 0 & 0 & 0 & 0 & 0\\
0 & 0 & 0 & 0 & 1 & 0 & 0\\
0 & 0 & 0 & -1 & 0 & 0 & 0\\
0 & 0 & 0 & 0 & 0 & 0 & 1\\
0 & 0 & 0 & 0 & 0 & -1 & 0
\end{pmatrix},
\begin{pmatrix}
0 & 0 & 0 & 0 & 0 & 0 & 0\\
0 & 0 & 0 & 0 & 0 & 0 & 0\\
0 & 0 & 0 & 0 & 0 & 0 & 0\\
0 & 0 & 0 & 0 & 0 & 1 & 0\\
0 & 0 & 0 & 0 & 0 & 0 & -1\\
0 & 0 & 0 & -1 & 0 & 0 & 0\\
0 & 0 & 0 & 0 & 1 & 0 & 0
\end{pmatrix},
\begin{pmatrix}
0 & 0 & 0 & 0 & 0 & 0 & 0\\
0 & 0 & 0 & 0 & 0 & 0 & 0\\
0 & 0 & 0 & 0 & 0 & 0 & 0\\
0 & 0 & 0 & 0 & 0 & 0 & 1\\
0 & 0 & 0 & 0 & 1 & 0 & 0\\
0 & 0 & 0 & 0 & 0 & 1 & 0\\
0 & 0 & 0 & -1 & 0 & 0 & 0
\end{pmatrix},
\begin{pmatrix}
0 & 0 & 0 & 0 & 0 & 0 & 0\\
0 & 0 & 0 & 0 & 0 & -1 & 0\\
0 & 0 & 0 & 0 & 0 & 0 & 1\\
0 & 0 & 0 & 0 & 0 & 0 & 0\\
0 & 0 & 0 & 0 & 0 & 0 & 0\\
0 & 1 & 0 & 0 & 0 & 0 & 0\\
0 & 0 & -1 & 0 & 0 & 0 & 0
\end{pmatrix},
\right.$$

$$\begin{pmatrix}
0 & 0 & 0 & 0 & 0 & 0 & 0\\
0 & 0 & 0 & 0 & 0 & 0 & 1\\
0 & 0 & 0 & 0 & 0 & 1 & 0\\
0 & 0 & 0 & 0 & 0 & 0 & 0\\
0 & 0 & 0 & 0 & 0 & 0 & 0\\
0 & 0 & -1 & 0 & 0 & 0 & 0\\
0 & -1 & 0 & 0 & 0 & 0 & 0
\end{pmatrix},
\begin{pmatrix}
0 & 0 & 0 & 0 & 0 & 0 & 1\\
0 & 0 & 0 & 0 & 0 & 0 & 0\\
0 & 0 & 0 & 0 & 1 & 0 & 0\\
0 & 0 & 0 & 0 & 0 & 0 & 0\\
0 & 0 & -1 & 0 & 0 & 0 & 0\\
0 & 0 & 0 & 0 & 0 & 0 & 0\\
-1 & 0 & 0 & 0 & 0 & 0 & 0
\end{pmatrix},
\begin{pmatrix}
0 & 0 & 0 & 0 & 0 & 0 & 0\\
0 & 0 & 0 & 0 & -1 & 0 & 0\\
0 & 0 & 0 & 1 & 0 & 0 & 0\\
0 & 0 & -1 & 0 & 0 & 0 & 0\\
0 & 1 & 0 & 0 & 0 & 0 & 0\\
0 & 0 & 0 & 0 & 0 & 0 & 0\\
0 & 0 & 0 & 0 & 0 & 0 & 0
\end{pmatrix}
\left.\vphantom{\begin{pmatrix}0\\0\\0\\0\\0\\0\\0\end{pmatrix}}\right\}$$

Exp of Rarenas PhD's $\mathfrak{g}_2$ basis [3]:

```
Table[ MatrixExp[z * gbasis2[[1]][[i]]] // MatrixForm, {i, 1, 7}]
```

$$
\left\{
\begin{pmatrix}
1 & 0 & 0 & 0 & 0 & 0 & 0 \\
0 & \cos[z] & \sin[z] & 0 & 0 & 0 & 0 \\
0 & -\sin[z] & \cos[z] & 0 & 0 & 0 & 0 \\
0 & 0 & 0 & \cos[z] & -\sin[z] & 0 & 0 \\
0 & 0 & 0 & \sin[z] & \cos[z] & 0 & 0 \\
0 & 0 & 0 & 0 & 0 & 1 & 0 \\
0 & 0 & 0 & 0 & 0 & 0 & 1
\end{pmatrix},
\right.
$$

$$
\begin{pmatrix}
\cos[z] & 0 & -\sin[z] & 0 & 0 & 0 & 0 \\
0 & 1 & 0 & 0 & 0 & 0 & 0 \\
\sin[z] & 0 & \cos[z] & 0 & 0 & 0 & 0 \\
0 & 0 & 0 & \cos[z] & 0 & -\sin[z] & 0 \\
0 & 0 & 0 & 0 & 1 & 0 & 0 \\
0 & 0 & 0 & \sin[z] & 0 & \cos[z] & 0 \\
0 & 0 & 0 & 0 & 0 & 0 & 1
\end{pmatrix},
\begin{pmatrix}
\cos[z] & \sin[z] & 0 & 0 & 0 & 0 & 0 \\
-\sin[z] & \cos[z] & 0 & 0 & 0 & 0 & 0 \\
0 & 0 & 1 & 0 & 0 & 0 & 0 \\
0 & 0 & 0 & \cos[z] & 0 & 0 & -\sin[z] \\
0 & 0 & 0 & 0 & 1 & 0 & 0 \\
0 & 0 & 0 & 0 & 0 & 1 & 0 \\
0 & 0 & 0 & \sin[z] & 0 & 0 & \cos[z]
\end{pmatrix},
$$

$$
\begin{pmatrix}
\cos[z] & 0 & 0 & 0 & -\sin[z] & 0 & 0 \\
0 & \cos[z] & 0 & 0 & 0 & \sin[z] & 0 \\
0 & 0 & 1 & 0 & 0 & 0 & 0 \\
0 & 0 & 0 & 1 & 0 & 0 & 0 \\
\sin[z] & 0 & 0 & 0 & \cos[z] & 0 & 0 \\
0 & -\sin[z] & 0 & 0 & 0 & \cos[z] & 0 \\
0 & 0 & 0 & 0 & 0 & 0 & 1
\end{pmatrix},
\begin{pmatrix}
\cos[z] & 0 & 0 & \sin[z] & 0 & 0 & 0 \\
0 & \cos[z] & 0 & 0 & 0 & 0 & \sin[z] \\
0 & 0 & 1 & 0 & 0 & 0 & 0 \\
-\sin[z] & 0 & 0 & \cos[z] & 0 & 0 & 0 \\
0 & 0 & 0 & 0 & 1 & 0 & 0 \\
0 & 0 & 0 & 0 & 0 & 1 & 0 \\
0 & -\sin[z] & 0 & 0 & 0 & 0 & \cos[z]
\end{pmatrix},
$$

$$
\begin{pmatrix}
\cos[z] & 0 & 0 & 0 & 0 & 0 & -\sin[z] \\
0 & \cos[z] & 0 & \sin[z] & 0 & 0 & 0 \\
0 & 0 & 1 & 0 & 0 & 0 & 0 \\
0 & -\sin[z] & 0 & \cos[z] & 0 & 0 & 0 \\
0 & 0 & 0 & 0 & 1 & 0 & 0 \\
0 & 0 & 0 & 0 & 0 & 1 & 0 \\
\sin[z] & 0 & 0 & 0 & 0 & 0 & \cos[z]
\end{pmatrix},
\left.
\begin{pmatrix}
\cos[z] & 0 & 0 & 0 & 0 & \sin[z] & 0 \\
0 & \cos[z] & 0 & 0 & \sin[z] & 0 & 0 \\
0 & 0 & 1 & 0 & 0 & 0 & 0 \\
0 & 0 & 0 & 1 & 0 & 0 & 0 \\
0 & -\sin[z] & 0 & 0 & \cos[z] & 0 & 0 \\
-\sin[z] & 0 & 0 & 0 & 0 & \cos[z] & 0 \\
0 & 0 & 0 & 0 & 0 & 0 & 1
\end{pmatrix}
\right\}
$$

```
Table[ MatrixExp[z * gbasis2[[2]][[i]]] // MatrixForm, {i, 1, 7}]
```

$$
\left\{
\begin{pmatrix}
1 & 0 & 0 & 0 & 0 & 0 & 0 \\
0 & 1 & 0 & 0 & 0 & 0 & 0 \\
0 & 0 & 1 & 0 & 0 & 0 & 0 \\
0 & 0 & 0 & \text{Cos}[z] & \text{Sin}[z] & 0 & 0 \\
0 & 0 & 0 & -\text{Sin}[z] & \text{Cos}[z] & 0 & 0 \\
0 & 0 & 0 & 0 & 0 & \text{Cos}[z] & \text{Sin}[z] \\
0 & 0 & 0 & 0 & 0 & -\text{Sin}[z] & \text{Cos}[z]
\end{pmatrix},
\right.
$$

$$
\begin{pmatrix}
1 & 0 & 0 & 0 & 0 & 0 & 0 \\
0 & 1 & 0 & 0 & 0 & 0 & 0 \\
0 & 0 & 1 & 0 & 0 & 0 & 0 \\
0 & 0 & 0 & \text{Cos}[z] & 0 & \text{Sin}[z] & 0 \\
0 & 0 & 0 & 0 & \text{Cos}[z] & 0 & -\text{Sin}[z] \\
0 & 0 & 0 & -\text{Sin}[z] & 0 & \text{Cos}[z] & 0 \\
0 & 0 & 0 & 0 & \text{Sin}[z] & 0 & \text{Cos}[z]
\end{pmatrix},
\begin{pmatrix}
1 & 0 & 0 & 0 & 0 & 0 & 0 \\
0 & 1 & 0 & 0 & 0 & 0 & 0 \\
0 & 0 & 1 & 0 & 0 & 0 & 0 \\
0 & 0 & 0 & \text{Cos}[z] & 0 & 0 & \text{Sin}[z] \\
0 & 0 & 0 & 0 & \text{Cos}[z] & \text{Sin}[z] & 0 \\
0 & 0 & 0 & 0 & -\text{Sin}[z] & \text{Cos}[z] & 0 \\
0 & 0 & 0 & -\text{Sin}[z] & 0 & 0 & \text{Cos}[z]
\end{pmatrix},
$$

$$
\begin{pmatrix}
1 & 0 & 0 & 0 & 0 & 0 & 0 \\
0 & \text{Cos}[z] & 0 & 0 & 0 & -\text{Sin}[z] & 0 \\
0 & 0 & \text{Cos}[z] & 0 & 0 & 0 & \text{Sin}[z] \\
0 & 0 & 0 & 1 & 0 & 0 & 0 \\
0 & 0 & 0 & 0 & 1 & 0 & 0 \\
0 & \text{Sin}[z] & 0 & 0 & 0 & \text{Cos}[z] & 0 \\
0 & 0 & -\text{Sin}[z] & 0 & 0 & 0 & \text{Cos}[z]
\end{pmatrix},
\begin{pmatrix}
1 & 0 & 0 & 0 & 0 & 0 & 0 \\
0 & \text{Cos}[z] & 0 & 0 & 0 & 0 & \text{Sin}[z] \\
0 & 0 & \text{Cos}[z] & 0 & \text{Sin}[z] & 0 & 0 \\
0 & 0 & 0 & 1 & 0 & 0 & 0 \\
0 & 0 & 0 & 0 & 1 & 0 & 0 \\
0 & 0 & -\text{Sin}[z] & 0 & 0 & \text{Cos}[z] & 0 \\
0 & -\text{Sin}[z] & 0 & 0 & 0 & 0 & \text{Cos}[z]
\end{pmatrix},
$$

$$
\begin{pmatrix}
\text{Cos}[z] & 0 & 0 & 0 & 0 & 0 & \text{Sin}[z] \\
0 & 1 & 0 & 0 & 0 & 0 & 0 \\
0 & 0 & \text{Cos}[z] & 0 & \text{Sin}[z] & 0 & 0 \\
0 & 0 & 0 & 1 & 0 & 0 & 0 \\
0 & 0 & -\text{Sin}[z] & 0 & \text{Cos}[z] & 0 & 0 \\
0 & 0 & 0 & 0 & 0 & 1 & 0 \\
-\text{Sin}[z] & 0 & 0 & 0 & 0 & 0 & \text{Cos}[z]
\end{pmatrix},
\begin{pmatrix}
1 & 0 & 0 & 0 & 0 & 0 & 0 \\
0 & \text{Cos}[z] & 0 & 0 & -\text{Sin}[z] & 0 & 0 \\
0 & 0 & \text{Cos}[z] & \text{Sin}[z] & 0 & 0 & 0 \\
0 & 0 & -\text{Sin}[z] & \text{Cos}[z] & 0 & 0 & 0 \\
0 & \text{Sin}[z] & 0 & 0 & \text{Cos}[z] & 0 & 0 \\
0 & 0 & 0 & 0 & 0 & 1 & 0 \\
0 & 0 & 0 & 0 & 0 & 0 & 1
\end{pmatrix}
\right\}
$$

# Appendix B

We assume all strings and programs are binary coded.

**Definition B.1**: The Kolmogorov Complexity $C_\mathcal{U}(x)$ of a string x with respect to a universal computer (Turing Machine) $\mathcal{U}$ is defined as

$$C_\mathcal{U}(x) = \min_{p:\mathcal{U}(p)=x} l(p)$$

the minimum length program p in $\mathcal{U}$ which outputs x.

**Theorem B.1 (Universality of the Kolmogorov Complexity)**: *If $\mathcal{U}$ is a universal computer, then for any other computer $\mathcal{A}$ and all strings x,*

$$C_\mathcal{U}(x) \le C_\mathcal{A}(x) + c_\mathcal{A}$$

where the constant $c_\mathcal{A}$ does not depend on x.

**Corollary B.1**: $\lim_{l(x)\to\infty} \frac{C_\mathcal{U}(x) - C_\mathcal{A}(x)}{l(x)} = 0$ *for any two universal computers.*

**Remark B.1**: *Therefore we drop the universal computer subscript and simply write C(x).*

**Definition B.2**: Self-delimiting string (or program) is a string or program which has its own length encoded as a part of itself i.e. a Turing machine reading Self-delimiting string knows exactly when to stop reading.

**Definition B.3**: The Conditional or Prefix Kolmogorov Complexity of self-delimiting string x given string y is

$$K(x \mid y) = \min_{p:\mathcal{U}(p,y)=x} l(p)$$

The length of the shortest program that can compute both x and y and a way to tell them apart is

$$K(x, y) = \min_{p:\mathcal{U}(p)=x,y} l(p)$$

**Remark B.2**: *x, y can be thought of as concatenation of the strings with additional separation information.*

**Theorem B.2**: $K(x) \le l(x) + 2\log l(x) + O(1), \quad K(x \mid l(x)) \le l(x) + O(1)$.

**Theorem B.3**: $K(x, y) \le K(x) + K(y)$.

**Theorem B.4**: $K(f(x)) \le K(x) + K(f)$ , $f$ a computble function